\documentclass[letterpaper]{JHEP3}
\usepackage{nicefrac}

\usepackage{amsmath,epsfig}
\usepackage{amssymb,amsfonts}

\usepackage{graphics}
\usepackage{epsfig}
\usepackage{amsfonts}

\makeatletter

\title{Diagrams for Symmetric Product Orbifolds}

\preprint{Brown-HET-1573 \\ YITP-SB-09-11}

\author{
Ari Pakman\footnote{Email: ari$\_$pakman@brown.edu}$^{~1}$, Leonardo Rastelli\footnote{Email: leonardo.rastelli@stonybrook.edu}$^{~2}$, and
Shlomo S. Razamat\footnote{Email: razamat@max2.physics.sunysb.edu}$^{~2}$
\\ \\ \\
\it $^1$ Department of Physics,\\ Brown University,\\
Providence, RI 02912, USA
\\
\\
\it $^2$ C.N. Yang Institute for Theoretical Physics,\\
\it Stony Brook University, \\
\it Stony Brook, NY 11794-3840, USA}

\abstract{

\medskip

We develop a diagrammatic language for
symmetric product orbifolds of two-dimensional conformal field theories. 
Correlation functions of twist operators are written as sums of diagrams:
each diagram corresponds to a branched covering map from a surface where the fields
are single-valued to the base sphere where twist operators are inserted. This
diagrammatic language facilitates the study of the large $N$ limit and  makes more transparent the analogy
between  symmetric product orbifolds and {\it free} non-abelian gauge theories.
We give a general algorithm to calculate
  the leading large $N$ contribution  to four-point correlators of twist fields.
}

\def\d{\partial}

\def\s{\sigma}

\def\a{\alpha}

\def\h{\eta}

\def\half{{\frac12}}

\def\IC{\relax\hbox{$\inbar\kern-.3em{\rm C}$}}

\def\IC{{\bf C}}

\def\bea{\begin{eqnarray}}
\def\eea{\end{eqnarray}}
\def\be{\begin{equation}}
\def\ee{\end{equation}}
\def\ba{\begin{align}}
\def\ea{\end{align}}
\def\bse{\begin{subequations}}
\def\ese{\end{subequations}}
\def\1F1{{}_1\!F_1}
\def\2F0{{}_2\!F_0}

\def\a{\alpha}

\def\h3{$\textrm{H}_3^+$}

\def\d{{\partial}}

\def\IC{{\mathbb C}}

\def\lbldef#1#2{\expandafter\gdef\csname #1\endcsname {#2}}

\def\href#1#2{#2}

\newcommand{\beq}{\begin{equation}}
\newcommand{\eeq}{\end{equation}}
\newcommand{\ber}{\begin{eqnarray}}
\newcommand{\eer}{\end{eqnarray}}

\def\be{\begin{eqnarray}}
\def\ee{\end{eqnarray}}

\def\({\left(}
\def\){\right)}
\def\[{\left[}
\def\]{\right]}
\def\<{\langle}
\def\>{\rangle}
\def\d{\partial}

\def\gg{\mathsf g }

\keywords{CFT, Large N, AdS/CFT}

\begin{document}

\section{Introduction}

Symmetric product orbifolds
are ubiquitous  in theoretical physics. They arise in the  ``second-quantization'' of a configuration space --
the procedure of forming  products of identical copies of the space, and
imposing equivalence under permutation of the copies.
Symmetric product orbifolds of two-dimensional conformal field theories (CFTs) ~\cite{Klemm:1990df}
appear in  many related contexts:
as instanton moduli spaces~\cite{Vafa:1994tf}, in the counting problem
of  black hole microstates~\cite{Strominger:1996sh}, in matrix string theory \cite{Dijkgraaf:1996xw,Motl:1997th,Dijkgraaf:1997vv,Dijkgraaf:1998zd, Cove:2007kz},  in the AdS$_3$/CFT$_2$ correspondence~\cite{Maldacena:1997re,deBoer:1998ip,Dijkgraaf:1998gf,Seiberg:1999xz,
Larsen:1999uk,Lunin:2002fw,Gomis:2002qi,Gava:2002xb,
Gaberdiel:2007vu, Dabholkar:2007ey, Pakman:2007hn, Taylor:2007hs,
David:2008yk}. 
See also \cite{Dijkgraaf:1999za,Bantay:1999us,Halpern:2007vj,Recknagel:2002qq} for more examples.
The calculation of correlation functions in symmetric orbifold CFTs has been discussed before;
a partial list of references includes~\cite{Arutyunov:1997gt,Arutyunov:1997gi, Jevicki:1998bm, Lunin:2000yv,Lunin:2001pw}.

Our main motivation to reconsider the subject
comes from  the holographic
correspondence~\cite{Maldacena:1997re}.
 The field theory dual to IIB on $AdS_3 \times S^3 \times {\cal M}_4$  (with $ {\cal M}_4$ hyperk\"ahler)
is the symmetric orbifold of $N$ copies of the  $2d$ sigma model with ${\cal M}_4$ target.
This  is to be contrasted with the AdS$_5$/CFT$_4$ instance
of the duality, where the conformal field theory is an ordinary gauge theory. An intuitive picture of how the gauge/string duality arises 
for ordinary gauge theories is due to 't Hooft \cite{'tHooft:1973jz}, and is based on a simple topological
analysis of Feynman diagrams:  the large $N$ expansion
of a gauge theory can be viewed as the perturbative expansion of a dual closed string theory, with coupling $g_s \sim 1/N$.
In this work we will  
gain a similar understanding for
symmetric orbifolds. The basic intuition was provided by Lunin and Mathur \cite{Lunin:2000yv,Lunin:2001pw}, who observed
that correlation functions of twist operators admit a genus expansion, since they can be evaluated on the covering surface(s) where the fields are single-valued: 
the genus of the covering surfaces
controls the large $N$ counting, albeit in a more complicated way than for $U(N)$ gauge theories.
In this  paper we  make their observations systematic, 
by defining a diagrammatic expansion for symmetric product orbifolds akin to the usual Feynman diagram expansion of gauge theories.

The simplest symmetric product orbifolds  are obtained by taking $N$ copies of a free conformal theory and gauging the $S_N$ symmetry.
Let us contrast such simple orbifolds with the free field limit of conventional gauge theories.
The projection onto $S_N$ invariant states is analogous
to the projection onto gauge-invariant states (Gauss law constraint) that one must perform
even in  a free gauge theory. However the orbifold theory also contains
twisted sectors, and calculations involving   twist operators appear
at first qualitatively different from calculations in a free gauge theory.
In a free gauge theory any correlator of gauge invariant composite operators is evaluated as a sum of a finite number of Feynman diagrams, with
each diagram given (in position space) by  a simple product of propagators.  
A correlator of twist operators is considerably more involved.
It  can be evaluated by going to the covering surface, where it reduces to a vacuum partition function, but
determining explicitly the covering map is  a non-trivial task.
Nevertheless, as we show in this paper, it is still possible
to regard a correlator of twist operators as a finite sum of appropriate {\it diagrams}.
The diagrammatic language that we introduce  makes the structure of correlators
more intuitive and  the analogy with free gauge theories more transparent.

The genus expansion of correlators in ${\rm Sym}^N {\cal M}_4$ parallels the genus expansion
of the dual IIB string theory on $AdS_3 \times S^3 \times {\cal M}_4$. It is tempting to identify
the auxiliary covering surfaces that enter the calculation of orbifold correlators with the worldsheets
of the dual closed string theory.  On the orbifold side, a correlator receives contributions
from a finite number of covering Riemann surfaces, with {\it fixed} complex structure (function of the spacetime positions
of the twist operators).
On the string theory side, we are instead instructed to {\it integrate} over complex structures.
Recall that the orbifold point  is dual to a strongly coupled, possibly topological point\footnote{
See~\cite{Berkovits:2008ga} for a concrete suggestion in the $AdS_5/CFT_4$ case.}  of the string theory moduli space.
 We speculate that at this special point
  the integration over worldsheet moduli localizes to the discrete set of surfaces seen on the orbifold side, perhaps by a mechanism similar
 to the ones at work in topological \cite{Distler:1989ax, Verlinde:1990ku} and in minimal \cite{Belavin:2006ex} string theories.

There is an extensive  mathematical literature on associating graphs to the enumeration
of branched covering maps between Riemann surfaces (the so-called Hurwitz problem),
 see {\it e.g}.~\cite{Lando:2003gx,Lando2,Okounkov:2000gx,Mironov:2009cj}.  
We found it more useful to develop from first principles a diagrammatic
language  designed for  concrete applications to CFT, and did not attempt  to connect
in detail our graphic construction with those of \cite{Lando:2003gx,Lando2,Okounkov:2000gx,Mironov:2009cj},
though undoubtedly connections exist.

A brief outline of the paper is as follows.
 In Section \ref{symmprodCFT} we show that a correlator of twist operators can be expanded
as a sum over different terms, which can be grouped into equivalence classes, and we develop a method to associate a diagram to each class in the expansion.
For each correlator the number of diagrams is finite.  This number
is a (generalized) Hurwitz number\footnote{
Hurwitz numbers have appeared before in another duality between a $2d$ theory ($2d$ pure Yang-Mills)
and
a string theory \cite{Gross:1992tu,Gross:1993hu,Gross:1993yt} (see \cite{Cordes:1994fc} for a review). 
}:
each diagram corresponds to a branched covering 
map from a Riemann surface where the fields are single-valued to the
the $2d$ space-time of the CFT (assumed to be a sphere) where the twist operators are inserted.
We show how to compute the $N$ dependence of a generic correlator.
The actual computation of the correlators needs an explicit knowledge of the covering maps.
In Section 3  we present a method to obtain the genus zero covering 
 map for general four-point functions. 
We  illustrate in some examples how the map encodes  the ideas of Section \ref{symmprodCFT}.
We conclude with a speculation: the covering surfaces that appear in the calculation
of symmetric product orbifold correlators should be identified with the worldsheets of the dual string theory formulation.
Two Appendices contain technical results and connections to previous work \cite{Lunin:2000yv} on four-point functions.

Some explicit calculations of extremal correlators in the
 $(4,4)$ superconformal symmetric product orbifold of $T^4$, which is dual to type IIB string theory on $AdS_3 \times S^3 \times T^4$~\cite{Maldacena:1997re}, 
 reveal a very direct  connection with the Hurwitz problem. They  are reported in a companion paper~\cite{PRR2}.


\section{Diagrams}\label{symmprodCFT}

We begin by recalling basic facts and notations about symmetric orbifold conformal field theories.
For definiteness, consider a sigma model of the form
\be\label{bosaction}
S=\frac{1}{2\pi}\int d\s d\tau\,  G_{ij} (X) \left(\d_\s X^i \d_\s X^j -\d_\tau X^i \d_\tau X^j\right) + \dots \, ,
\ee
where $X^i$, $i\in \{ 1,\dots, D \}$, are the coordinates of the target space manifold ${\cal M}$, with metric $G_{ij}(X)$,
and the dots indicate possible fermionic terms.
We assume that the sigma model is conformal invariant at the quantum level. Important special cases are
 ${\cal M} = T^4$, $K3$, and $\mathbb{R}^8$, when the theory (with the appropriate fermionic completion) is in fact $(4, 4)$ superconformal.
The symmetric orbifold CFT   is defined by
 taking $N$ copies of the target space ${\cal M}$,  identified up to permutations,
\be
{\rm Sym}^N({\cal M}) \equiv \otimes^N {\cal M} / {S_N} \, .
\ee
Concretely, we endow the coordinates
with an extra ``color'' index    $I\in \{1,\dots, N \}$ to label the different  {copies}, and
 make the orbifold  identification
\be\label{inv}
\, X^i_I \cong X^i_{h(I)} \qquad \forall h\in S_N \,.
\ee
 The internal structure of the manifold ${\cal M}$ will play little role in the following,
and we will often omit the index $i$. Indeed  our general considerations would apply  to the symmetric
product orbifold of an abstract CFT with no geometric interpretation.

The orbifold theory has twisted sectors with boundary conditions
\be\label{trans}
X^i_I(\s+2\pi)=\,X^i_{g(I)} \,.
\ee
From (\ref{inv}) and (\ref{trans}) it follows
 that each twisted sector corresponds to a conjugacy class  $[g]$ of the symmetric group.
 The twist field $\sigma_{[g]} (z)$, defined  in the $z$ plane $z=\exp(\tau+i \s)$,
 is  the local operator associated (in the state/operator correspondence) to
  the twisted sector vacuum labeled by $[g]$.
 Let us first introduce twist operators $\s_g(z)$, labeled by individual elements $g$ of $S_N$, 
such that the fields $X_I$ have monodromy
\be
X^i_I(e^{2\pi i}\,z)\s_g(0)= X^i_{g(I)}( z)\s_g(0) \,.
\ee
Clearly the $\s_g$'s are {\it not} $S_N$--invariant.  The proper 
``gauge-invariant''  twist field $\s_{[g]}$, labeled by a conjugacy class,
is obtained by  summing over the group orbit,
\be  \label{gaugeinvariant}
\s_{[g]} \equiv
\sum_{h\in S(N)}\s_{h^{-1}gh} \, .
\ee
We are  interested in correlators of gauge-invariant twist operators,
\be\label{corr}
\langle\prod_{j=1}^{s} \s_{[g_j]} (z_j, \bar z_j) \rangle=
\langle \prod_{j=1}^s \, \sum_{h_j\in S(N)} \s_{h_jg_jh_j^{-1}} (z_j, \bar z_j) \rangle \, .
\ee
Their computation is reduced to evaluating individual {\it terms} of the form
\be \label{term}
\langle   \s_{\hat g_1}  (z_1, \bar z_1)  \dots \s_{\hat g_s} (z_s, \bar z_s)\rangle \, ,
\ee
where we have set $\hat g_j=h_jg_jh_j^{-1}$. We will restrict to correlators
defined on the plane, or Riemann sphere (henceforth the {\it base sphere} $S^2_{base}$).
 In the operator formalism  correlators on the sphere are written as radial ordered vacuum expectation values.
We can always assume (by renaming  the coordinates if needed) that $|z_1 |  \leq |z_2|  \leq  \dots \leq |z_s|$. 
Then
\be \label{termradial}
\langle   \s_{\hat g_1}  (z_1, \bar z_1)  \dots \s_{\hat g_s} (z_s, \bar z_s)\rangle  = 
\langle 0 |  \s_{\hat g_s}  (z_s, \bar z_s)  \dots \s_{\hat g_1} (z_1, \bar z_1)  | 0 \rangle \,.
\ee
Each term is specified by an ordered sequence $(\hat g_1 \dots \hat g_s)$
of  $s$ group elements of $S_N$, with the ordering dictated by the radial
ordering of the coordinates. A necessary condition
for (\ref{termradial}) to contribute to (\ref{corr}) is that 
\be\label{nontrivcond}
\hat g_1 \hat g_2 \dots \hat g_s = 1 \,.
\ee
A sequence $(\hat g_1 \dots \hat g_s)$ and the corresponding term (\ref{termradial}) will be called {\it non-trivial} if this condition is obeyed.
From now on, all sequences will be assumed to be non-trivial.

Two ordered sequences of $n$ group elements of $S_N$ are said to be {\it equivalent}
if they are related by a global $S_N$ transformation,
\be \label{equrelation}
 ( \hat g_1    \dots \hat g_n  )  \sim   ( h \hat g_1 h^{-1}    \dots  h \hat g_n h^{-1}  ) \,  ,
\ee
that is, by an overall relabeling of the color indices.
Terms (\ref{term}) specified by equivalent sequences are numerically equal, so it is sufficient to evaluate a representative for each class
and multiply by the number of elements in the class.

Our goal is to associate a {\it diagram} to each non-trivial equivalence {\it class},
and to regard the  gauge-invariant correlator (\ref{corr}) as a sum of such diagrams.

Henceforth we shall restrict to   twist fields $\sigma_{[g]}$
corresponding to {\it single-cycle} permutations. Recall that
 each element $g \in S_N$ is the   product of mutually commuting cyclic permutations. The number $N_k$
 of cyclic permutations of length $k$ is the same for each element
 of a  conjugacy class $[g]$. Conjugacy classes are thus in correspondence with
 {\it partitions}  of $N$, ordered sequences 
 of non-negative integers $(N_1 \dots N_N )$ obeying
 \be
 \sum_{k=1}^N k \, N_k = N \,.
 \ee
A single-cycle  permutation of length $s >1$ corresponds to $N_s = 1$, $N_1 =N-s$ and 
$N_k = 0$ for $k \neq 1,\, s$.  We will use the notation
\be
g = (i_1 \dots i_s) \, ,  \quad i_k \in \{1, 2, \dots, N \} \, ,
\ee
for the single-cycle permutation $i_1 \to i_2$, $i_2 \to i_3$, \dots $i_s \to i_1$.

The restriction to single-cycle permutations is not essential,  since
twist fields with more complicated cycle structure can be obtained by considering single-cycle twist fields at separated
points and taking the OPE limit.  Moreover, single-cycle permutations play a preferred
role in physical applications -- for instance, they are associated to single-particle
states in the AdS$_3$/CFT$_2$ duality. By a slight abuse of notation, we denote
single-cycle twist operators with $\sigma_{[k]}$, where $k$ is the length of the cycle.
Recalling (\ref{gaugeinvariant}), we have 
\be
\sigma_{[k]} = \sum_{h \in S_N}     \sigma_{  ( h(1)   \dots h(k) )} \,.
\ee

\subsection{``Feynman'' rules}  \label{diagconst}

Let us consider a very simple example of correlator, 
\be \label{simpleexample}
\langle \sigma_{[3]} (z_a, \bar z_a) \sigma_{[2]} (z_b, \bar z_b) \sigma_{[2]} (z_c, \bar z_c ) \rangle\, .
\ee
We assume that $|z_a|  < |z_b |  < |z_c|$. We have\footnote{To avoid cluttering we will often drop the dependence on the coordinates,
implicit in the ordering of the operators (always $a, b, c, \dots$). Sometimes the dependence
will be indicated schematically, as in $\sigma_{[s]} (i) = \sigma_{[s]} (z_i , \bar z_i) $. Another short
hand  will be to drop  ``$\sigma$'' and write {\it e.g.} $(324)_i  \equiv \sigma_{(324)}(z_i, \bar z_i)  $. }
\be
\langle \sigma_{[3]} (a) \sigma_{[2]} (b) \sigma_{[2]} (c) \rangle  = 
\sum_{h_a, h_b, h_c \in S_N} \langle \s_{(h_a(1) h_a(2) h_a(3)  ) } \; \s_{(h_b(1) h_b(2)   ) }  \; \s_{(h_c(1) h_c(2)   ) }   \rangle \,.
\ee
Most of the terms do not contribute because the product of permutations (in the given ordering $a\, b \,c$) is different from the identity.
A non-trivial term is
\be
\langle \sigma_{(123)} (a)  \, \sigma_{(12)} (b) \,   \sigma_{(23)}(c)  \rangle  \equiv (123)_a \, (12)_b  (23)_c \, ,
\ee
where on  the right hand side we have introduced a convenient short-hand notation. There are three ``active'' colors (1, 2 and 3).
It is easy to see that this is in fact the only non-trivial term, up to renaming of the active colors (accomplished
by a global $S_N$ transformation (\ref{equrelation})).\footnote{
An example of equivalent term would be  $(352)_a \, (35)_b \, (52)_c$, 
obtained by relabeling $1 \to 3$, $2 \to 5$, $3\to 2$.  
} 
In this simple case there is only one equivalence class. We now give our prescription to draw the corresponding 
diagram.
\begin{figure}[htbp]
\begin{center}
$\begin{array}{c@{\hspace{0.55in}}c}
  \epsfig{file=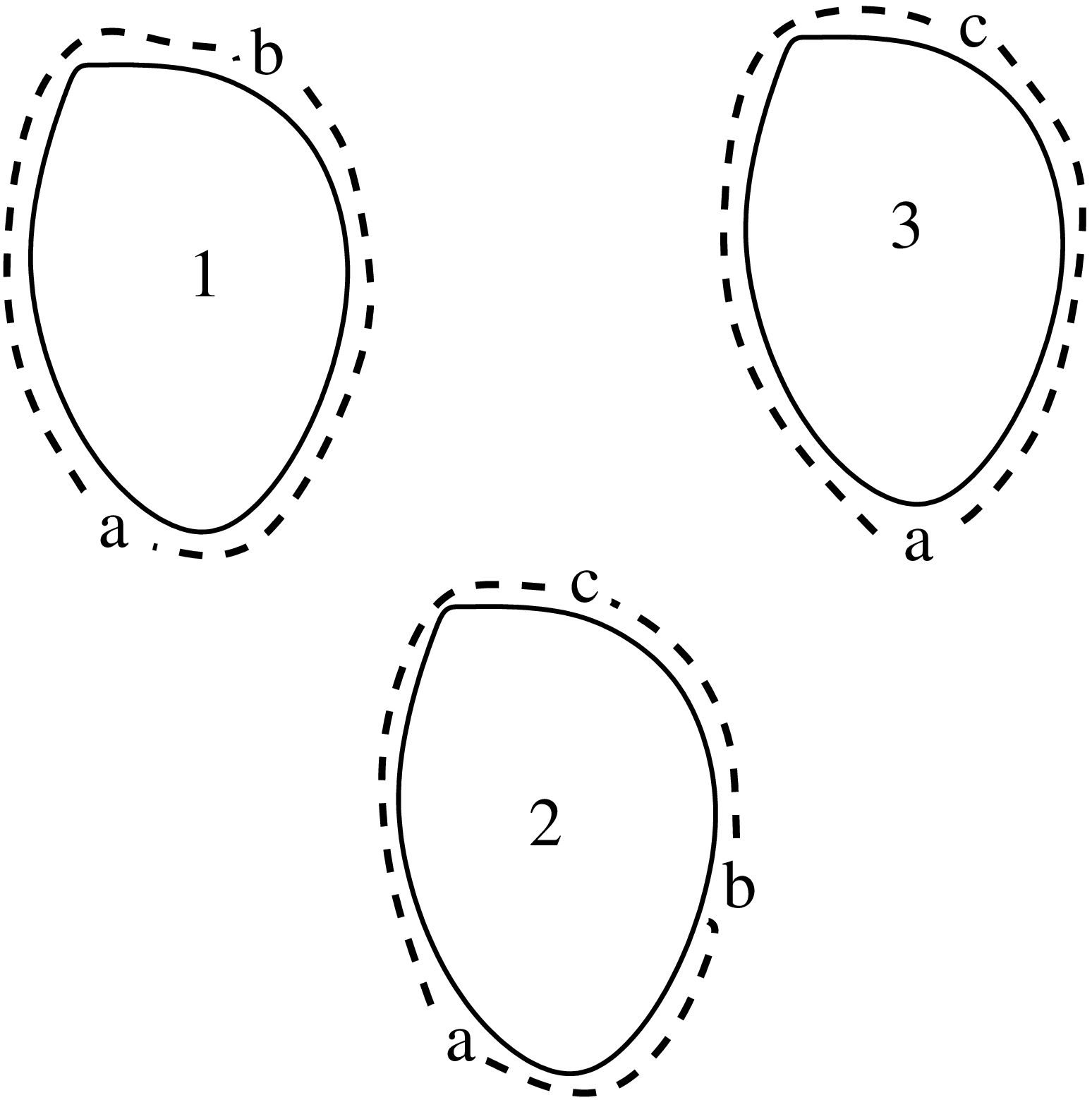,scale=0.35}& \epsfig{file=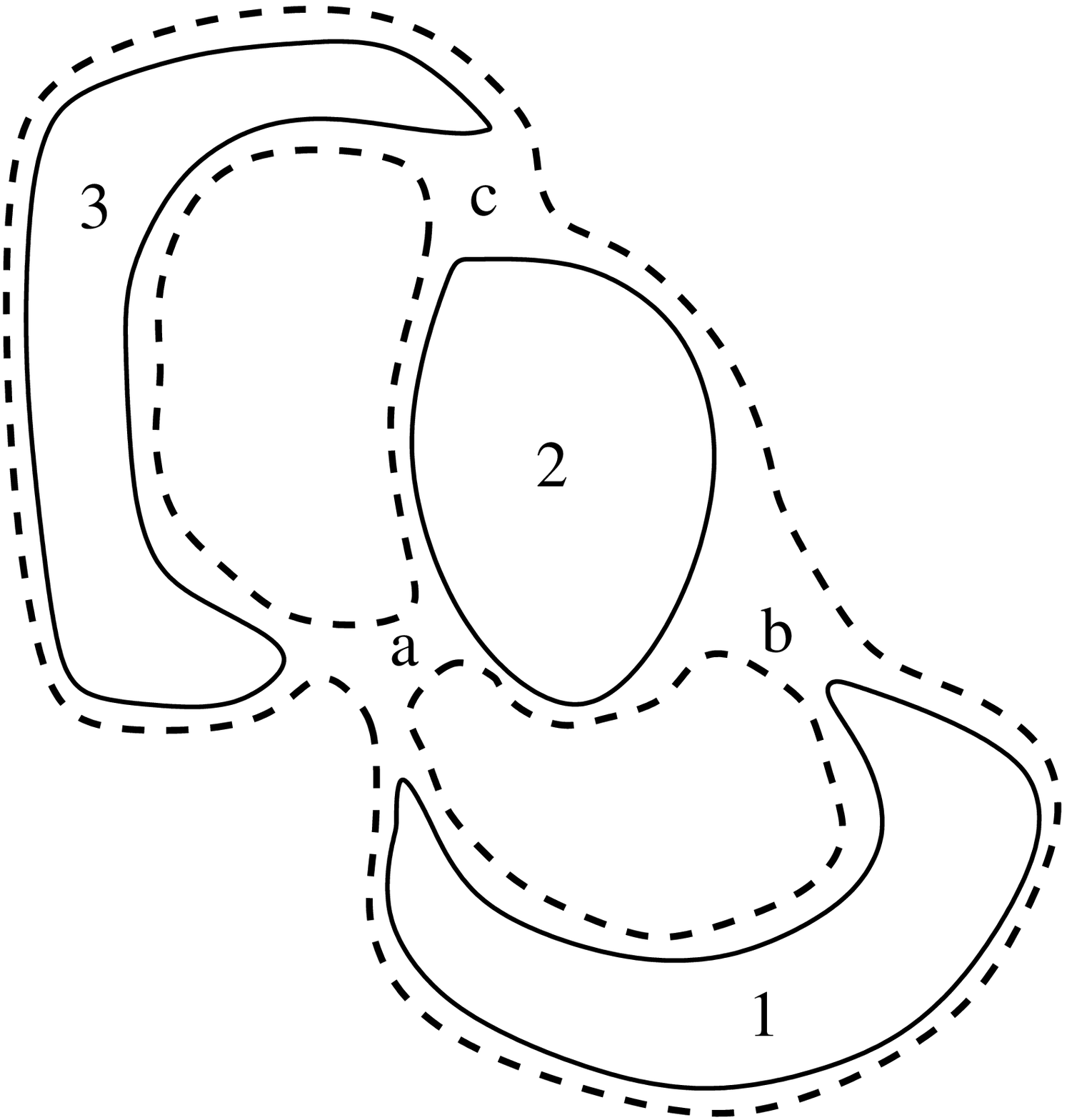,scale=0.35}  \\
\\ [0.2cm]
\end{array}$
\end{center}
  \caption{We illustrate the construction of the diagram for the term
$(1\,2\,3)_a(1\,2)_b(2\,3)_c$. On the left we have the first step of the construction: we draw fatgraphs loops for
each of the indices (active colors), 
marking the appropriate vertices (letters) on the outer side of the loops.  On the right we glue the vertices to obtain the diagram.} \label{combdiag}
\end{figure}

For each active color we draw a ``fatgraph'' loop (see Figure \ref{combdiag}), writing the corresponding index inside the inner circle.
The two sides of the fatgraph are inequivalent -- the inner circle is drawn with a solid line and the outer circle with a dashed line. We will refer to the solid line as the ``color line''.
We mark the external (dashed) line of each fatgraph with the  labels
of the twist fields that contain the corresponding color.
 (So for  example, in (\ref{simpleexample}) the twist field $b$ contains colors $1$ and $2$ and is represented by the letter $b$ 
 on  fatgraphs 1 and 2). The cyclic ordering of the twist fields on each loop  is determined by the radial 
 ordering  ($a  b  c$ in the example). Finally (right side of Figure \ref{combdiag})
we glue the non-color loops together at the positions of the twist fields, in such way that the order of the loops at each vertex (circling the vertex counterclockwise)
corresponds to the cycle structure of the corresponding twist field. 

\begin{figure}[htbp]
\begin{center}
$\begin{array}{c@{\hspace{0.45in}}c}
  \epsfig{file=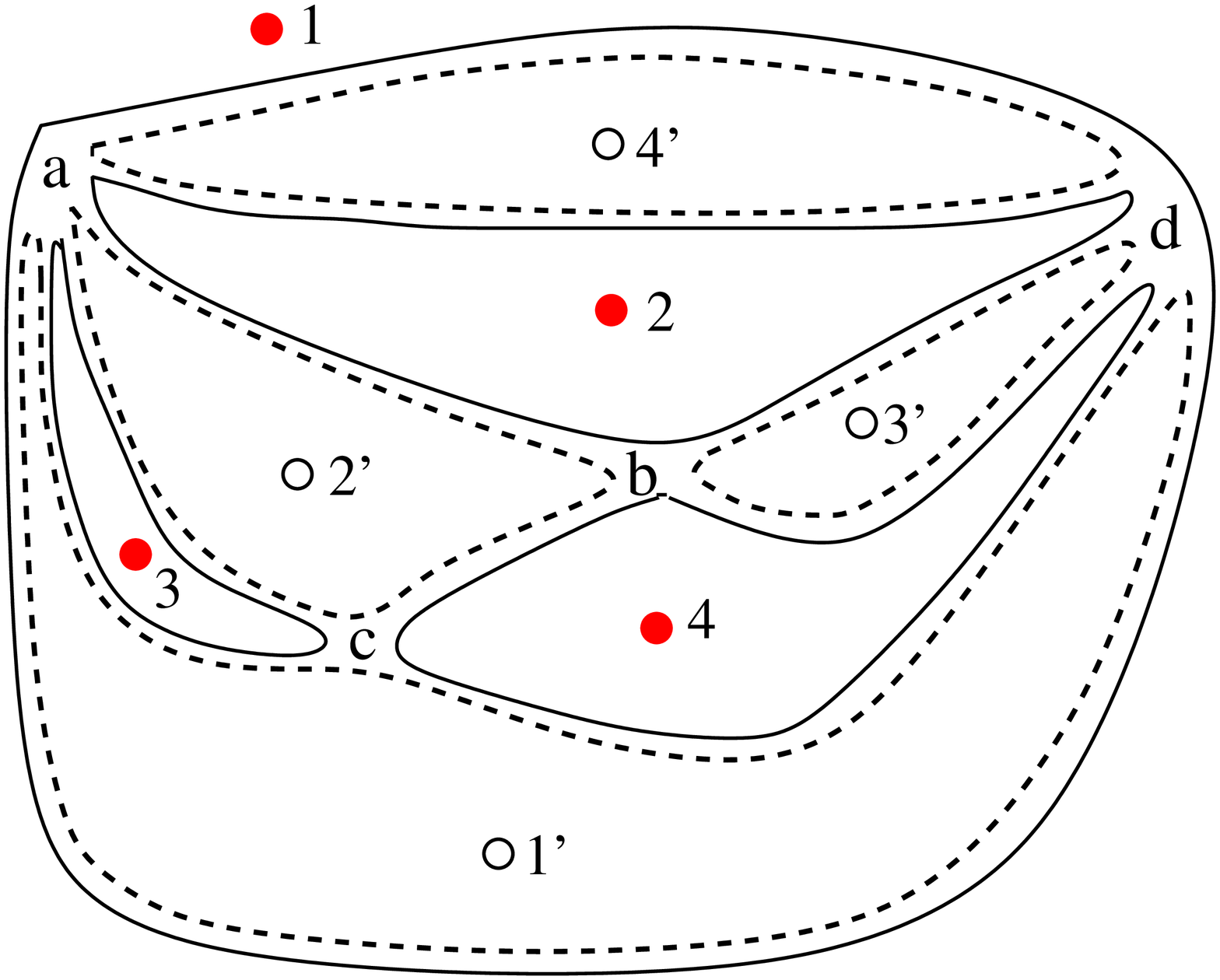,scale=0.26}& \epsfig{file=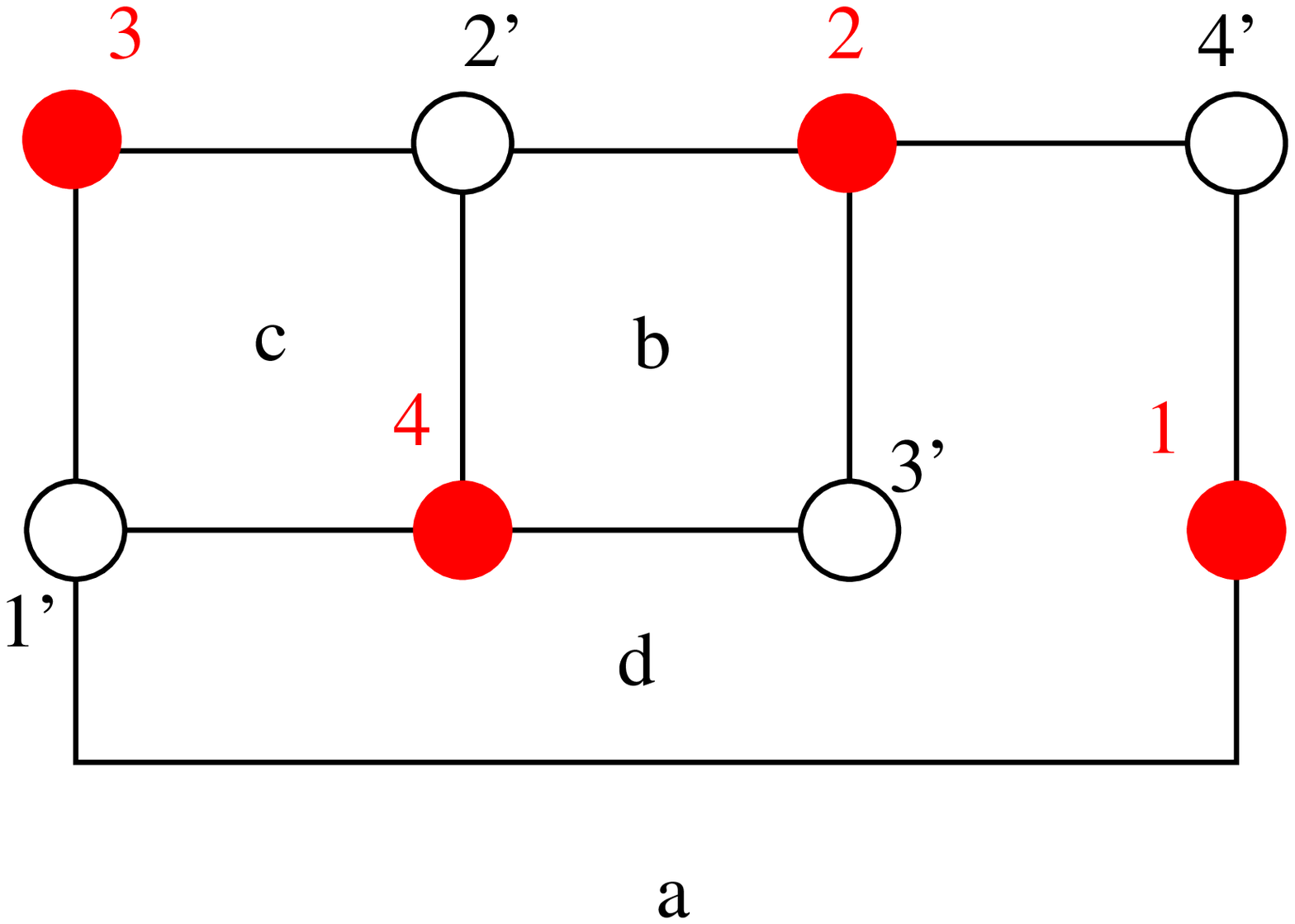,scale=0.4}  \\
\end{array}$
\end{center}
  \caption{{\it On the left}, the diagram corresponding to $(132)_a (24)_b (34)_c (241)_d$.
  A red (solid) dot is drawn for clarity on the inside of each color (solid) loop and is labeled by a  color index.
  Each {\it vertex} (letter) corresponds to a twist field: going around
  the vertex counterclockwise one reads off the color indices of
  the corresponding cyclic permutation.
{\it On the right},
the (graph theoretic) dual diagram, obtained as usual by dualizing  vertices into faces. 
Each {\it loop} in the dual graph corresponds to a twist field.} \label{okexample}
\end{figure}

As another illustration of this construction, the left side of Figure \ref{okexample} depicts
a specific term appearing in the expansion of $\langle \sigma_{[3]}(a) \sigma_{[2]}(b) \sigma_{[2]}(c) \sigma_{[3]}(d) \rangle$.
The procedure is completely general and allows to associate a diagram to any (non-trivial) term  appearing  in the expansion of a generic correlator. 
It is also clear that equivalent terms give rise to topologically equivalent diagrams, differing only by
a relabeling of the color indices. 

A {\it term}  is said to be {\it reducible}
if the group elements  $ \{ \hat g_j \}$  can be split into two sets so that the elements in each set
act trivially on the elements of the other set.  (Another way to state this condition is to say that the group elements $ \{ \hat g_j \}$  of an irreducible term generate a {\it transitive} subgroup of $S_N$.)
A reducible term factorizes into irreducible components.
If a term is reducible, all the terms in the same class are reducible, so we may
speak of reducible and irreducible classes.
It is clear that our procedure associates irreducible classes 
to connected diagrams, and reducible classes to disconnected diagrams.
The usual combinatorial arguments apply: the generating functional of all diagrams is the exponential
of the generating functional of irreducible diagrams.\footnote{Note however the following subtlety: the $N$ dependence of a term which splits into 
several irreducible components is not equal to the product of $N$ dependencies of each of the components, because
in contrast to ordinary gauge
 theories, there should not 
  be joint colors between different irreducible components of a given term.}
We may thus restrict our analysis to connected diagrams.

\begin{figure}[htbp]
\begin{center}
\epsfig{file=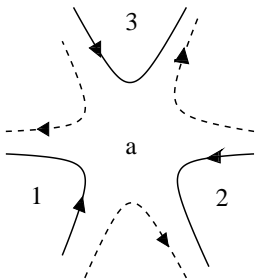,scale=0.3}
\caption{The vertex corresponding to $\sigma_{(123)}(z_a, \bar z_a)$. The solid lines (color lines) are numbered counter-clockwise
in the cyclic ordering $(123)$. The letter in the center labels the coordinate of the twist operator.
} 
\label{vertexsym}
\end{center}
\end{figure}

\begin{figure}[htbp]
\begin{center}
$\begin{array}{c@{\hspace{0.45in}}c}
  \epsfig{file=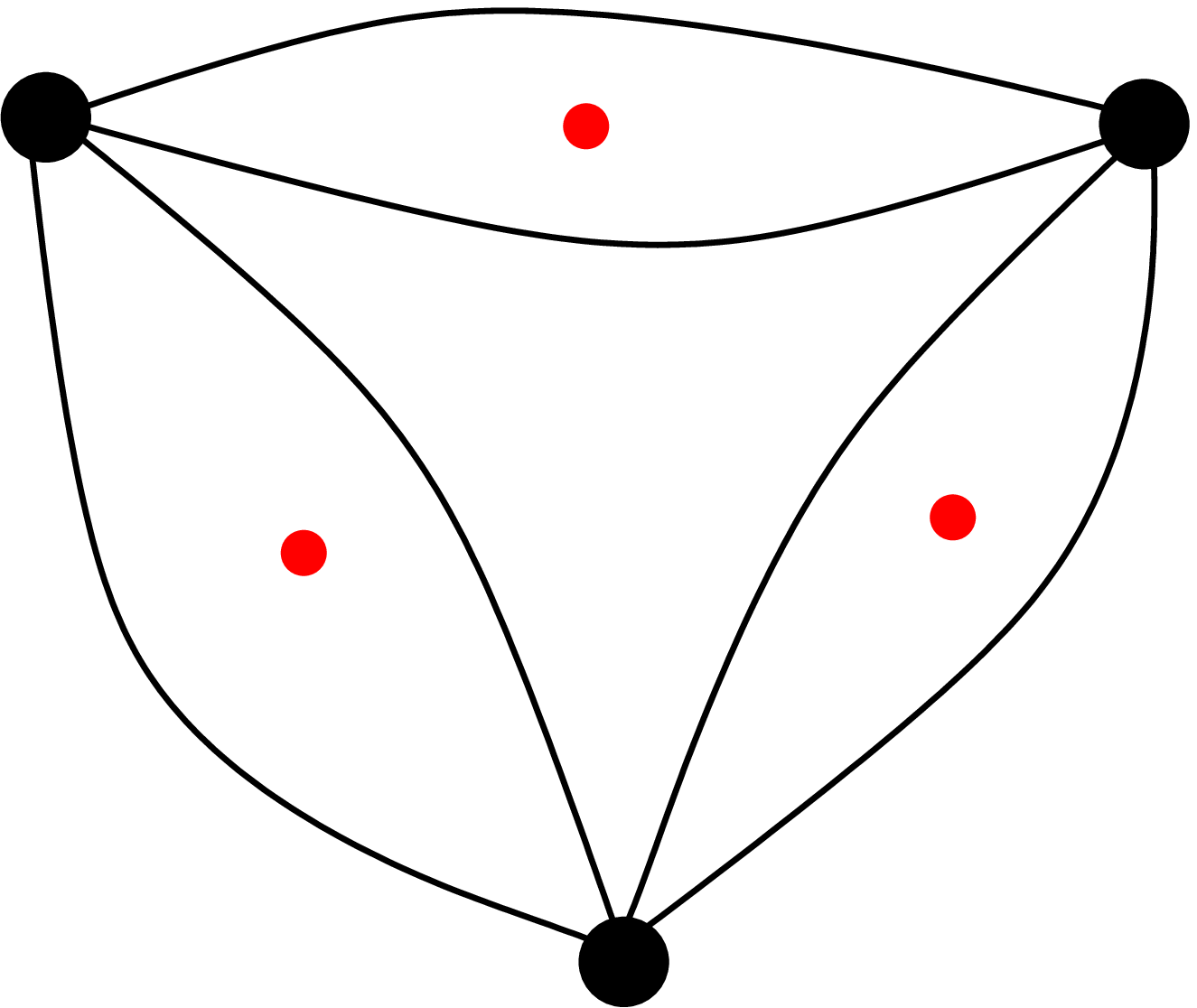,scale=0.25}& \epsfig{file=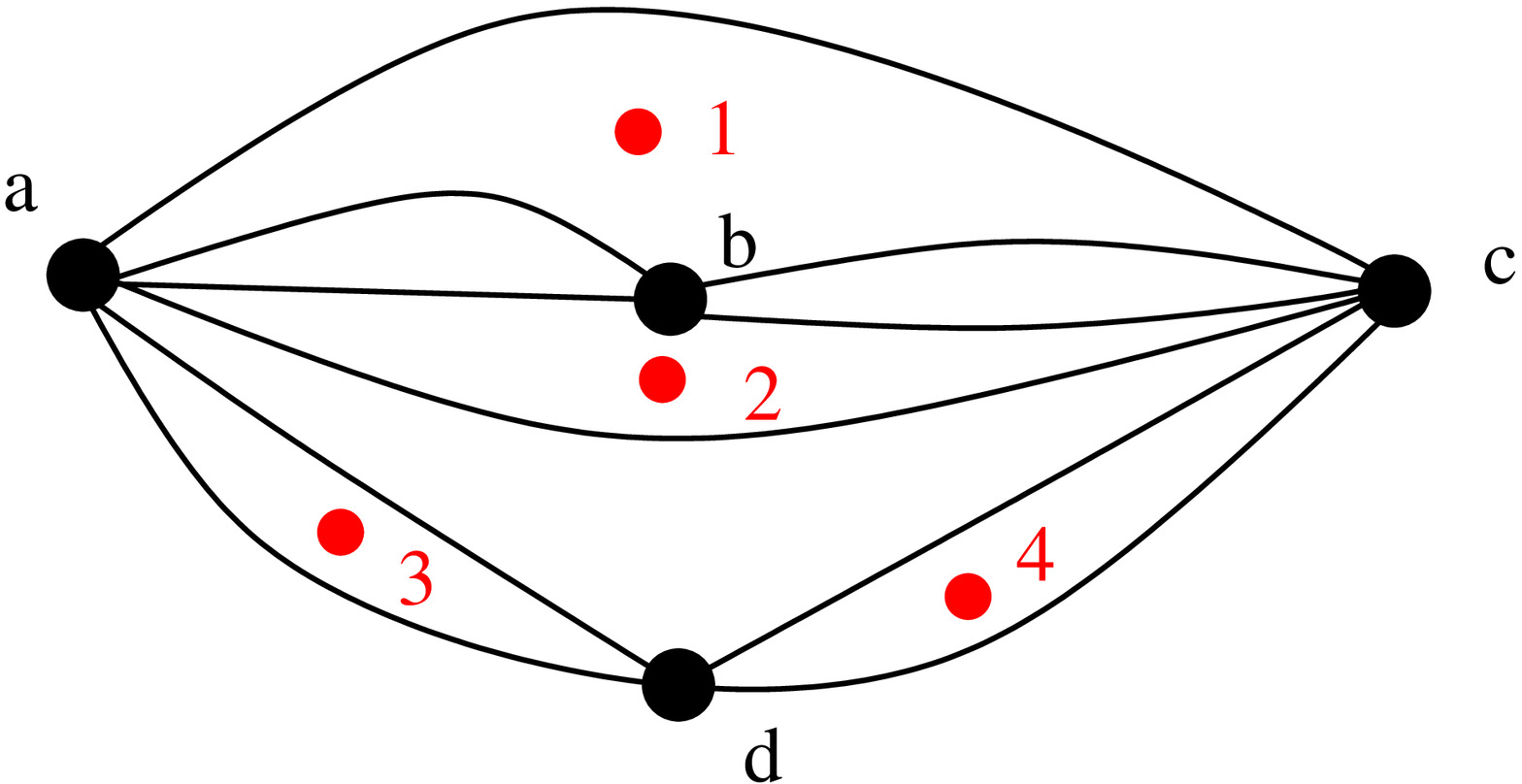,scale=0.3}  \\
\\ [0.2cm]
\end{array}$
\end{center}
  \caption{Two examples of illegal diagrams. (To avoid cluttering of the Figures we draw the fatgraph propagators
  with a single line, and use red dots to denote the ``color'' (solid) sides of the propagators.)
   On the left the numbers of two types of loops do not coincide (three color loops
and two non-color loops).  On the
right the partial orderings
defined by the color loops are incompatible: color $1$ defines partial ordering on vertices $abc$, and color
$2$ defines the inverse ordering $bac$.} \label{orderexample}
\end{figure}

In our diagrams, twist fields correspond to vertices. An $s$-cycle twist field  corresponds
to a vertex with $2s$ fatgraph propagators emating from it: $s$ solid (``color'') and $s$ dashed 
 oriented lines,  in alternating order, as shown in Figure~\ref{vertexsym}.  Each vertex is labeled by the coordinate
 where the corresponding twist field is inserted.
 The diagrams generated by our procedure are {\it not} the most general diagrams 
that we may draw starting with a set of vertices and connecting the fat graph propagators in all possible ways.
 Indeed all diagrams are subject to two restrictions:
 \begin{enumerate}
\item
 { The number of color (solid) loops is equal to the number of non-color (dashed) loops.  }
\item
The solid and dashed loops define  partial cyclic orderings of the vertices.
By convention the  solid loops are oriented counterclockwise 
and the dashed loops are oriented clockwise.
{\it  All these partial orderings  must be compatible with the radial ordering of the vertices.}
\end{enumerate}
Figure \ref{orderexample} gives two examples of  diagrams {\it violating} these restrictions.

To understand the first restriction, we can view
 the dashed loops  (with clockwise orientation) as  the ``trajectories'' of each index. 
Consider the example  of Figure \ref{okexample},
\be
\begin{array}{c@{\hspace{0.1in}}c@{\hspace{0.1in}}c@{\hspace{0.1in}}c}
(1\, 3\, 2)_a&(2 \,4)_b&(3\, 4)_c&(1\, 2\, 4)_d.
\end{array}
\ee 
There are four active colors and thus four color loops.  The  four ``trajectories'' are
\be
&&(1')\;=\;1\to^a3\to^c4\to^d1,\qquad  (4')\;=\;2\to^a1\to^d 2,\\
&&(2')\;=\;3\to^a2\to^b4\to^c3,\qquad(3')=4\to^b 2\to^d 4.\nonumber
\ee 
The superscripts on the arrows correspond to the vertices. 
One can read off the ``trajectories'' from the diagrams by going clockwise along the dashed loops.
Since the product of the cycles multiplies to the identity, 
 the number of trajectories is always equal to the number of active colors.
Thus the numbers of the two types of loops are equal.

The second restriction holds by construction for the partial orderings associated to the color loops.
It  holds for the non-color loops because  the 
trajectories of the indices follow  the ordering of the group elements, which coincide by construction with the
radial ordering of the vertices.

Viceversa, given a diagram\footnote{In speaking of a ``diagram'', we mean ``a diagram with labelled vertices'' 
(by the coordinates of the corresponding twist operators).} 
 obeying the two restrictions,
we can uniquely associate to it a non-trivial  equivalence class of terms.
 We just label the color loops with indices from $1$ to $c$ (= number of active colors),
  and associate to each vertex the twist field obtained by reading the color indices
counterclockwise around the vertex. (The way indices are assigned to the color loops is immaterial,
as different choices are related by a global $S_N$ transformation.) Thus the correspondence
between diagrams and equivalence classes of terms is one-to-one.

We are finally in the position to quote our  ``Feynman'' rules to write a correlator as a formal sum of diagrams.
Given a generic correlator of gauge-invariant twist fields,
\be\label{defccorr}
\langle \s_{[n_1]} (a_1)  \dots \s_{[n_s]} (a_s) \rangle \, ,  \qquad |z_{a_1}| < |z_{a_2}| < \dots |z_{a_a}| \, ,
\ee 
to compute its connected part we draw all  connected diagrams having $s$ vertices of type $n_k$, $k=1,\dots s$,
with no self-contractions at each vertex,
and obeying the two restrictions discussed above. We can write 
\be\label{pertexp00}
\langle \s_{[n_1]} (a_1)  \dots \s_{[n_s]} (a_s) \rangle_{conn} \,=\sum_{\a} C_{\a}(N,\{n_j\})\,
\langle\prod_{j=1}^s\s_{g_j^{(\a)}}(a_j)\rangle \, ,
\ee where the index  $\a$ runs over all the contributing (connected) diagrams, 
which by construction are in one-to-one correspondence  with the equivalence classes of (connected) terms.
The ordered sequence of group elements $g_1^{(\a)} \dots g_s^{(\a)} $ is a representative of the class.
The numerical factor $ {\mathcal N}_{\a}(N,\{n_j\})$ counts the number of terms in each class and
 will be determined shortly.

Given a diagram, we can construct its graph-theoretic {\it dual}  by the usual procedure of dualizing
vertices into faces, as illustrated in Figure \ref{okexample}. The dual diagrams are bipartite graphs,
with red (solid) and white (empty) nodes, corresponding respectively
to the color (solid) loops  and the non-color (dashed) loops of the diagram before dualization.
The twist operators map to the faces of the dual diagram. Since dual diagrams are perhaps easier to draw,
we will mostly use them in the rest of the paper.

\begin{figure}[htbp]
\begin{center}{\footnotesize
$\begin{array}{c@{\hspace{0.25in}}c@{\hspace{0.25in}}c}
\epsfig{file=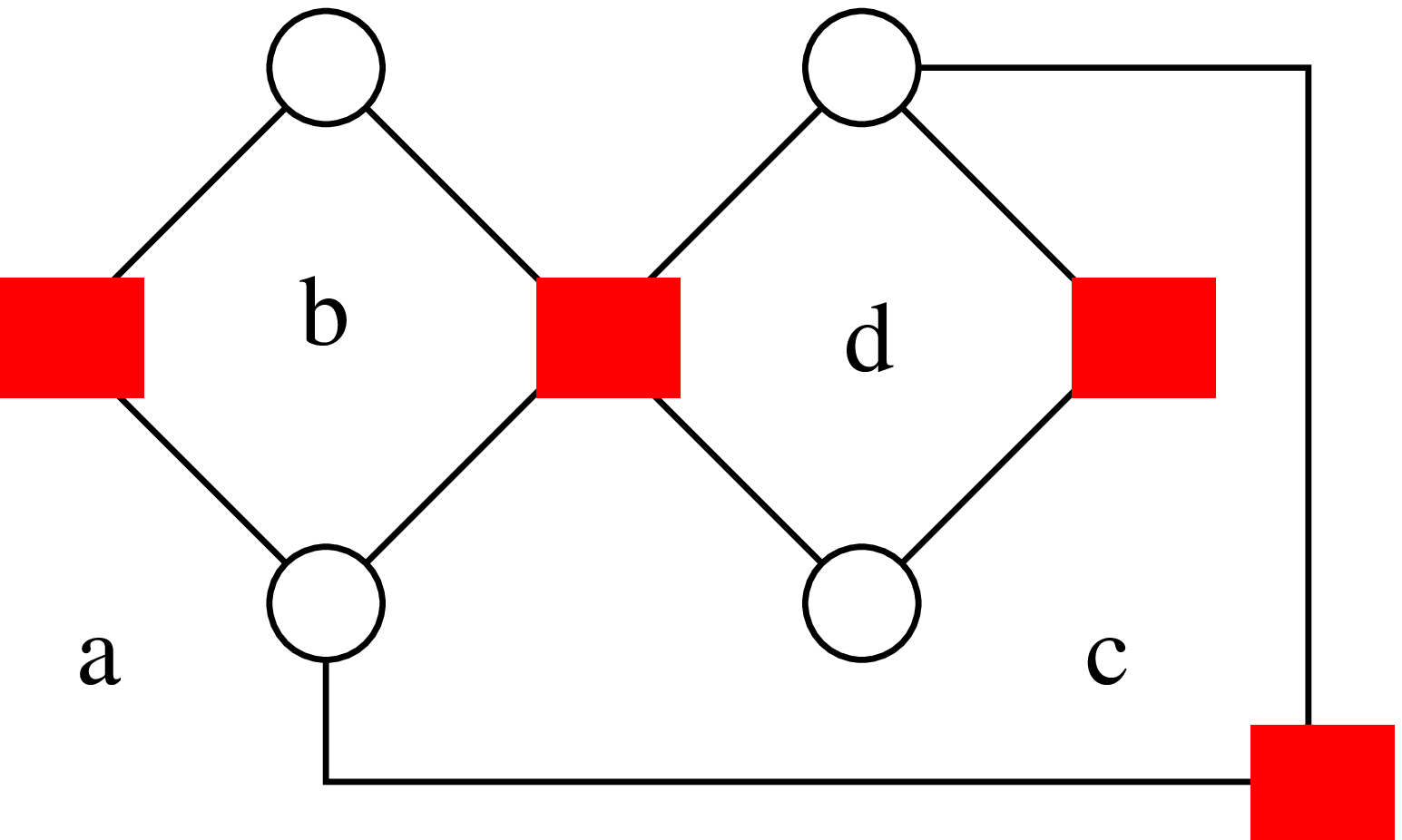,scale=0.15} & \epsfig{file=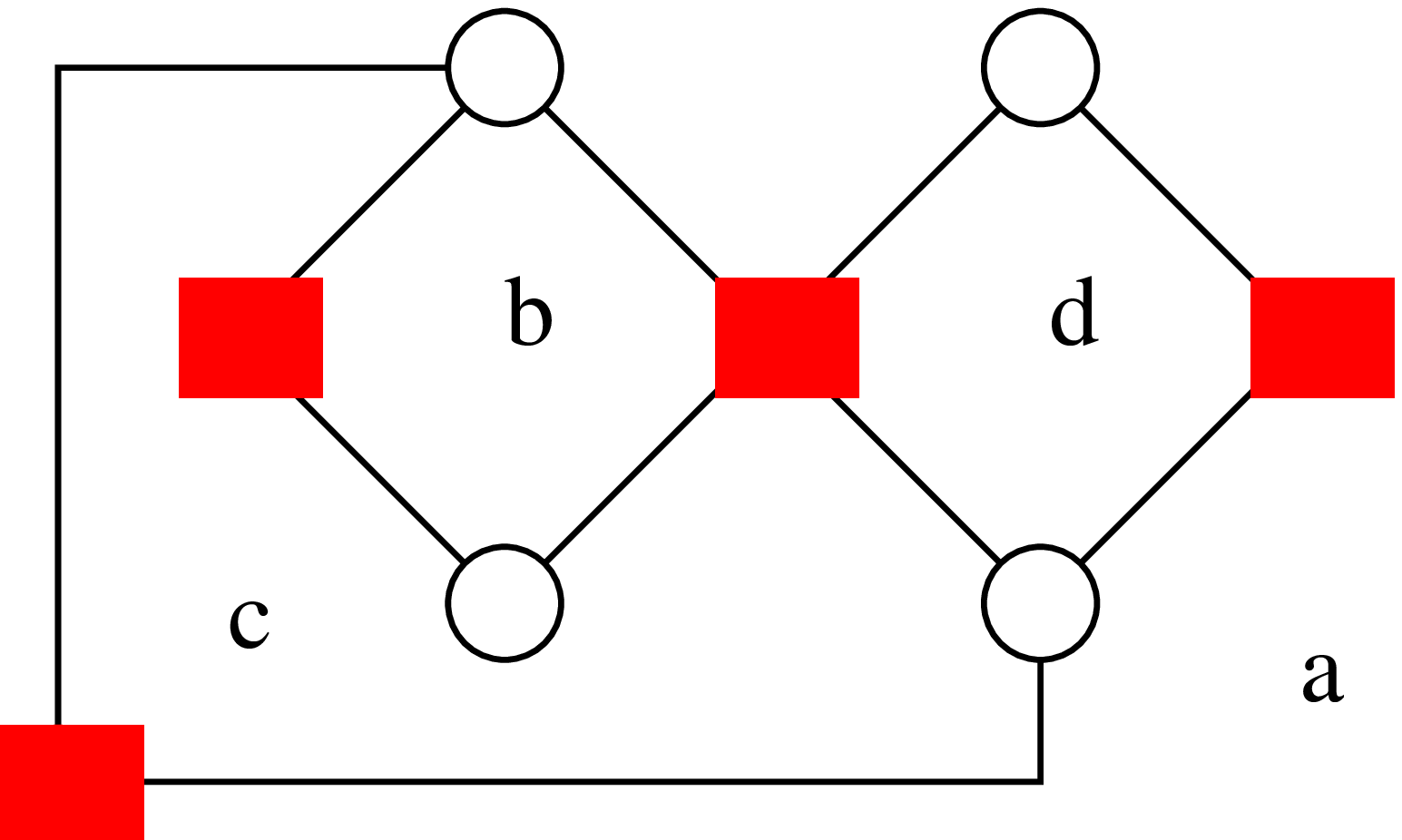,scale=0.15} &
\epsfig{file=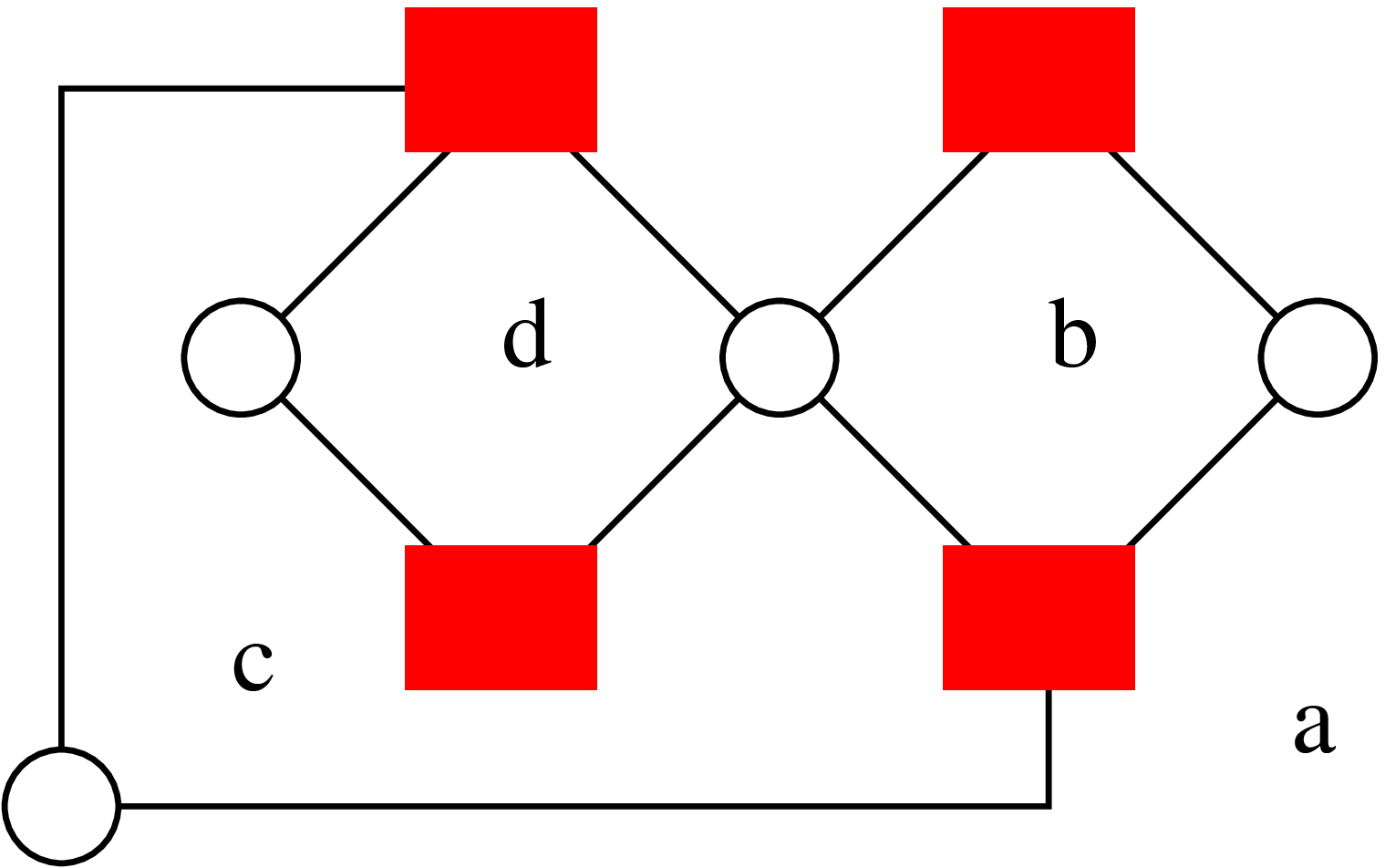,scale=0.15}\\
\a_1=(4\,3\,2)_a(2\,3)_b(4\,1\,2)_c (1\,2)_d&
\a_2=(1\,4\,2)_a(2\,3)_b(3\,4\,2)_c (1\,2)_d&
\a_3=(4\,1\,2)_a(2\,4)_b(4\,3\,1)_c (1\,3)_d\\
\epsfig{file=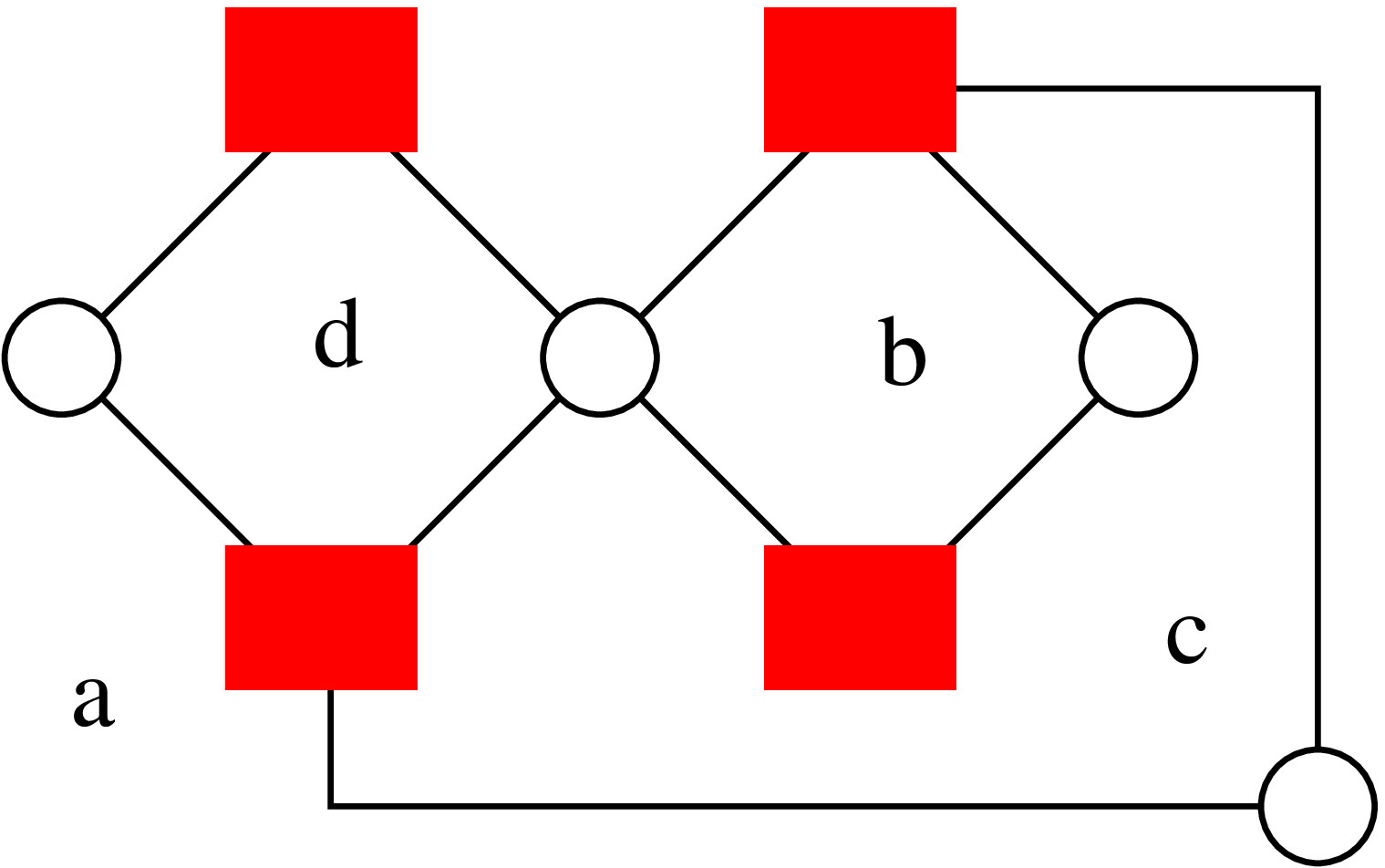,scale=0.15} & \epsfig{file=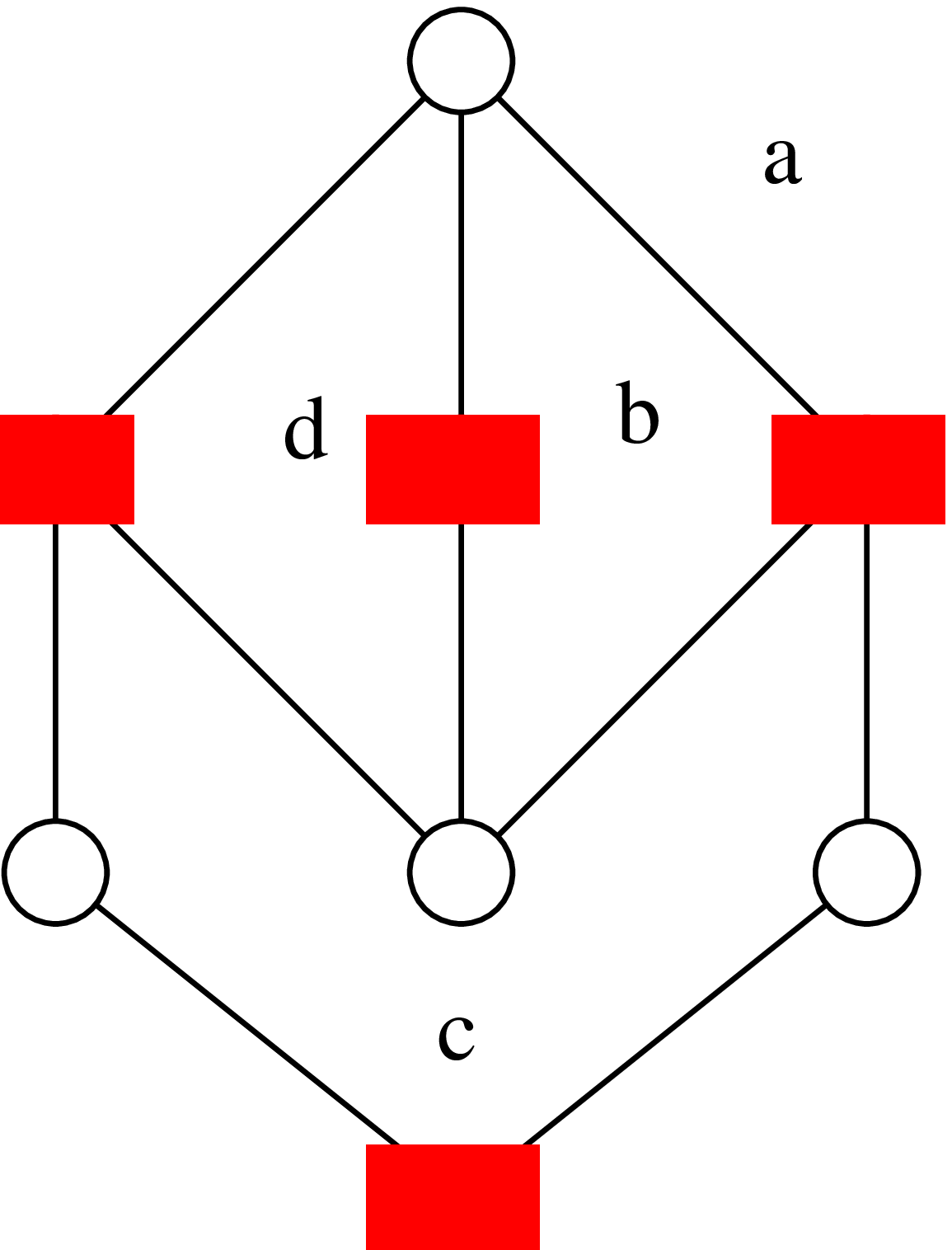,scale=0.15} &
\epsfig{file=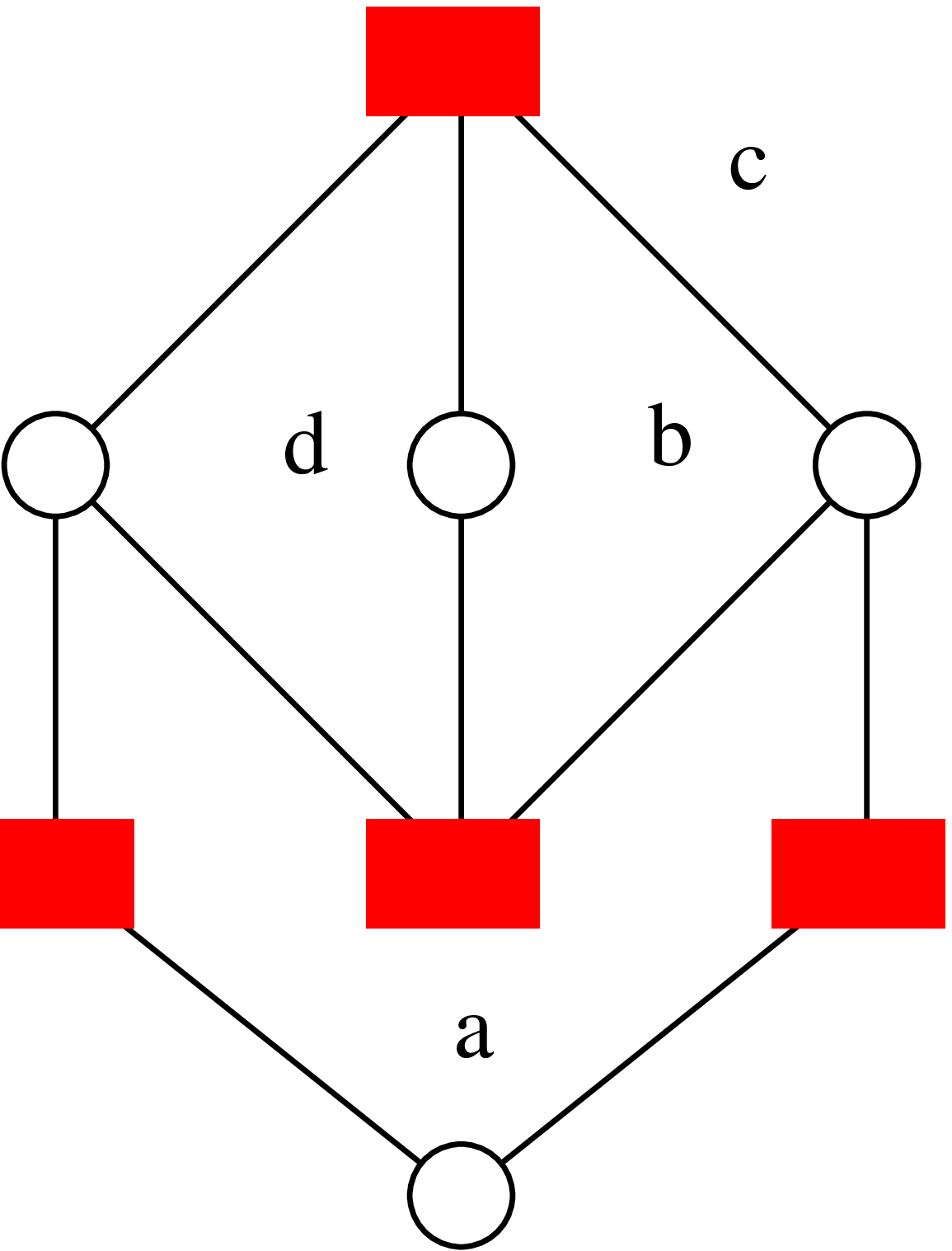,scale=0.15}\\
\a_4=(2\,3\,1)_a(2\,4)_b(2\,4\,3)_c (1\,3)_d&
\a_{5}=(4\,1\,3)_a(2\,3)_b(1\,4\,3)_c (1\,2)_d&
\a_{6}=(4\,2\,3)_a(1\,4)_b(1\,3\,2)_c (1\,4)_d\\
\epsfig{file=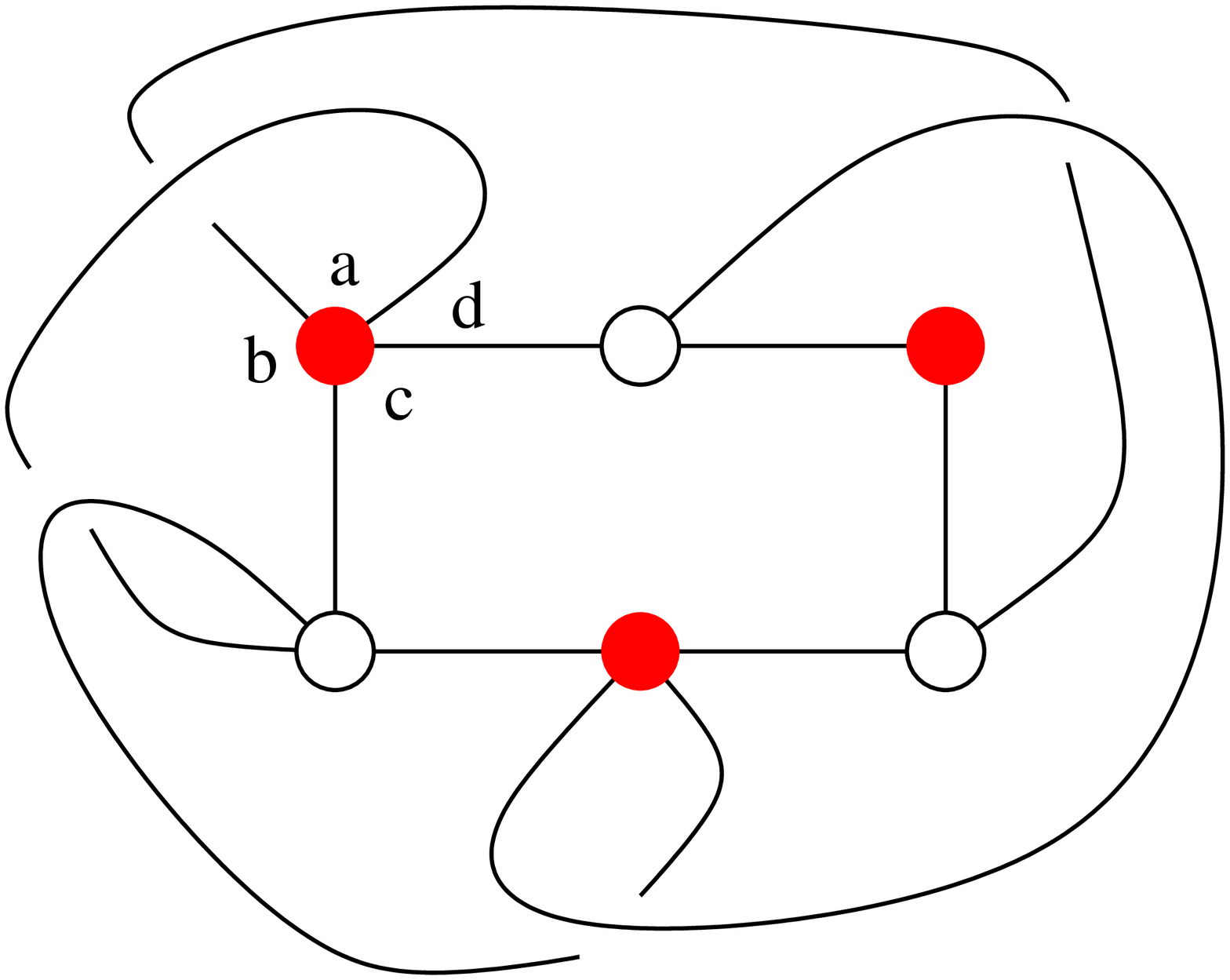,scale=0.15}&\epsfig{file=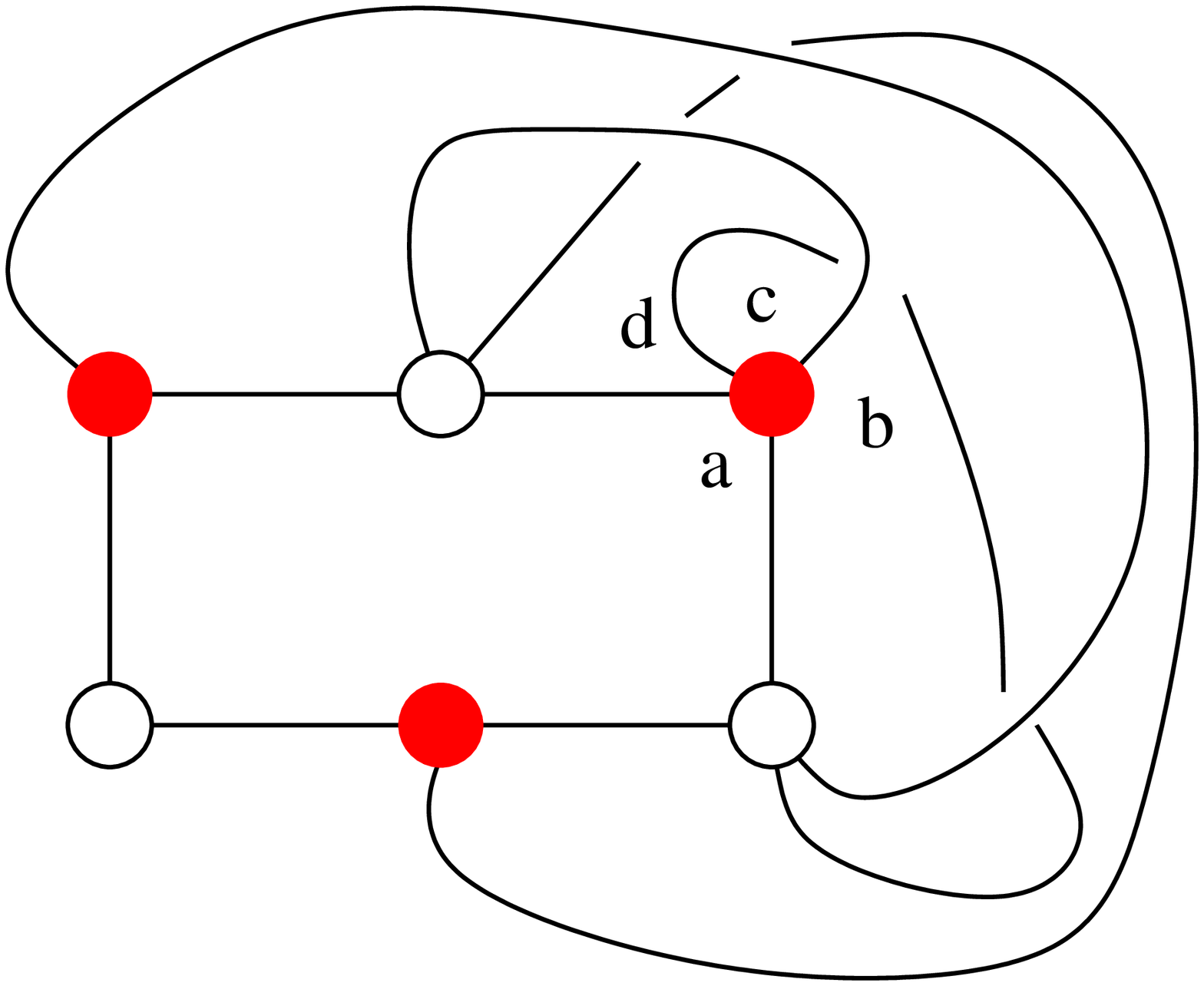,scale=0.15}&\\
\a_{7}=(1\,2\,3)_a(1\,3)_b(1\,2\,3)_c (1\,3)_d&
\a_{8}=(1\,2\,3)_a(1\,3)_b(3\,2\,1)_c (2\,3)_d&
\\ [0.2cm]
\end{array}$}
\end{center}
\begin{center}
\caption{Connected diagrams contributing to   $\langle \sigma_{[3]}(a) \sigma_{[2]}(b) \sigma_{[3]}(c) \sigma_{[2]}(d) \rangle$ when $ |z_a|<|z_b|<|z_c|<|z_d|$.
}
 \label{2233all1} 
\end{center}
\end{figure}

\begin{figure}[htbp]
\begin{center}{\footnotesize
$\begin{array}{c@{\hspace{0.25in}}c@{\hspace{0.25in}}c}
\epsfig{file=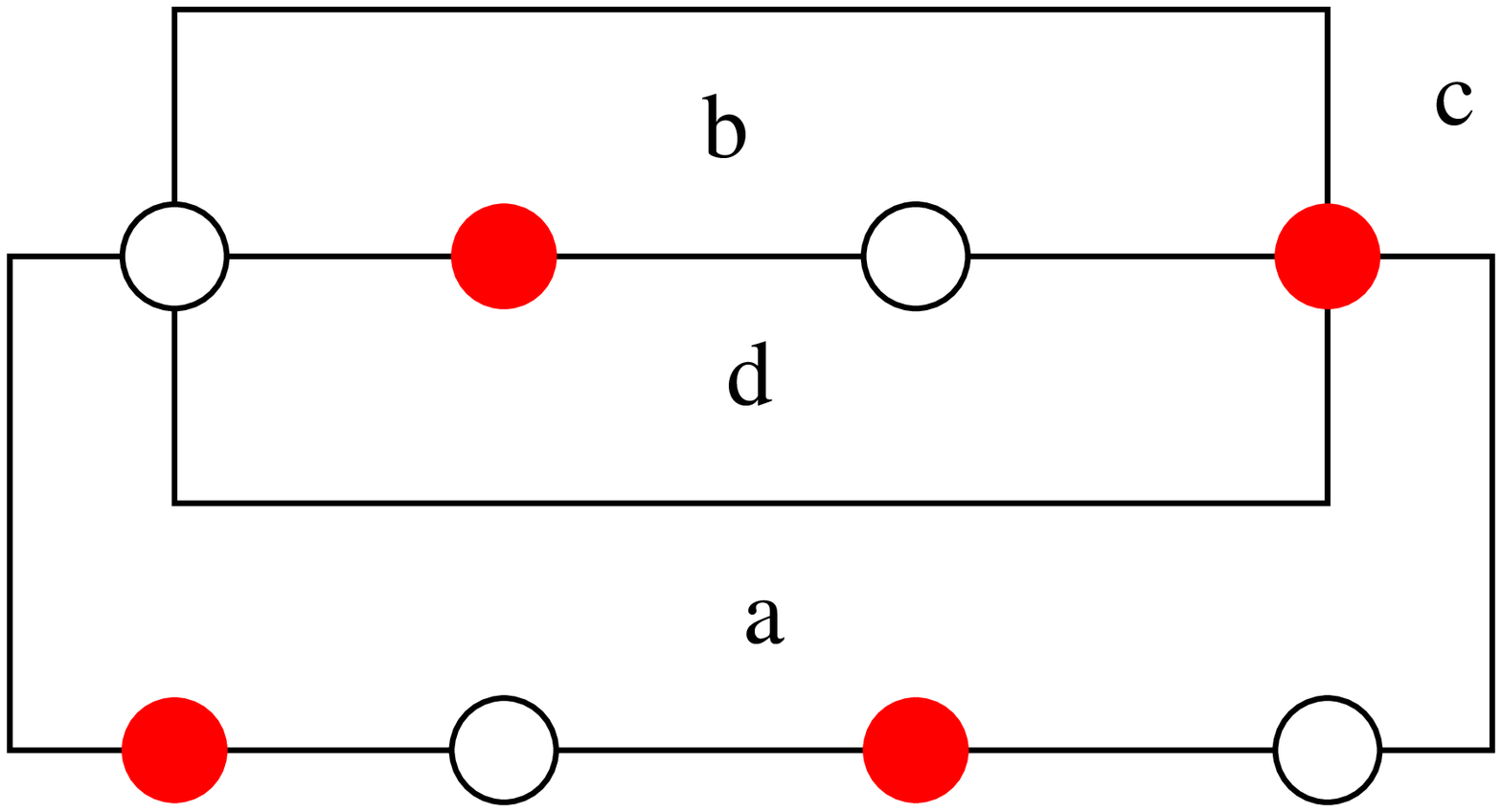,scale=0.15}& \epsfig{file=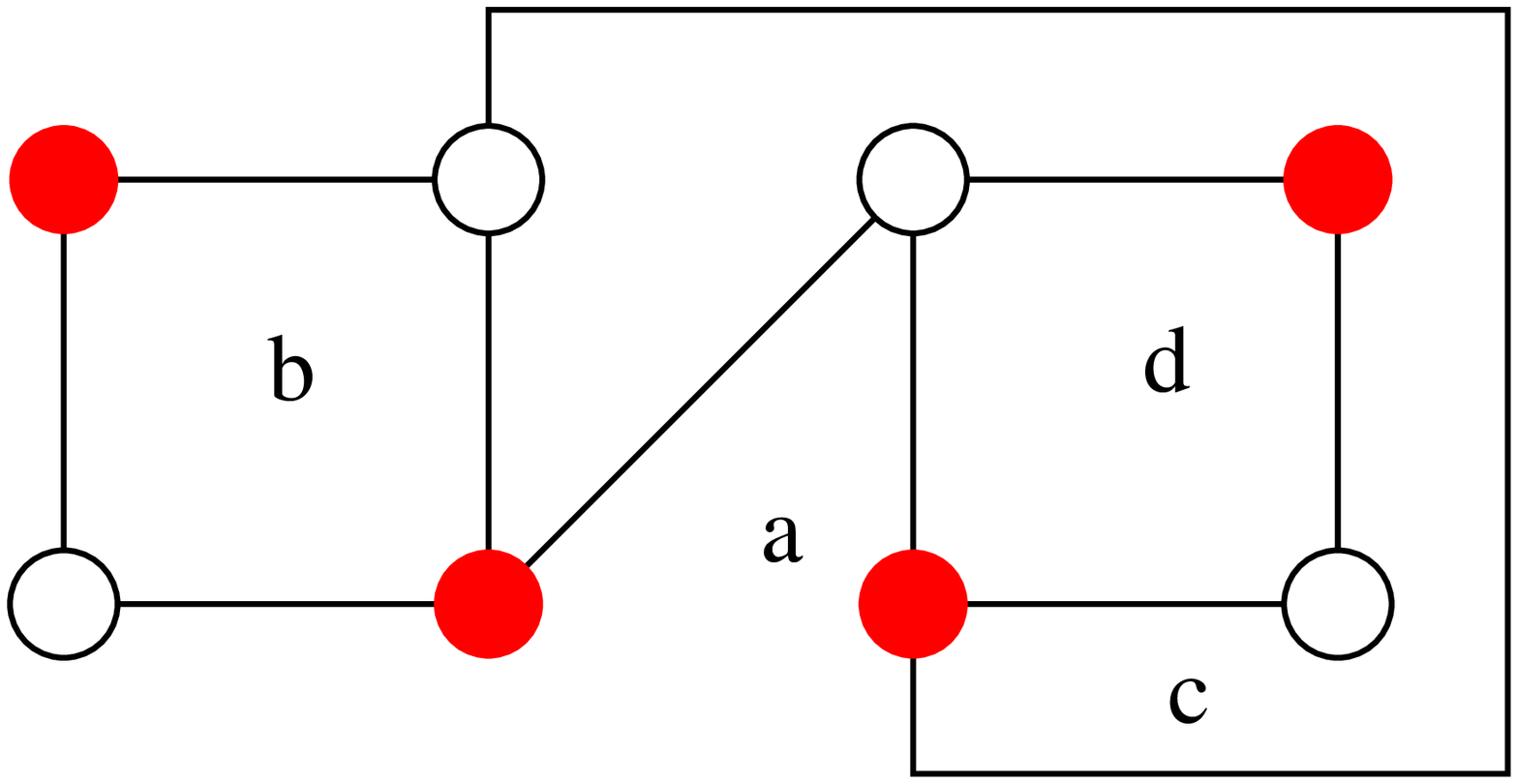,scale=0.15} &
\epsfig{file=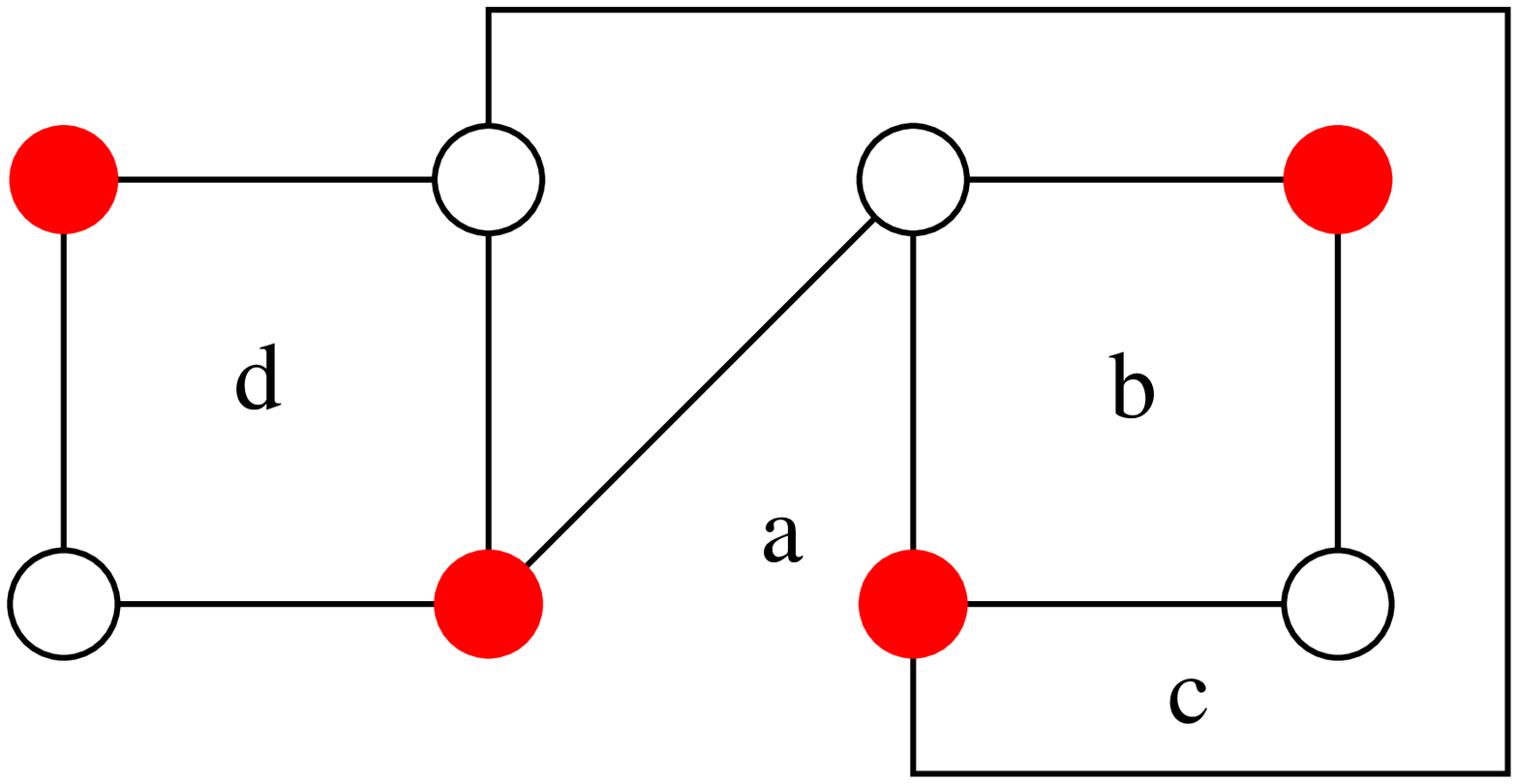,scale=0.15}\\
\a_{9}=(1\,3\,4)_a(4\,3\,1)_c(1\,2)_b(2\,1)_d&
\a_{10}=(2\,1\,3)_a(4\,3\,2)_c(1\,2)_b(3\,4)_d&
\a_{11}=(2\,1\,3)_a(4\,3\,2)_c(3\,4)_b(1\,2)_d\\
\epsfig{file=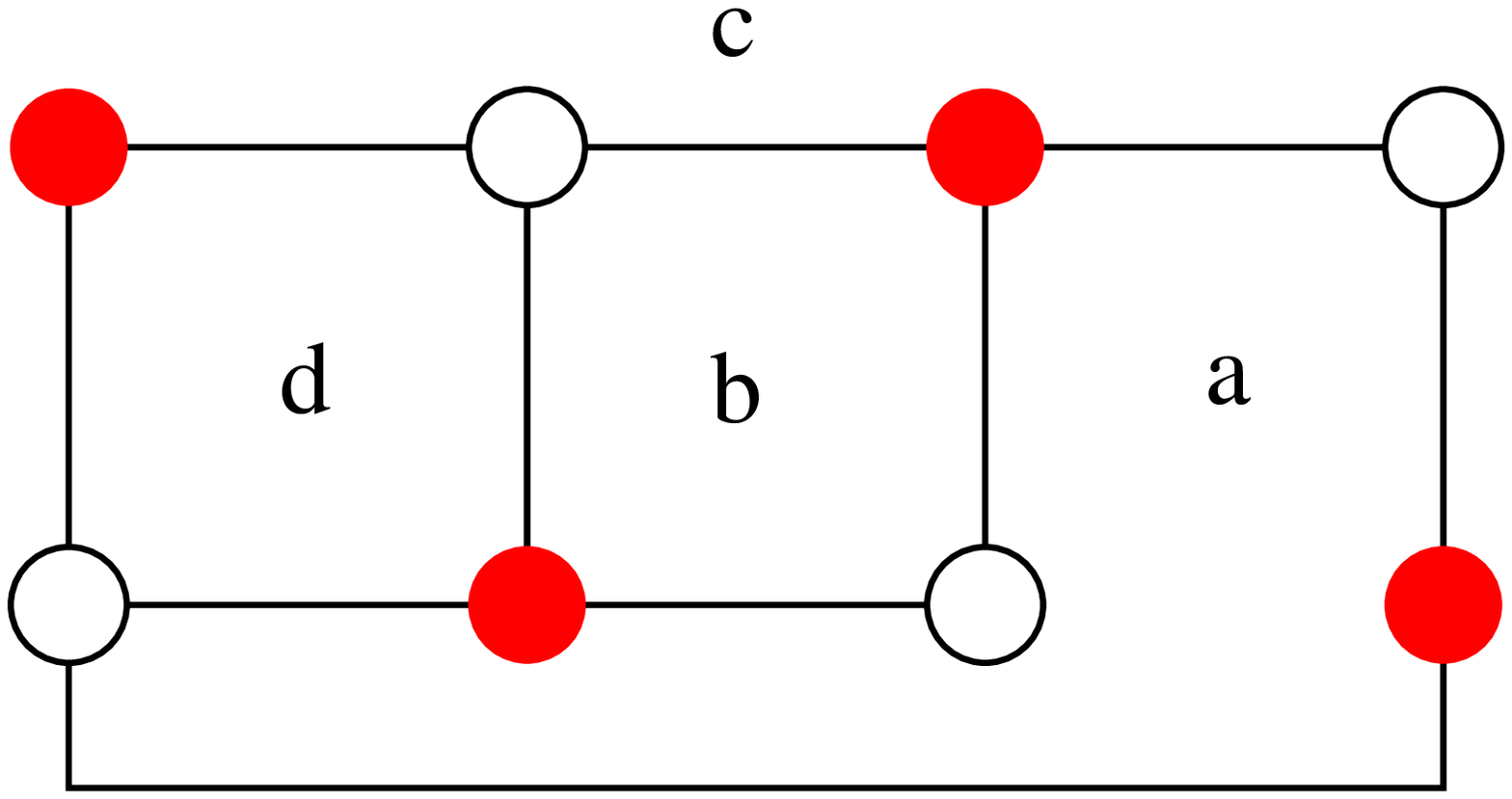,scale=0.15} & \epsfig{file=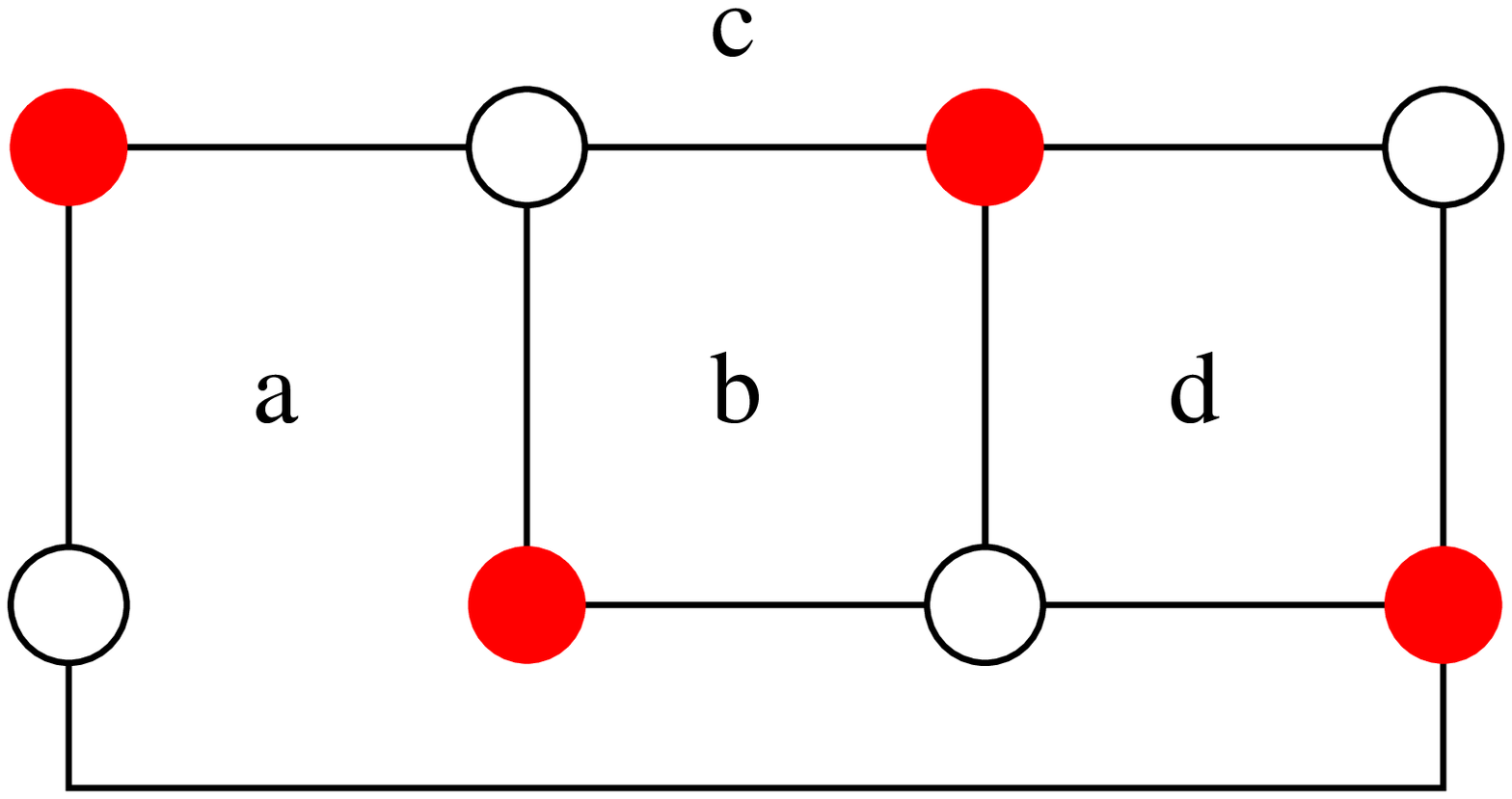,scale=0.15}&\epsfig{file=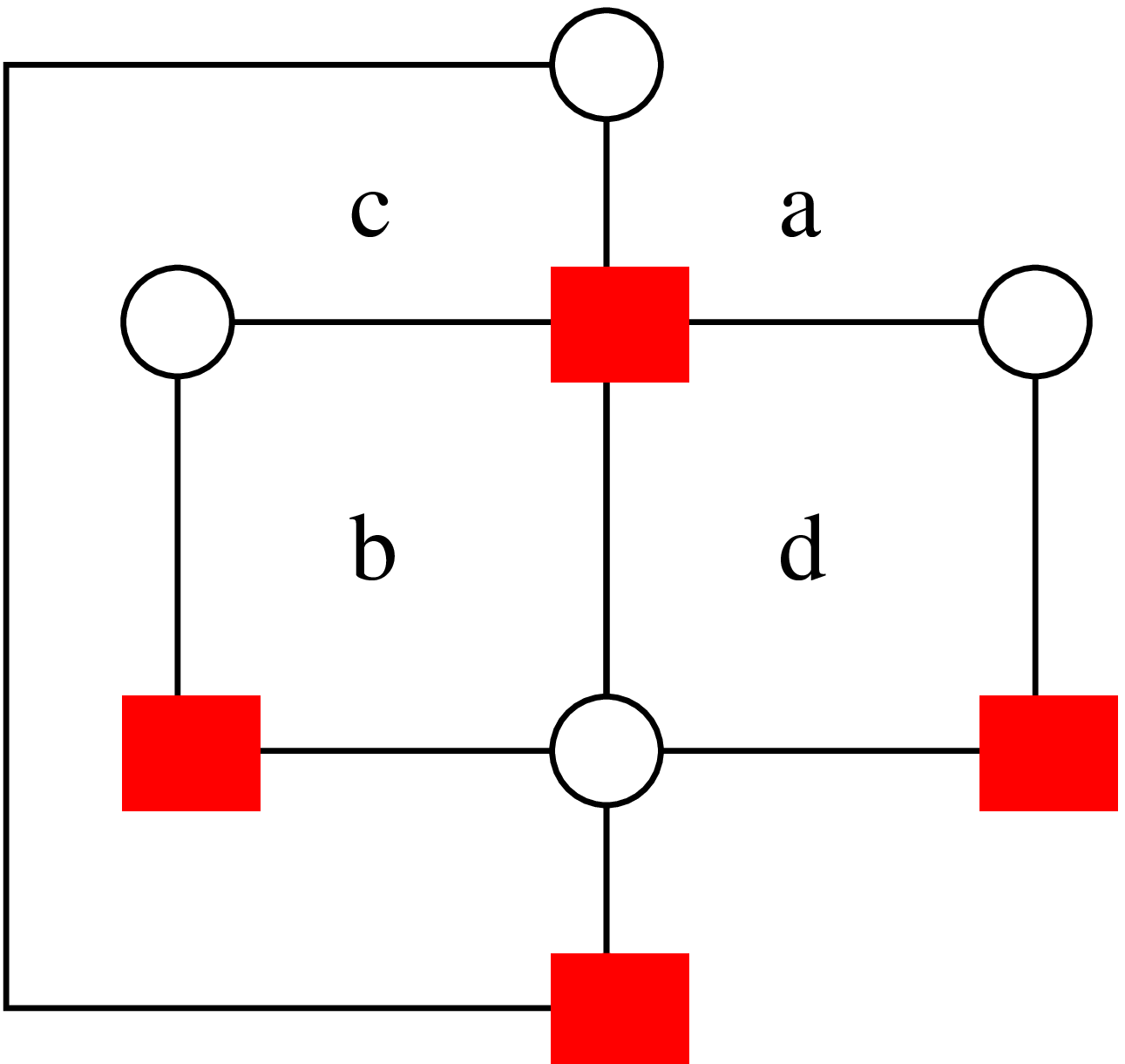,scale=0.15}\\
\a_{12}=(4\,3\,2)_a(3\,4\,1)_c(3\,2)_b(1\,2)_d&
\a_{13}=(1\,4\,3)_a(2\,3\,4)_c(1\,2)_b(3\,2)_d&
\a_{14}=(3\,4\,1)_a(1\,4\,2)_c(1\,2)_b(1\,3)_d\\
\epsfig{file=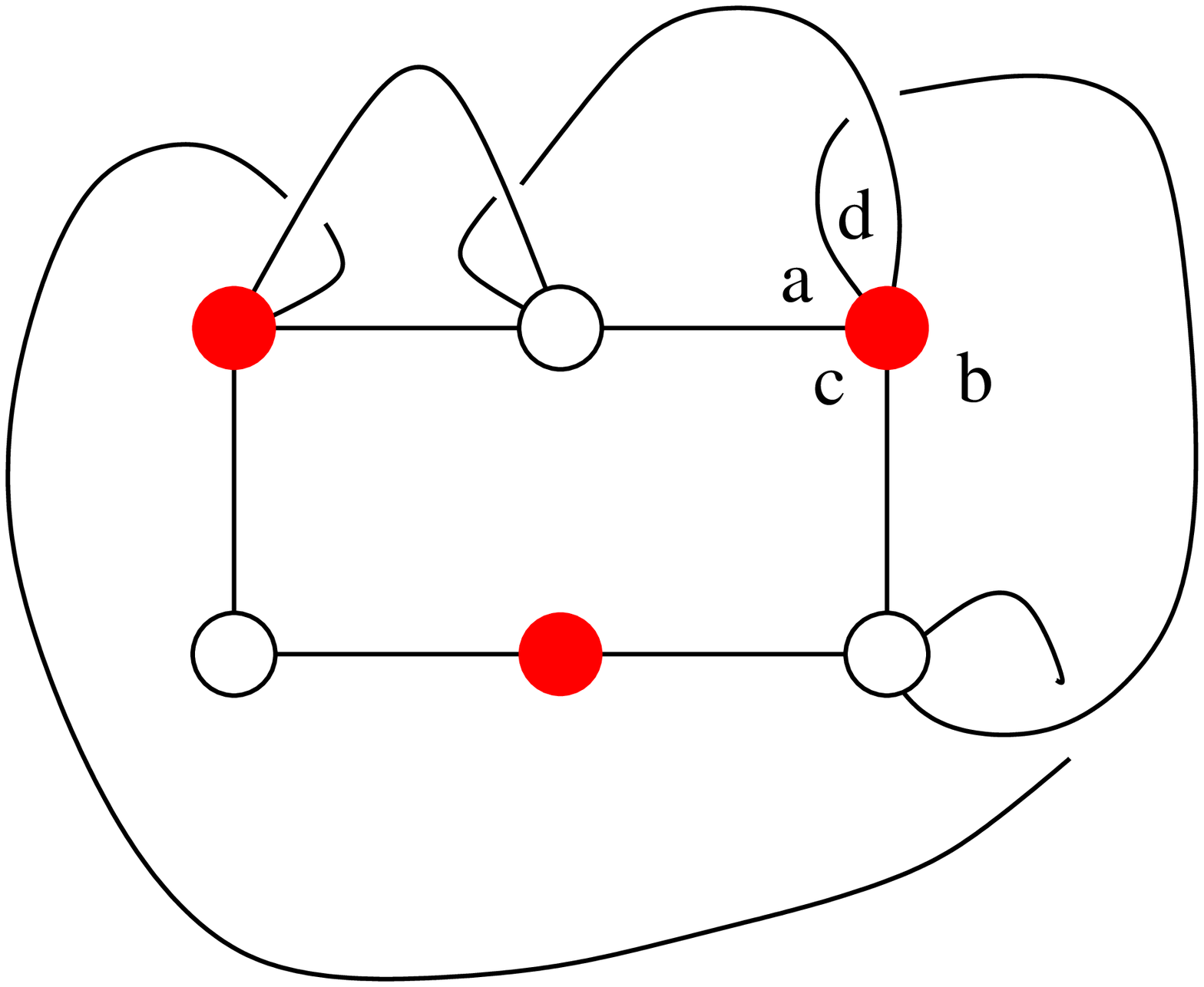,scale=0.15}&\epsfig{file=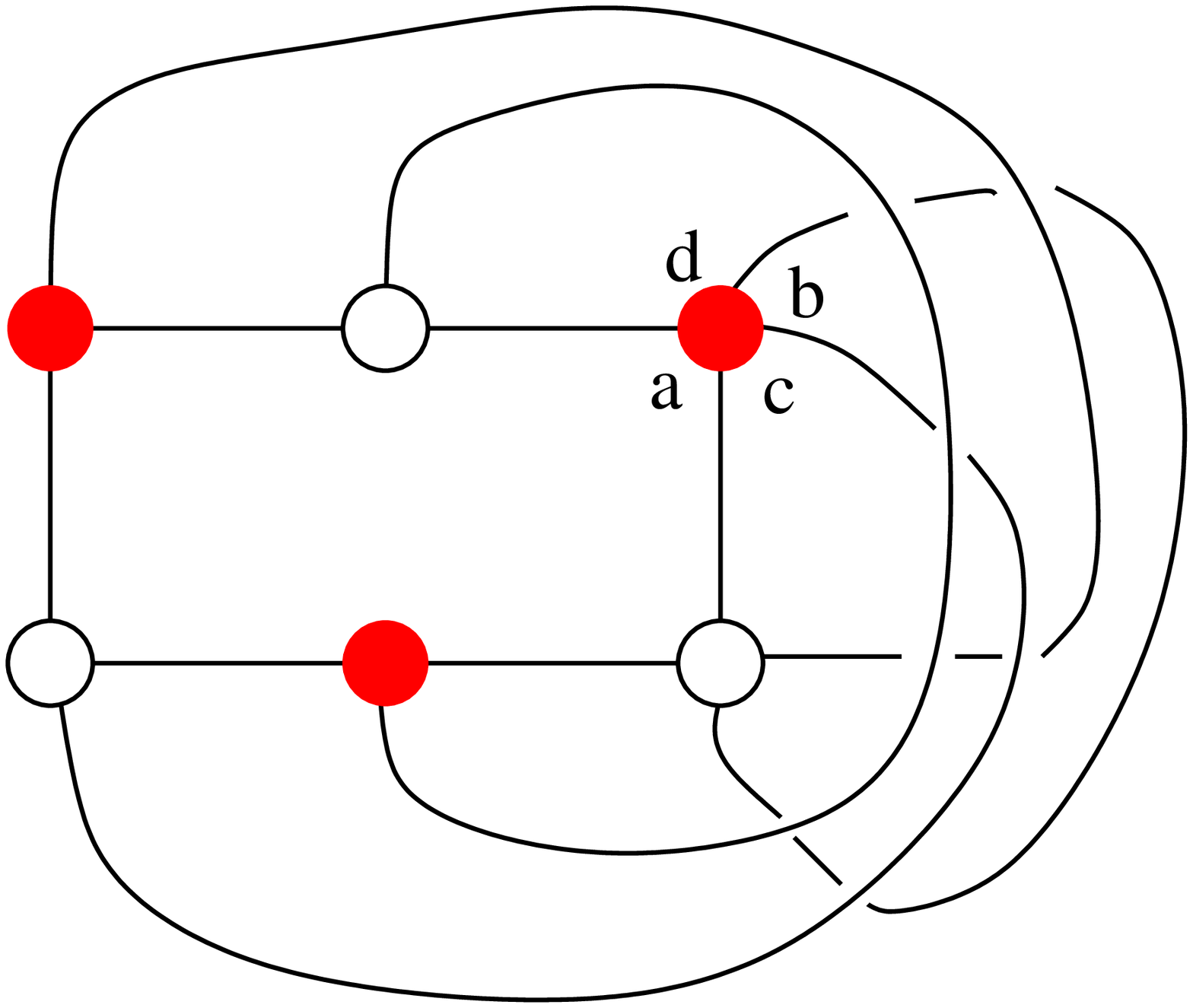,scale=0.15}&\\
\a_{15}=(1\,2\,3)_a(3\,2\,1)_c(2\,3)_b(2\,3)_d&
\a_{16}=(1\,2\,3)_a(1\,2\,3)_c(1\,3)_b(3\,2)_d&\\
[0.2cm]
\end{array}$}
\caption{Connected diagrams contributing to $\langle \sigma_{[3]}(a) \sigma_{[2]}(b) \sigma_{[3]}(c) \sigma_{[2]}(d) \rangle$ when $ |z_a|<|z_c|<|z_b|<|z_d|$.
} 
\label{2233all2}
\end{center}
\end{figure}

As a concrete application of the Feynman rules, let us consider the correlator
\be \label{corr1}
\langle \sigma_{[3]}(a) \sigma_{[2]}(b) \sigma_{[3]}(c) \sigma_{[2]}(d) \rangle \, ,\qquad |z_a|<|z_b|<|z_c|<|z_d|\,.
\ee  
All contributing (dual) diagrams are depicted in  Figure \ref{2233all1}.
There are six genus-zero and two genus-one diagrams and no higher 
genus contributions. If we consider instead the same correlator (\ref{corr1}) but with a different ordering of the coordinates, 
\be
\label{corr2}
\langle \sigma_{[3]}(a) \sigma_{[2]}(b) \sigma_{[3]}(c) \sigma_{[2]}(d) \rangle \, ,\qquad |z_a|<|z_c|<|z_b|<|z_d|\,,
\ee
another set of diagrams contributes to the calculation. They are depicted in Figure \ref{2233all2}.
There are again six genus zero and two genus one diagrams. As we are going to explain shortly, this is a general
property: the number of diagrams of given genus is the same for different radial orderings.

The genus of a diagram is easily computed. A diagram contributing to the $s$-point correlator of twist operators $\sigma_{[n_j]}$,
$j=1,\dots s$, and containing $c$ active colors, is a polygon with $v=s$ vertices, $f=2c$ faces and $e=\sum_{j=1}^s n_j$ edges. Its genus 
is then
\be\label{rh1}
\gg=\half\left(e-v-f+2\right)=\half \sum_{j=1}^s \left(n_j-1\right)-c+1 \,.
\ee 
We recognize  the {\it Riemann-Hurwitz} relation, which determines the genus of a $c$-sheeted covering of the sphere 
with $s$ ramification points of order $n_j$. The relation between diagrams and ramified coverings of the sphere
will be made explicit in the next subsection.

We can re-write the expansion of a correlator making manifest the sum over genera,
\be\label{pertexp2}
\langle \s_{[n_1]} (a_1)  \dots \s_{[n_s]} (a_s) \rangle_{conn} =
\sum_{\gg=0}^{\gg_{max}}\, \sum_{\a_\gg} C_{\gg,\a_\gg}(N,\{n_j\})\,\langle\prod_{j=1}^s\s_{g_j^{(\a_\gg)}}(a_j)\rangle_{\gg} \, ,
\ee 
where the index $ \a_\gg$ runs over the diagrams of genus $\gg$. We can now see more clearly the analogy
between twist correlators in the symmetric orbifold theory and correlators of gauge-invariant composite operators
in a conventional free gauge theory. In both cases any given correlator is given by a finite 
sum over diagrams,  with  the genus of the diagrams bounded by some
$\gg_{max}$ (which depends on the correlator). 
A curious difference is that  while in a gauge theory the basic vertex is cubic,
   in a symmetric orbifold the basic vertex is  quartic, associated to a twist-two field.
 So in a gauge theory  the dual
  of  a generic Feynman diagram gives a triangulation of a Riemann surface,
  but in a symmetric orbifold it gives a quadrangulation.

\subsection{Correspondence between diagrams and branched coverings}

We have discussed a combinatorial algorithm to associate a diagram to each non-trivial equivalence class of terms
 in the expansion of a correlator of gauge-invariant twist fields.  The diagrams have a nice geometric interpretation as well.
The actual computation of correlators  requires finding the covering surface(s) where the fields  $X_I$ are single-valued.
Each covering surface is a $c$-sheeted ramified covering\footnote{We recall the definition (see {\it e.g.} \cite{Lando2}): 
a continuous map $f: {\cal C} \to S^2$ from an oriented compact surface ${\cal C}$ to the sphere
is called a $c$-sheeted {\it ramified covering} of the sphere 
if: (i) the image of $f$ contains a finite
subset of points $\{ z_1, \dots z_s \}$, such that the map $f$ is a $c$ to 1 covering over the complement of this set; (ii) in a neighborhood
of each point $z_i$ one can introduce a complex coordinate, and in the neighborhood of each of the pre-images of this point
one can also introduce a complex coordinate $x$, such that the map takes the form $f(x) = x^{n_i}$. Here $n_i$ is an integer, called
the order of ramification of the point $z_i$.  A theorem of Riemann states that chosen
a complex structure on the $S^2$, there is a unique complex structure
on ${\cal C}$ such that $f$ is a meromorphic function.}
of the base sphere, with a ramification point at
 each insertion of a twist field: a twist field $\sigma_{[n]} (z, \bar z)$ corresponds to a ramification point
of order $n$ at $z$.  The computation of the correlator 
\be \label{corragain}
\langle \sigma_{[n_1]} (z_1, \bar z_1) \dots   \sigma_{[n_s]}(z_s, \bar z_s) \rangle
\ee
requires finding {\it all} the ramified coverings of the sphere  with 
ramification points of order $n_i$ at $z_i$. It turns out that for any given gauge-invariant correlator,
{\it the  different diagrams   correspond to the different ramified coverings.}

The enumeration of branched coverings of the sphere with specified ramification type is 
a classic mathematical problem, known as the {\it Hurwitz problem}.  There is a 
reformulation of the Hurwitz problem in terms of subgroups of the  symmetric group (see {\it e.g.} \cite{Lando2}).
Let us define $H(n_1, \dots, n_s)$ to be the number of different ramified coverings of the  sphere with $s$ ramification points of order $n_i$, $i=1,\dots s$.
Consider
 $s$-tuples $(g_1, g_2, \dots , g_s )$, where $g_k \in S_N$ is a {\it single-cycle} permutation of length $n_k$, and
define the equivalence relation
\be
(g_1, g_2, \dots , g_s ) \sim (g'_1, g'_2, \dots , g'_s )\;  \leftrightarrow \; 
 \exists h\in S_N: \; g'_k = h g_k h^{-1} \,\;\;  {\rm for} \; k=1\dots s\,.
\ee
A fundamental theorem in Hurwitz theory states that $H(n_1, \dots, n_s)$  is equal to the number of equivalence classes of $s$-tuples  $(g_1, g_2, \dots , g_s )$
such that 
\begin{enumerate} 
\item the   subgroup of $S_N$ generated
by the $g_k$ is transitive, and 
\item
$g_{\pi(1)} g_{\pi(2)} \dots g_{\pi(s)} =1$, \\
where $\pi \in S_s$ is some {\it arbitrary} ordering of the ramification points $\{1, 2, \dots, s\}$.
\end{enumerate}
It  follows that $H(n_1, \dots, n_s)$ is equal to the number of non-trivial equivalence classes 
of terms (as defined in the previous subsection) in the expansion of the connected part of the correlator (\ref{corragain});
 thus the number of ramified coverings with ramification type $\{ n_i \}$ is equal to the number of connected diagrams that contribute to the correlator (\ref{corragain}),
 as we had claimed.
 
 In the previous subsection it was natural to choose the arbitrary ordering $\pi$ 
to be the radial ordering of the coordinates $z_i$. We see that any other ordering would give the same
number of equivalence classes (=diagrams) -- this explains  why changing the ordering of the coordinates
gave the same number of diagrams in the example of Figure \ref{2233all2}. This means
for a given ramified covering,  there are  really $(s-1)!$ diagrams, 
corresponding to the different choices of $\pi$ (taking into account that cyclically related choices yield the same diagram):
but precisely one diagram for each ramified covering appears in the expansion
of a correlator of gauge-invariant twist fields, namely the one that corresponds to the choice of $\pi$ as radial ordering.

We have established that the {\it number} of diagrams in the expansion of a correlator
equals the {\it number} of contributing ramified coverings: we now proceed to associate
a particular diagram to  each ramified covering. We discuss two equivalent methods
 in the rest of this subsection.

\medskip
{\it  \noindent Cut-picture method}

\medskip

For a given branched covering of the sphere, we can think of the covering surface ${\cal C}$
as a union of $c$ Riemann sheets, each corresponding to a copy of the base sphere, and with 
 a system of cuts defined on each copy. It is always possible to deform the cuts in such a way
 that  a cut line  emanates from each ramification point and goes to infinity (in some direction).  Given such a picture,
 there is a natural cyclic
 ordering defined on the ramification points, obtained by  ordering the cut lines at infinity in counterclockwise fashion.
  The cyclic orderings on the different Riemann sheets are consistent with each other. 
 We can also assume that the point    $z=\infty$ on the base sphere is  a regular point  (no twist field insertion), and thus the point  of infinity in each sheet is also not ramified.

\begin{figure}[htbp]
 \begin{center}
$\begin{array}{|@{\hspace{0.1in}}c@{\hspace{0.3in}}|@{\hspace{0.3in}}c@{\hspace{0.1in}}|}
\hline
\epsfig{file=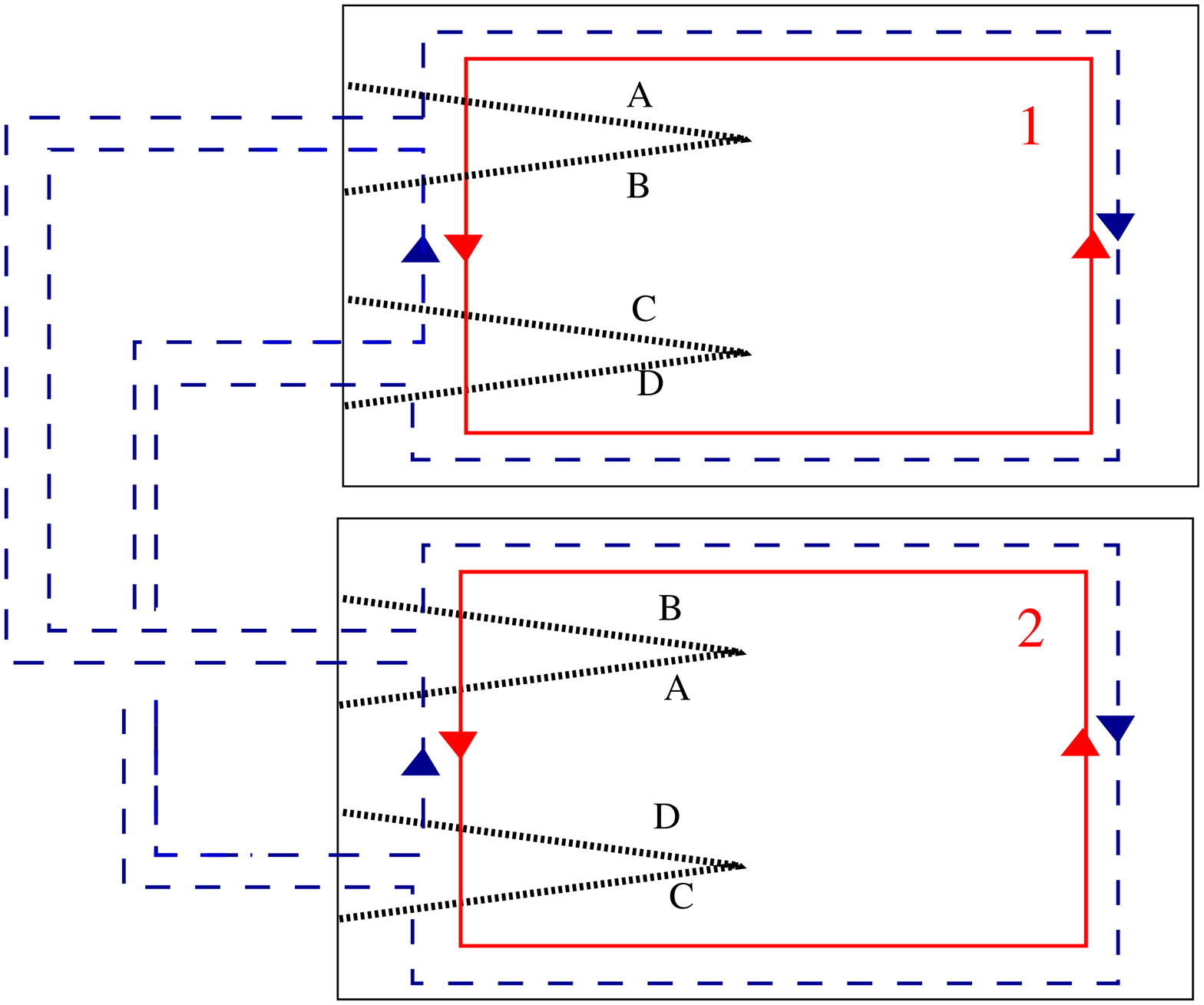,scale=0.3} &    \epsfig{file=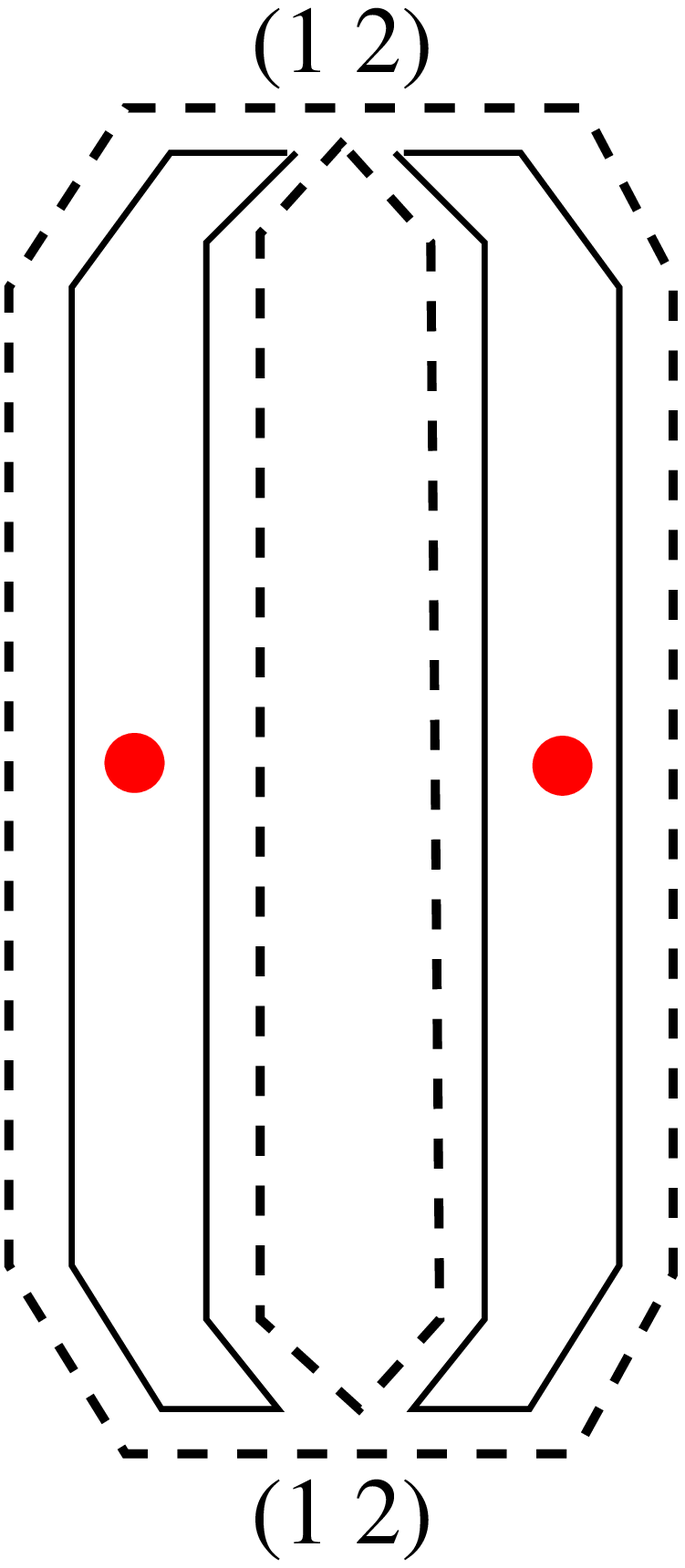,scale=0.3}
\\ [0.2cm]
\hline
\end{array}$
\caption{From the cut-picture to the diagram: example of a two-point correlator. 
The dashed and not dashed lines are the non-color and color loops respectively.
The two sides of cuts are labeled by capital letters. On the left we draw the cut picture and
show how  the diagram can be drawn on the different sheets. The color loops are associated
with each Riemann sheet and the non-color loops follow the ``trajectories'' around infinity, {\it i.e.} they follow
the trajectory of each color under the action of the twist operators.
Each trajectory encircles $z=\infty$ exactly once.  } \label{2p}
\end{center}
\end{figure}

\begin{figure}[htbp]
 \begin{center}
$\begin{array}{|@{\hspace{0.1in}}c@{\hspace{0.1in}}|@{\hspace{0.1in}}c@{\hspace{0.1in}}|}
\hline
\epsfig{file=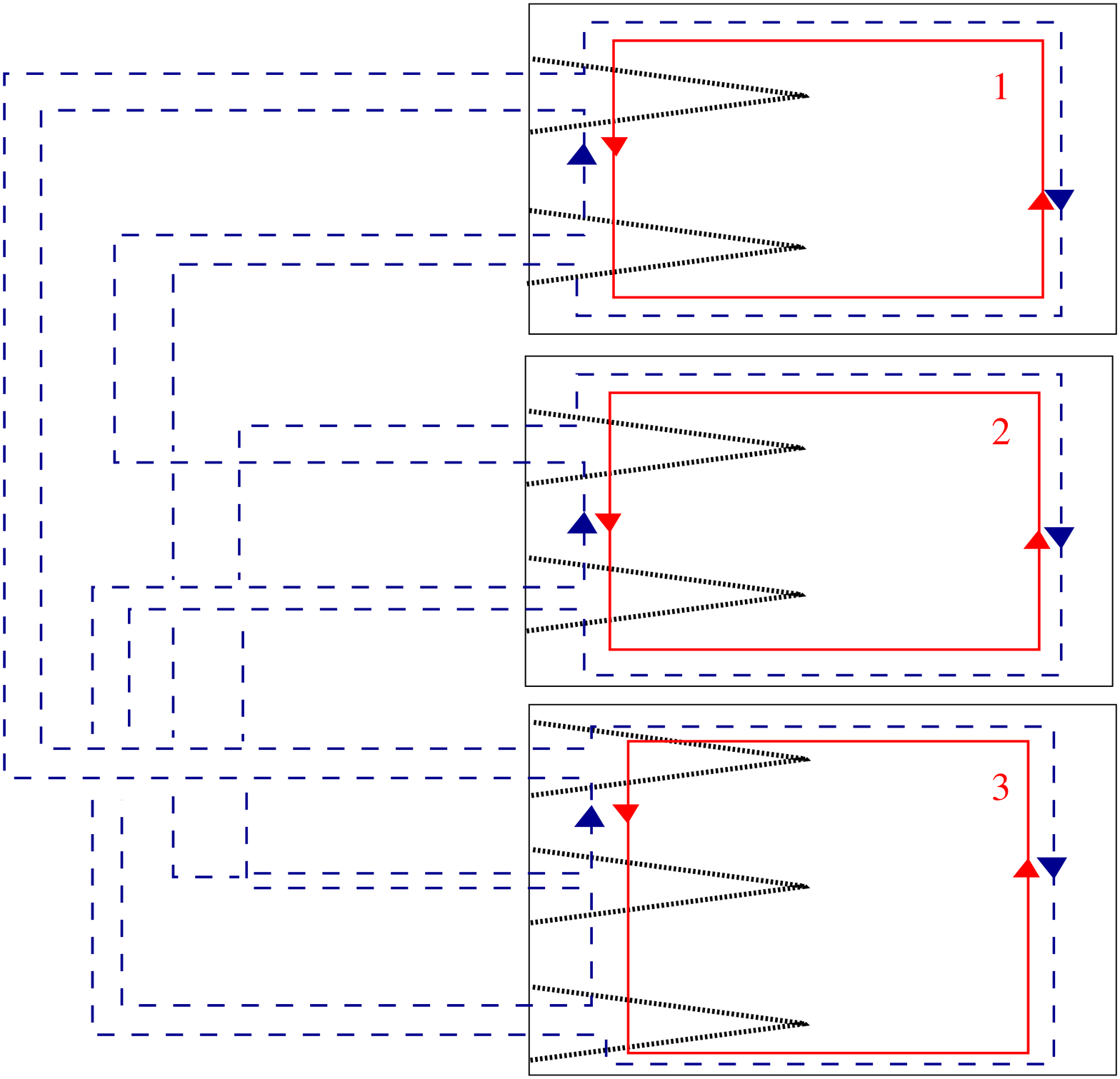,scale=0.2} &    \epsfig{file=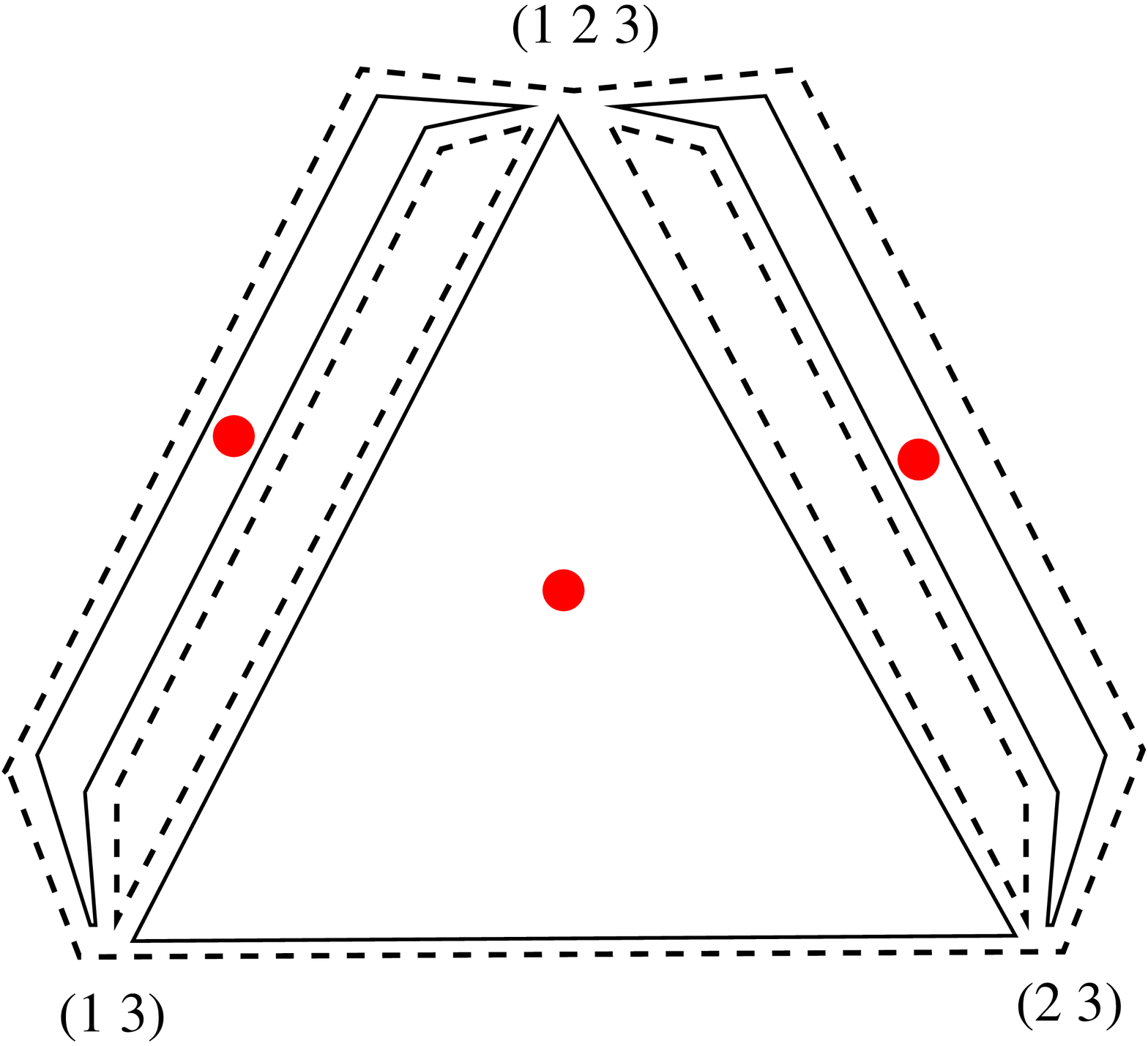,scale=0.2}
\\ [0.2cm]
\hline
\end{array}$
\caption{From the cut-picture to the diagram: example of  a genus zero branched covering contributing for a three-point correlator.} \label{3p}
\end{center}
\end{figure}

\begin{figure}[htbp]
 \begin{center}
$\begin{array}{|@{\hspace{0.1in}}c@{\hspace{0.1in}}|@{\hspace{0.1in}}c@{\hspace{0.1in}}|}
\hline
\epsfig{file=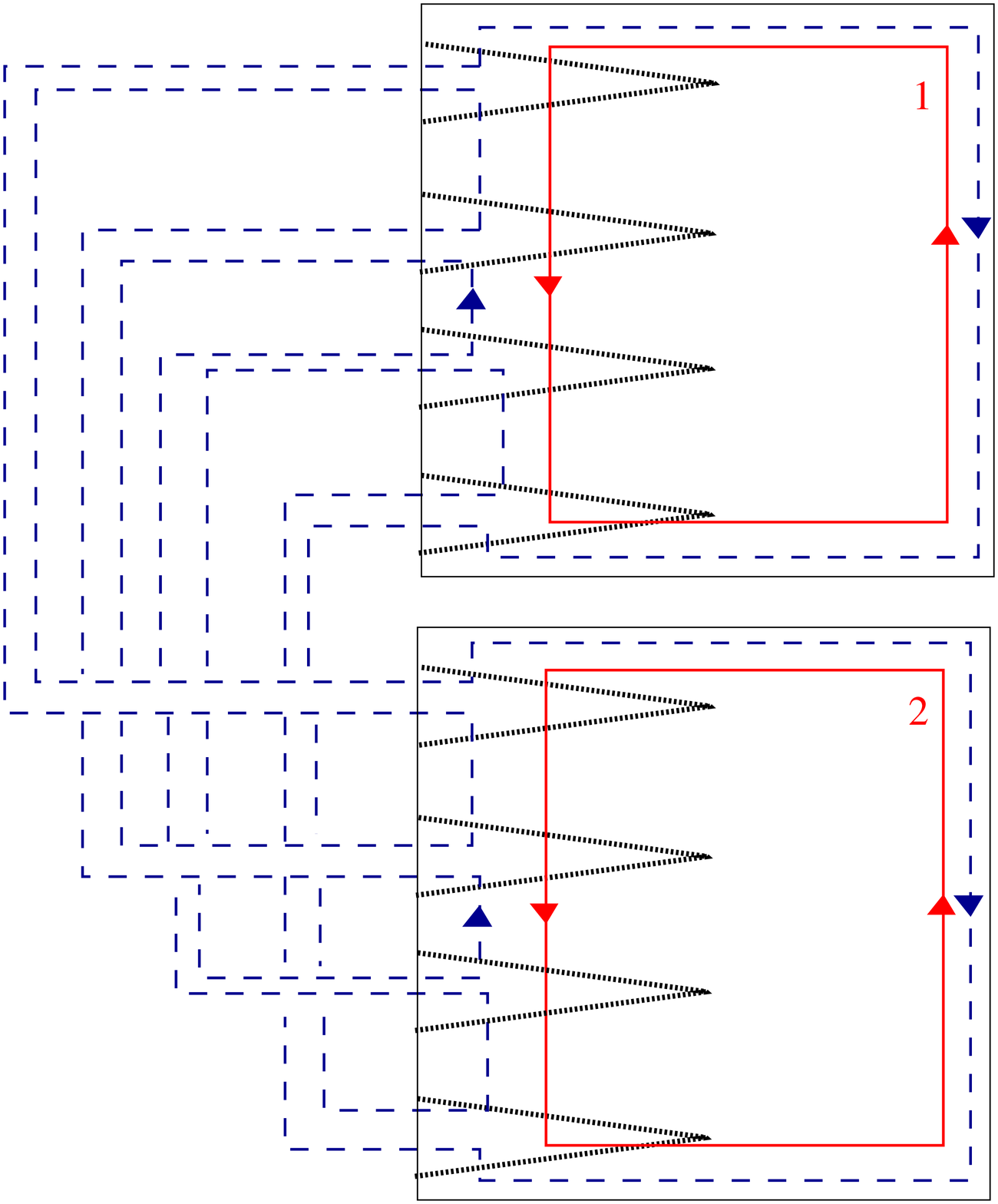,scale=0.2} &    \epsfig{file=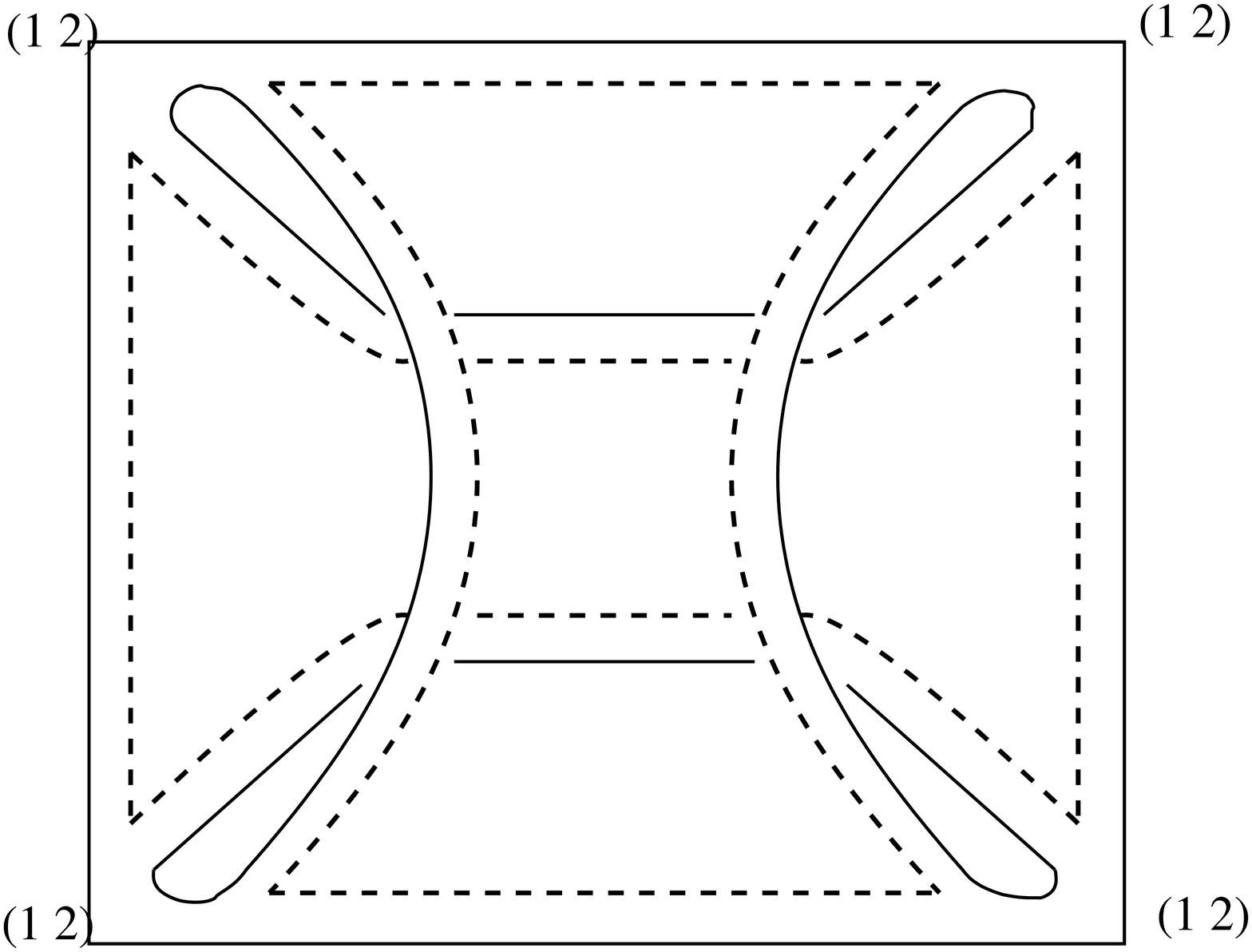,scale=0.2}
\\ [0.2cm]
\hline
\end{array}$
\caption{From the cut-picture to the diagram: example of a genus one branched covering contributing to  a four-point correlator.} \label{4p}
\end{center}
\end{figure}

To build the diagram from such a cut picture, we draw an oriented  loop on each Riemann
sheet  (with counterclockwise orientation as a matter of convention) -- this is the \textit{color} (solid) loop.
Next, beginning with one of the sheets, we start drawing another line just outside of the color loop, in clockwise direction; as we encounter
a cut, we move to a new sheet and keep going till we come back to the original sheet, and finally close the loop.
We repeat the procedure for each sheet. These are the \textit{non-color} loops.
Finally we smoothly deform the color and non-color lines to obtain the diagram. This procedure is illustrated in  few examples in Figures \ref{2p}, \ref{3p} and \ref{4p}.

The circular ordering and the fact that the  point at infinity 
on each sheet is a regular point imply that the diagrams are legal diagrams satisfying the two restrictions discussed in the previous subsection.


\medskip

{\noindent \it  Inverse-image method }

\medskip

\begin{figure}[htbp]
\begin{center}
$\begin{array}{c@{\hspace{0.25in}}c}
  \epsfig{file=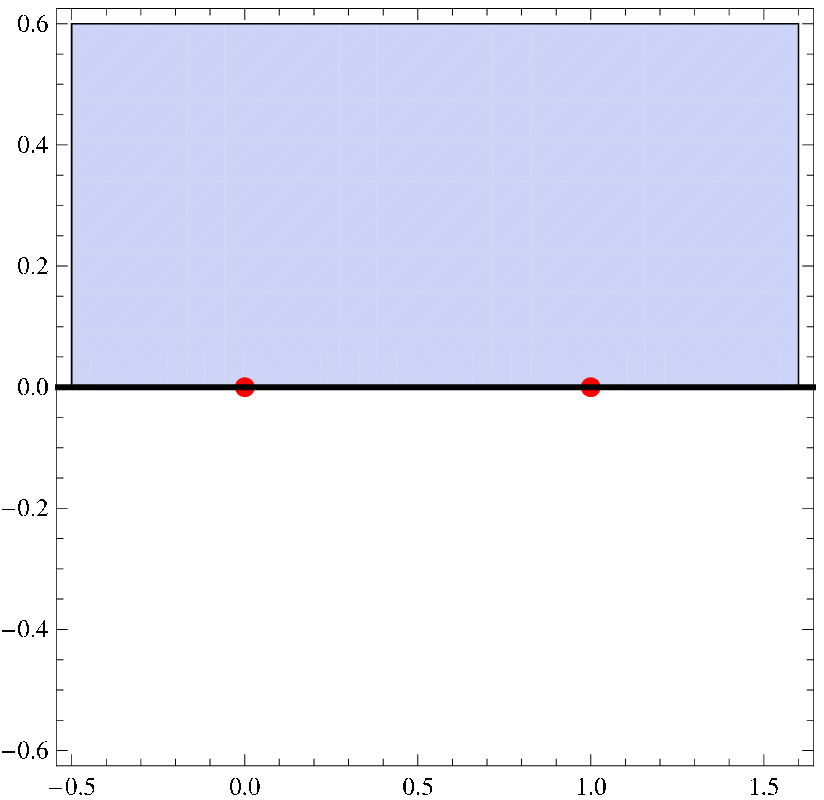,scale=0.55}& \epsfig{file=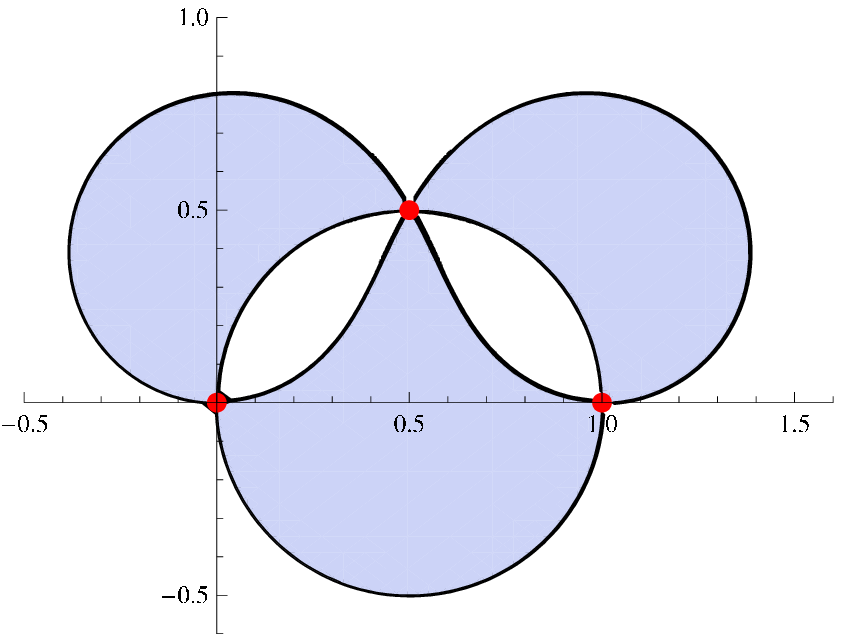,scale=0.7}  \\
\\ [0.2cm]
\end{array}$
\end{center}
  \caption{An example of the inverse-image method, applied to the correlator $\langle\s_{[2]}\s_{[2]}\s_{[3]}\rangle$.
On the left we show the base sphere with the twist -wo fields inserted at $z=0,1$ and the twist-three field at $z=\infty$. We connect the insertions
by a line going through the real axis. On the right we show the pre-image of this line on the covering sphere. The insertions are
now at $t=0,1$ and $t=\half+\half i$ respectively. The explicit branched covering map  is 
$z(t)=t^2 \frac{-3 i+(1 +3 i) t}{(-i+(1+i)t)^3}$. One finds  the same diagram as the one obtained from
the cut picture in Figure 8.} \label{math3}
\end{figure}

There is an even  simpler way to obtain the diagram from the branched covering.
We draw a closed loop without self-crossings on the base sphere, touching the positions of the twist fields. 
The closed loop divides the base sphere into two regions: after choosing an orientation for the loop, by convention
the ``color'' region is to the left  of the loop and  the ``non-color'' region to the right.  
The inverse image of this loop on the the covering space  defines the diagram.   
Figure \ref{math3} illustrates this procedure in a simple example.

Both methods to associate a diagram to a branched covering require to make some implicit choices.  In the first method, for a given branched covering there is some freedom in drawing the cuts,
implying different cyclic ordering on the operators. Likewise, in the second method the closed loop can be chosen to connect the locations of the twist fields in different orders.
In either method, different choices will result in a different diagram (for fixed branched covering).
This ambiguity is precisely the freedom in the choice of the arbitrary ordering $\pi$ in the theorem quoted above. In the application
to  $2d$ CFT, the natural ordering is radial ordering. {\it Any} ordering would lead to the same result, but
an ordering must be chosen to avoid overcounting. In Appendix B we  look at OPE limits of four-point functions and perform a check 
that our prescription of choosing one ordering (as opposed to summing over all possible orderings) gives indeed the correct normalization of correlators.

In Section \ref{modspace} we will expand on the relation between diagrams and ramified coverings. In particular we will
show that the different diagrams contributing to a given correlator can be connected to each other by a certain
``channel-crossing'' procedure.

\subsection{$N$ dependence.}\label{largeNsec}

Following \cite{Jevicki:1998bm, Lunin:2000yv}, we now determine the $N$ dependence of correlation functions. The first step
is to introduce normalized twist operators $\hat \sigma_{[n]} (z)$ with unit
two-point functions,
\be
\langle \hat \s_{[m]}(0) \hat \s_{[n]}(z)\rangle =\frac{ \delta_{mn}}{{|z|^{2 \Delta_n} } }\,.
\ee  
The two-point function
\be
\langle\s_{[m]}(0)\s_{[n]}(z)\rangle = \sum_{h \in S_N} \sum_{k \in S_N} \, \langle  \sigma_{h (1\dots m) h^{-1} }  (0) \, \sigma_{k (1\dots n) k^{-1} } (z) \rangle
\ee
 vanishes unless $n=m$, since the orders of  the two ramification
points must coincide for the covering surface to exist:
diagrammatically we can say that self-contractions of propagators at each vertex are not allowed.
There are $n$ fatgraph propagators joining the two vertices and thus $c = n$ active colors.
The graph is planar, as one can confirm from the Riemann-Hurwitz relation (\ref{rh1}), which gives $\gg = 0$ 
with $s=2$, $n_1 = n_2 = c= n$. 
There are $N!/(N-c)!$ possibilities of choosing the active colors. Moreover
for each of the gauge-invariant twist operators, we  can permute
the $N-n$ non-participating colors in all possible ways, contributing a factor of $((N-n)!)^2$. Finally
there is an extra  factor of $n$ which accounts for the freedom to make
a cyclic permutation of the chosen colors; diagrammatically it counts the number
of Wick contractions. All in all,
\be
\langle\s_{[m]}(0)\s_{[n]}(z)\rangle =n\,N!(N-n)! \, \frac{\delta_{nm}}{|z|^{\Delta_n}}\, ,
\ee 
which implies
\be\label{nor}
\hat\s_{[n]}\equiv\frac{1}{\sqrt{n\,N!(N-n)!}}\, \s_{[n]}.
\ee
We are interested in the  $N$ dependence of $s$-point functions of normalized operators. Let us re-write
the expansion (\ref{pertexp2}) for normalized operators,
\be\label{pertexp3}
\langle \hat \s_{[n_1]} (a_1)  \dots \hat \s_{[n_s]} (a_s) \rangle_{conn} =
\sum_{\gg=0}^{\gg_{max}}\, \sum_{\a_\gg} \hat C_{\gg,\a_\gg}(N,\{n_j\})\,\langle\prod_{j=1}^s\s_{g_j^{(\a_\gg)}}(a_j)\rangle_{\gg} \, ,
\ee 
where we recall that $\{ \alpha_\gg \}$ is the set of diagrams contributing to the correlator at genus $\gg$.
 We find
\be\label{largeNexp}
\hat {C}_{\gg,\a_\gg}(N,\{n_k\})
=A( \{ n_k \} )  \left[\prod_{j=1}^s\sqrt{\frac{(N-n_j)!}{n_j\,N!}}
\right]\,\frac{N!}{(N-c)!} \, , \;\; c = 1-\gg -\frac{s}{2} + \frac{1}{2} \sum_{j=1}^s n_j \,. 
\ee 
The coefficient $A( \{ n_i \} )$ is the  $N$-independent combinatorial factor that arises from Wick contractions;
it simply accounts for the freedom of cyclic re-ordering of each vertex, so
\be
A( \{ n_k \} ) = \prod_{j=1}^s n_j \,.
\ee
The term in square brackets  comes from the normalization factors of each operator and  from the number of permutations of the colors not participating in the given cycle.  Finally the last factor counts the number of ways to pick the $c$ (ordered) active colors; for given
$\{ n_i \}$ and given genus, $c$ is determined by the Riemann-Hurwitz relation.  Using the Stirling formula, we can
expand $\hat {C}_{\gg,\a_\gg}(N,\{n_k\})$ for large $N$,
\be
\hat {C}_{\gg,\a_\gg}(N,\{n_k\}) =  N^{1-\gg - \frac{s}{2}}  \left( a_0  + \frac{a_1}{N} + \dots \right) \,.
\ee
The leading $N$ dependence is very simple, but for given $\gg$ and $s$ there is a whole infinite series of subleading terms.
This is to be contrasted with the standard case of a $U(N)$ gauge theory, where the $N$ dependence
of correlators of normalized single trace operators is precisely $N^{2-2\gg - s}$. 

The functions $\hat C_{\gg,\a_\gg}(N,\{n_k\})$ are  independent of the specific diagram
$\a_\gg$: for a given correlator they are functions only of $\gg$ and $N$. This property can be understood
heuristically by recalling that
 a correlator of gauge-invariant twist fields must be single-valued
as a function of the  coordinates of the twist fields. However, as we will see, the contributions of the individual diagrams
 are in general not single valued. The different diagrams at a given genus 
 correspond to  the different zeros of a meromorphic function, and  to cancel
 the branch cuts we must  take the sum of such roots with equal weight. The
 diagrams should then have  a common $N$ dependence for their sum to produce a single-valued
 correlator.

If we wish to deform the symmetric orbifold CFT while maintaing
a sensible large $N$ limit,  the coupling of the deformation term should be scaled appropriately.
For example we may add to the action a two-cycle term (a blow-up mode of the orbifold),
schematically\footnote{In the case of ${\rm Sym}^N ({\cal M}_4)$, the precise form of the deformation 
that preserves $(4,4)$ superconformal invariance can be found for example in \cite{Gava:2002xb}.} 
\be
\delta S = f \, \int d^2z\; \hat \s_{[2]}(z) \,.
\ee 
We deduce from \eqref{largeNexp}  that
a sensible large $N$ limit requires 
\be
N\to\infty,\qquad f \, N^{-\half}\equiv\lambda =\text{fixed} \, .
\ee 
The combination $\lambda$ plays the role of  the 't Hooft coupling. 

\subsection{Computing the correlators}

To evaluate correlators of twist fields, we can use the covering surface(s) in at least two  ways: the stress-tensor
method of Dixon et al. \cite{Dixon:1986qv} and the path-integral method of Lunin and Mathur \cite{Lunin:2000yv}.

The standard approach to the calculation of twist correlators is the stress-energy tensor method~\cite{Dixon:1986qv},
which is  applicable to $s$-point functions with $s>3$.  To evaluate, say, a four-point function, we consider
 the quantity
\be
g(z,u)=\frac{\langle T(z)\phi_1(0)\phi_2(1)\phi_3(u)\phi_4(\infty) \rangle}{\langle \phi_1(0)\phi_2(1)\phi_3(u)\phi_4(\infty) \rangle} \, ,
\ee 
where $T(z)$ is the stress-energy tensor and $\phi_i(z)$ denotes schematically the holomorphic part of a  primary operator in a twisted sector.
 As we have seen, several covering surfaces (one for each diagram $\alpha_\gg$) contribute to the correlator.
For each $\alpha_\gg$, we can find  $g(z,u)$ by map to the covering surface, taking into account the well-known
transformation properties of $T$ and of the primaries $\phi$. 
Using
the OPE of $T(z)$ with
$\phi(u)$,
\be
T(z)\phi_2(u)=\frac{\Delta_{\phi_3}}{(z-u)^2}\phi_3(u)+\frac{1}{z-u}\d \phi_3(u)+\dots 
\ee 
 we deduce
\be
\d_u \ln G(u)_{\alpha_\gg}=\left\{g(z,u)_{\alpha_\gg  }\right\}_{\frac{1}{z-u}} \,.
\ee 
Here  $G(u)_{\alpha_\gg} \equiv \langle   \phi_1(0)  \phi_3(1) \phi_3(u)\phi_4(\infty) \rangle \rangle_{\alpha_\gg}$ 
is the contribution to the holomorphic part of the correlator from the covering surface $\alpha_\gg$;
 on the right hand side we take the coefficient of
$\frac{1}{z-u}$ in the expansion of $g(z,u)$. This equations determine $G(u)_{\alpha_\gg}$
up to a normalization factor.   After repeating the same calculation for the anti-holomorphic part,  we sum the partial
contributions $G(u, \bar u)_{\alpha_gg}$ over all the covering surfaces $\{ \alpha_g \}$. The relative normalizations
can be fixed by requiring that the result is well-defined (single-valued) on the base sphere,
while the overall normalization can be fixed by looking at OPE limits. Three-point functions can be obtained indirectly by factorization of four-point functions.

Lunin and Mathur \cite{Lunin:2000yv} devised an alternative computational method that uses
 directly   the path integral definition of the theory. In going to the covering surface, we have to take  into account the transformation
of the measure of the path integral, which may be encoded in a certain  Liouville action.
This approach has the advantage of keeping track of  the absolute normalization of correlators and can
thus be directly applied to three-point functions, for which  the only
non-trivial piece of information is indeed the overall normalization. 
  In Appendix \ref{luninmathur} we  apply the results of \cite{Lunin:2000yv}
 to evaluate the intrinsic normalization of some four-point functions, and check the consistency of  our
 ``Feynman rules''  in various OPE limits.

\section{Planar covering surfaces for four-point correlators}

It is in general difficult to find explicit expressions for the branched covering maps. In this Section we focus on the simplest non-trivial class of branched coverings, 
the genus zero covering surfaces with four ramification points.
We will present a general algorithm to obtain them, in terms of polynomial solutions of Heun's differential equation.
We will also study in detail some simple examples, with the aim of gaining more insight into the relation
between covering maps and associated diagrams.

\subsection{Heun's equation}\label{heunsec}

Consider the four-point correlator
\be
\langle \sigma_{[n_1]}(z_1)   \sigma_{[n_2]}(z_2)  \sigma_{[n_3]}(z_3)  \sigma_{[n_4]}(z_4)   \rangle \, ,
\ee
defined on the base  sphere $S^2_{base}$. We will always use the letter $z$ to denote the uniformizing coordinate on  $S^2_{base}$. 
By an $SL(2, \mathbb{C})$ transformation, we fix
\be \label{zi}
z_1 = 0 \, ,\quad z_2 = 1 \, , \quad z_3 = u \, , \quad z_4 = \infty \,.
\ee
We will denote with $t$ the uniformizing coordinate on the covering surface, also taken to be a sphere,  $S^2_{cover}$
 The goal is to find  all the covering maps 
 \be
 t  \in S^2_{cover} \to z(t) \in S^2_{base}
 \ee
with  four ramification points $z_i$ of order $n_i$. The ramification points $z_i$ have unique pre-images $t_i$ on $S^2_{cover}$,
which by another $SL(2, \mathbb{C})$ transformation we fix to
\be \label{ti}
t_1 = 0 \, ,\quad t_2 = 1 \, , \quad t_3 = x \, , \quad t_4 = \infty \,.
\ee
At this stage the location $x$ of the pre-image of ramification at $z=u$  is a parameter of the map. We will
see that there is a discrete set of possible values for $x$  for fixed value of $u$.
The   Riemann-Hurwitz relation (\ref{rh1}) gives the number $c$ of sheets in the ramified covering,
\be \label{rh2}
c = \frac{n_1 +n_2 +n_3 +n_4}{2} -1 \,.
\ee
In CFT language, the $c$  copies (colors) of the field, 
  $X_I(z)$, $I = 1, \dots c$, are traded
for a single field $X(t_I(z)) \in S^2_{cover}$, where $t_I(z) \in S^2_{cover}$ are the pre-images
of the generic point  $z \in S^2_{base}$. As $z$ approaches a ramification point $z_i$, 
$n_i$ of its pre-images converge to the same point $t_i$ on $S^2_{cover}$.

In summary, we are  looking for a $c$-sheeted map $z: S^2_{cover} \to S^2_{base}$
with the following branching behavior:
\be \label{b0}
\lim_{t \rightarrow 0} z(t) &\sim &  b_1 t^{n_1},
\\ \label{b1}
\lim_{t \rightarrow 1} z(t) &\sim& 1 + b_2(t-1)^{n_2},
\\ \label{bx}
\lim_{t \rightarrow x} z(t) &\sim& u + b_3(t-x)^{n_3},
\\
\lim_{t \rightarrow \infty } z(t) &\sim& b_4 t^{n_4} \,.
\label{branch-inf}
\ee
We will generalize to our case the technique used in \cite{Lunin:2000yv}, where coverings with three branching points were considered.
We build the map as a quotient of two polynomials,
\be\label{mapf12}
z(t) = \frac{f_1(t)}{f_2(t)} \, ,
\ee
of degrees $d_1$ and $d_2$ respectively, which we can assume to have no common factor.
  From (\ref{b0}--\ref{bx}), we must have
$f_2(t)\neq 0$   for $t =0, 1, x$, 
while from (\ref{branch-inf}) we deduce
\be \label{d2=}
d_2 = d_1 -n_4 \,.
\ee
In particular $d_1>d_2$, and since $z(t)= z$ should have
generically $c$ solutions, we identity $c$ with the degree $d_1$.  Then
from  (\ref{rh2}),
\be
d_1 = c = \frac{n_1 +n_2 +n_3 +n_4}{2} -1 \, ,
\ee
and thus clearly
\be
d_2  = \frac{n_1 +n_2 +n_3 - n_4}{2} -1 \,.
\ee
Consider now the linear combination
\be
f(t) = \alpha f_1(t) + \beta f_2(t) \, ,
\label{lin-comb}
\ee
which satisfies
\be
\left|
\begin{array}{ccc}
f & f' & f'' \\
f_1 & f_1' & f_1'' \\
f_2 & f_2' & f_2''
\end{array}
\right| =0\,.
\ee
Expanding the determinant, we get the following equation for $f$
\be
W (t) f'' -W'(t) f' - c(t)f = 0 \, ,
\label{feq}
\ee
where we have defined
\bea
W(t)  & \equiv &  f_1'(t) f_2(t) - f_1(t) f_2'(t)
=f^2_2(t) \frac{dz(t)}{dt}
\label{wzp} \\
c(t) &  \equiv & f_2' f_1'' - f_1'f_2'' \,.
\eea
The strategy is to determine the functions  $W(t)$ and~$c(t)$ from the branching behavior (\ref{b0}--\ref{branch-inf}),
and then solve the differential equation (\ref{feq}) for $f$:  its two solutions will be
identified with~$f_1$ and~$f_2$. We claim that $W$ is given by
\be
W(t) &=& Ct^{n_1-1}(t-1)^{n_2-1}(t-x)^{n_3-1}
\label{wsing}
\ee
for some constant $C$. Indeed $W$ should be a polynomial of degree
$d_1 + d_2 -1 =  n_1 +n_2 +n_3 - 3$,
whose zeroes at $0,1,x$ are fixed from (\ref{wzp}) as
\begin{equation}
\lim _{t \rightarrow 0}\frac{dz(t)}{dt} \sim t^{n_1-1}\, \quad
\lim _{t \rightarrow 1}\frac{dz(t)}{dt} \sim (t-1)^{n_2-1}\, ,\quad
\lim _{t \rightarrow x}\frac{dz(t)}{dt} \sim  (t-x)^{n_3-1} \,.
\label{zerosz}
\end{equation}
The unique such polynomial 
is~(\ref{wsing}).
To obtain $c(t)$, we expand  (\ref{feq}) around  $t = 0$,
\be
-C (n_1-1) t^{n_1-2} f'(0) + c(t) f(0) + O(t^{n_1-1}) =0\,. 
\ee
Since $f(0)$ is in general non-vanishing (because $f_2(0) \neq 0$), we must have
\be
c(t) \sim t^{n_1-2} + O(t^{n_1-1}) \, \quad t \to 0\,.
\label{clim}
\ee
A similar analysis around the points $t=1$ and $t=x$, and the requirement that $c(t)$ should be a polynomial of degree
$n_1+n_2+n_3-5$, 
lead uniquely to
\be
c(t)&=&  t^{n_1-2}(t-1)^{n_2-2}(t-x)^{n_3-2} \left( t \tilde{\gamma} + \tilde{q} \right) \, ,
\ee
where $\tilde{\gamma}$ and $\tilde{q}$ are arbitrary constants.
It is convenient to write the derivative of $W$ (from  (\ref{wsing})) as
\be
W'(t)= C t^{n_1-2}(t-1)^{n_2-2}(t-x)^{n_3-2} P(t,x) \, ,
\ee
where we have defined
\be
P(t,x) \equiv (n_1-1)(t-1)(t-x) + (n_2 -1)t(t-x) + (n_3-1)t(t-1)\,.
\ee
The differential equation (\ref{feq}) for $f$ now  becomes, after dividing by $C t^{n_1-2}(t-1)^{n_2-2}(t-x)^{n_3-2}$,
\be
t(t-1)(t-x) f'' - P(t,x) f' + ( \gamma t + q ) f =0 \,,
\label{eqfg}
\ee
where $\gamma \equiv \tilde{\gamma}/C$ and $q \equiv \tilde{q}/C$.
We can fix $\gamma$ by taking the limit $t\rightarrow \infty$ in (\ref{eqfg}).
Assuming that $ f(t) \sim t^{d}$ for $t \to \infty$, we find
\be
d(d-1) - d(n_1+n_2+n_3-3) +\gamma =0\,.
\label{eqforgamma}
\ee
The two solutions to this equation are the degrees $d_1$ and $d_2$ of $f_1$ and $f_2$,
thus we learn
\be
\gamma = d_1 d_2.
\ee
The differential equation for $f$ is finally
\be\label{heun}
 f''  - \left[ \frac{n_1-1}{t }  + \frac{n_2-1}{(t-1)} + \frac{n_3-1}{(t-x)} \right] f'
  + \frac{( d_1 d_2 t + q )}{t(t-1)(t-x)} f =0\,.
\ee
This is Heun's equation.\footnote{For a comprehensive discussion of this differential equation see \cite{heun}.}

\subsection{Polynomial solutions of Heun's equation}\label{polheun}

The parameters of Heun's equation are known functions of $n_i$, $i=1,2,3,4$, except for  $q$ and $x$.
As we now proceed to show,  the parameters $q$ and $x$
are fixed by requiring that the two solutions $f_1(t)$ and $f_2(t)$ 
 are polynomial and that the map  $z(t) \equiv f_1(t)/f_2(t)$ sends $t=x$ into $z(x)=u$.

We are looking for two polynomial solutions of Heun's equation  (\ref{heun}), one of degree $d_1$ and another of degree $d_2$. Expanding either solution
in a power series,
\be
f(t)=\sum_{k=0}^\infty c_k\,t^k \, ,
\ee we demand that  $c_k=0$ for $k>d_i$. The differential equation (\ref{heun}) gives the recursion relation
\be\label{recurs}
c_{k+1}(k+1)(k-n_1+1)x&-&c_{k}(k\left[(k-n_1)(1+x)+x(1-n_2)+1-n_3\right]+q) \\&+&c_{k-1}(k-1-d_1)(k-1-d_2)=0\, ,\nonumber
\ee 
for $ k \geq 0$, with the understanding that $c_{-1} \equiv 0$. We may take $c_0$ and
 $c_{n_1}$  as input and solve for all other $c_k$'s from the recursion. The requirement that the recursion truncates
gives a first relation between the  parameters $q$ and $x$, according to the following algorithm. 
If the four cycles have generic lengths $n_a\leq n_b\leq n_c\leq n_d$, we can always use  $SL(2,\mathbb{C})$ to set
$n_a = n_4$ and $n_b = n_1$,  so that
\be
d_2-n_1=\half\left(n_c+n_d-n_a-n_b\right)-1\geq -1 \, .
\ee
There are  two  cases: (i)  $d_2-n_1\geq 0$, and (ii) $d_2 - n_1 = -1$.

Consider the first case,
\be \label{condpol}n_1\leq d_2<d_1\,  ,\ee
where the second inequality is true by construction (recall  (\ref{d2=})).
We find $f_1(t)$ by solving the recursion with
 $c_0=0$ and $c_{n_1}=1$, and demanding that 
 \begin{equation} \label{1rel}
 c_{d_1+1}(q, x) = 0 \, . 
 \end{equation}
  It is clear from the recursion that $c_{d_1 +1}(q,x)$
 is a polynomial in $q$ and $x$. This procedure yields $f_1(t)$
 as a polynomial of degree $d_1$  proportional to $t^{n_1}$.
To obtain $f_2(t)$, we switch on 
both $c_0  \neq 0$ and $c_{n_1}\neq0$, and fix
$c_{n_1 +1}/c_0$ 
by requiring that $c_{d_2+1}=0$. This determines $f_2(t)$ up to an overall constant; 
by construction it is  a polynomial of degree
$d_2$ which is non vanishing at $t=0$. Finally we fix the overall 
constant of $f_2$ by demanding that  $z(1) \equiv \frac{f_1(1)}{f_2(1)} =1$. 

In the second case,  $d_2-n_1=-1$,
 we constrain $q$ and find $f_1$ as above, and $f_2$ is simply the solution with $c_0=1$ and $c_{n_1}=0$,
as in this case $d_2=n_1-1$ and setting $c_{n_1}=0$ makes this solution polynomial regardless of the value of $q$.

This procedure yields a map $z(t ; q,x)$ depending on the two parameters $q$ and $x$. So far
  $q$ and $x$ are constrained by one polynomial relation, equ.(\ref{1rel}). A second relation
  arises by recalling
that $x$ was defined as the pre-image of $u$ on the covering sphere, hence
\be\label{2rel}
 z(t=x; q,x)  \equiv v(q,x) = u \, .
\ee 
The function $v(q,x ) \equiv z(t=x; q,x)$ is a rational function in $x$ and $q$.  So for fixed $u$,
 $q$ and $x$ are determined by the system of  two polynomial equations
\begin{eqnarray}\label{3rel}
c_{d_1+1} (q,x) & = & 0 \\
v(q,x ) & = & u \, , \nonumber
\end{eqnarray}
which has a finite set of solutions $\{ (x_i(u), q_i(u) \}$. Substituting back in $z(t; q, x)$, we
find for fixed $u$
a finite set of maps $\{ z_j (t) \}$, for $j=1, \dots M$. As discussed in Section 2, the number $M$ of maps
corresponds to the number of equivalence classes of terms in the expansion of the correlators, and there is
a 1-1 correspondence between the maps and the diagrams produced by the Feynman rules.

Let us examine in detail a specific example,
\be
n_1=n_4=n\, ,\quad n_2=n_3=2 \,\quad  \longrightarrow \quad d_1=n+1\,, \quad d_2=1 \,.
\ee 
Condition \eqref{condpol} is not satisfied,
 but  this case is simple enough that it can be  solved without invoking any $SL(2,\mathbb{C})$ transformations.\footnote{The map for this case was also
obtained in \cite{Lunin:2000yv}.}  In this case choosing $c_0=0$ leaves only $c_{n}$ and $c_{n+1}$ undetermined. Taking $k=n$ in (\ref{recurs})
we get
\be
c_{n+1}=-\frac{n(1+x)-q}{x(n+1)}c_n \,.
\ee With $k=n+1$ in (\ref{recurs}), demanding  $c_{n+2}=0$,
\be
-q\,c_{n+1}-(n-1)\,c_{n}=0 \,.
\ee Thus,  $q$ satisfies a simple quadratic equation,
\be \label{qquadratic}
q^2-n(x+1)\,q+x(n^2-1)=0 \, ,
\ee 
which gives
\be
q_\pm =\half n (1+x) \pm \half\sqrt{n^2(1+x)^2-4x(n^2-1)} \,.
\ee 
The function $f_1(t)$ is thus
\be
f_1^{\pm}(t)=\,t^n\,\left(1-\frac{n(1+x)\mp \sqrt{n^2 (1-x)^2 + 4 x}}{2x(n+1)}\,t\right) \,.
\ee To find the second solution we take $c_0=1$ and immediately find
\be
c_1=\frac{q}{x(1-n)} \,.
\ee Demanding the vanishing of $c_2$,
\be
(n(1+x)-q)c_1+(n+1)=0 \, ,
\ee which implies
\be
q^2-n(1+x)q+x(n^2-1)=0 \, ,
\ee the same condition as above. From here we obtain for $f_2$,
\be
f_2^{\pm}(t)=1-\frac{n(1+x)\pm \sqrt{n^2 (1-x)^2 + 4 x}}{2x(n-1)}\,t \,.
\ee Finally the map is given by
\be
z^\pm(t; x)=\left(\frac{f_2^\pm(1)}{f_1^\pm(1)}\right)\,\frac{f_1^\pm(t)}{f_2^\pm(t)} \,.
\ee  
For fixed $x$ there are two possibilities, corresponding to the two values of $q$.
Finally we require that $x$  is the pre-image of $u$,
\be\label{uofx1}
z^\pm(t=x; x) \equiv v^\pm(x)=u \, .
\ee 
Explicitly we obtain
\be
v^\pm(x)=
\half x^{n-1} \left(2 x + n^2 (x-1)^2  \mp n(x-1) \sqrt{n^2 (1-x)^2 + 4 x}\right) \,.
\ee Thus  \eqref{uofx1} has $2n$ solutions. Note that 
 for both choices the set of solutions to \eqref{uofx1}
will be the same: if we pick {\it either} $q^+$ or $q^-$ 
and  {\it all}  solutions for $x$ in \eqref{uofx1}, each map $z_j(t)$ is obtained once.

It is instructive to count the number of different diagrams/equivalence classes that we have in this simple example.
To count the equivalence classes we  count the number of ways we can satisfy 
\be\label{condcol2}
(n)_a\,(2)_b\,(2)_c \,(n)_d =1 
\ee modulo global $S_N$ transformations. The Riemann-Hurwitz relation implies that the number of  colors
 is $c=n+1$. Assuming $(n)_a$ and $(2)_b$ have {\it one} overlapping index, say
\be
(n)_a\,(2)_b=(1\, 2\,\dots\, n)_a(1 \, n+1)_b=(1\, 2\,\dots\, n\, n+1) \, ,
\ee  we get that $(n)_d$ and $(2)_c$ have must also have { one} overlapping index. Modulo global $S_N$ 
transformations there are exactly $n+1$ possibilities for this case, indeed
 fixing $S_N$ by choosing the common
 index of  $(n)_a$ and $(2)_b$ we have $n+1$ choices for the common index of $(n)_d$ and $(2)_c$. 
Assume now that $(n)_a$ and $(2)_b$ have {\it  two} overlapping indices, say
\be
(n)_a\,(2)_b=(1\, 2\,\dots\, n)_a(1 \, k)_b=(1\, 2\,\dots\, k-1)(k\,\dots\,n) \,.
\ee 
Now we must have either $k = n$ or $k=2$. Indeed if $k \neq 2$ and $k \neq n$,  $n$ different colors would appear in cycles $a$ and $b$ (and 
the same colors would have to appear in $c$ and $d$
cycles in order to satisfy \eqref{condcol2}),  in contradiction with the the fact that the total number of active colors is $c=n+1$. We can choose
 $k=n$ as $k=2$ choice is related to this by a global $S_N$ transformation.  The cycles $(n)_d$ and $(2)_c$ also have two overlapping indices,
\be
(2)_c\,(n)_d=(k\, n+1)_c(n+1\,k-1\,k-2 \,\dots 1\, n-1\, \dots\,k+1\, k)_d =(n-1\, n-2\,\dots\, 1)\,.\nonumber\\ 
\ee  After fixing the global $S_N$ by choosing the two cycles $(2)_b$ and $(n)_a$ we have $n-1$ possibilities
to specify $(2)_c$ and $(n)_d$ by choosing a common color of $(2)_c$ and  $(n)_a$.
In summary we have $(n+1)+(n-1)=2n$ equivalence classes, exactly as the number of different maps, {\it i.e.}
solutions to \eqref{uofx1}. The actual (six planar)  diagrams in the case of $n=3$ are depicted in Figure \ref{2233all2}.

\subsection{Polynomial case}\label{extremcasesec}

It is easy to solve for $z(t)$ when $d_2=0$: the map is just a polynomial
and there is no need to use Heun's equation. We will refer to the correlators whose with polynomial covering map 
as polynomial correlators.
Setting $d_2=0$  corresponds to taking
\be n_4=n_1+n_2+n_3-2 = d_1\,.\ee 
Then we also have $d_1=n_4$. 
From the monodromies around the twist operators, we must have
\be
 z'(t; x)=C\,t^{n_1-1}(t-1)^{n_2-1}(t-x)^{n_3-1} \, ,
\ee  which we can immediately integrate to get
\be
z(t; x)=y^{n_3}\sum_{k=0}^{n_1+n_2-2}a_k\, y^k+v(x) \, ,\quad  y\equiv t-x\,.
\ee 
Note that in this case there is no parameter $q$.
The coefficients $a_k$ can be explicitly computed (see Appendix \ref{details} for details). We find that $v(x)$ is given by
\be
&&  \hspace{2cm} v(x)  =  \frac{N_u}{D_u}\, , \\
N_u & = & \sum_{k=0}^{n_1-1}\sum_{l=0}^{n_2-1}\frac{(-1)^{k+l+n_3}}{k+l+n_3}
\left(
\begin{array}{c}
n_1-1  \\
k
\end{array}
\right)
\left(
\begin{array}{c}
n_2-1  \\
l
\end{array}
\right)
x^{n_3+l}(x-1)^{-l}\, \nonumber \\
D_u & = & \sum_{k=0}^{n_1-1}\sum_{l=0}^{n_2-1}\frac{(-1)^{k+l+n_3}}{k+l+n_3}
\left(
\begin{array}{c}
n_1-1  \\
k
\end{array}
\right)
\left(
\begin{array}{c}
n_2-1  \\
l
\end{array}
\right)
\left[x^{n_3+l}(x-1)^{-l}-x^{-k}(x-1)^{n_3+k}\right]\,.
\nonumber
\ee

From the explicit expression we see that in the polynomial case the equation $z(t=x)\equiv v(x) = u$
has exactly $n_4$ solutions and thus there are $n_4$ different maps for any polynomial four-point correlator.
Let us  reproduce this result diagrammatically. 
\begin{figure}[htbp]
\begin{center}
$\begin{array}{c@{\hspace{0.25in}}c@{\hspace{0.25in}}c}
 \epsfig{file=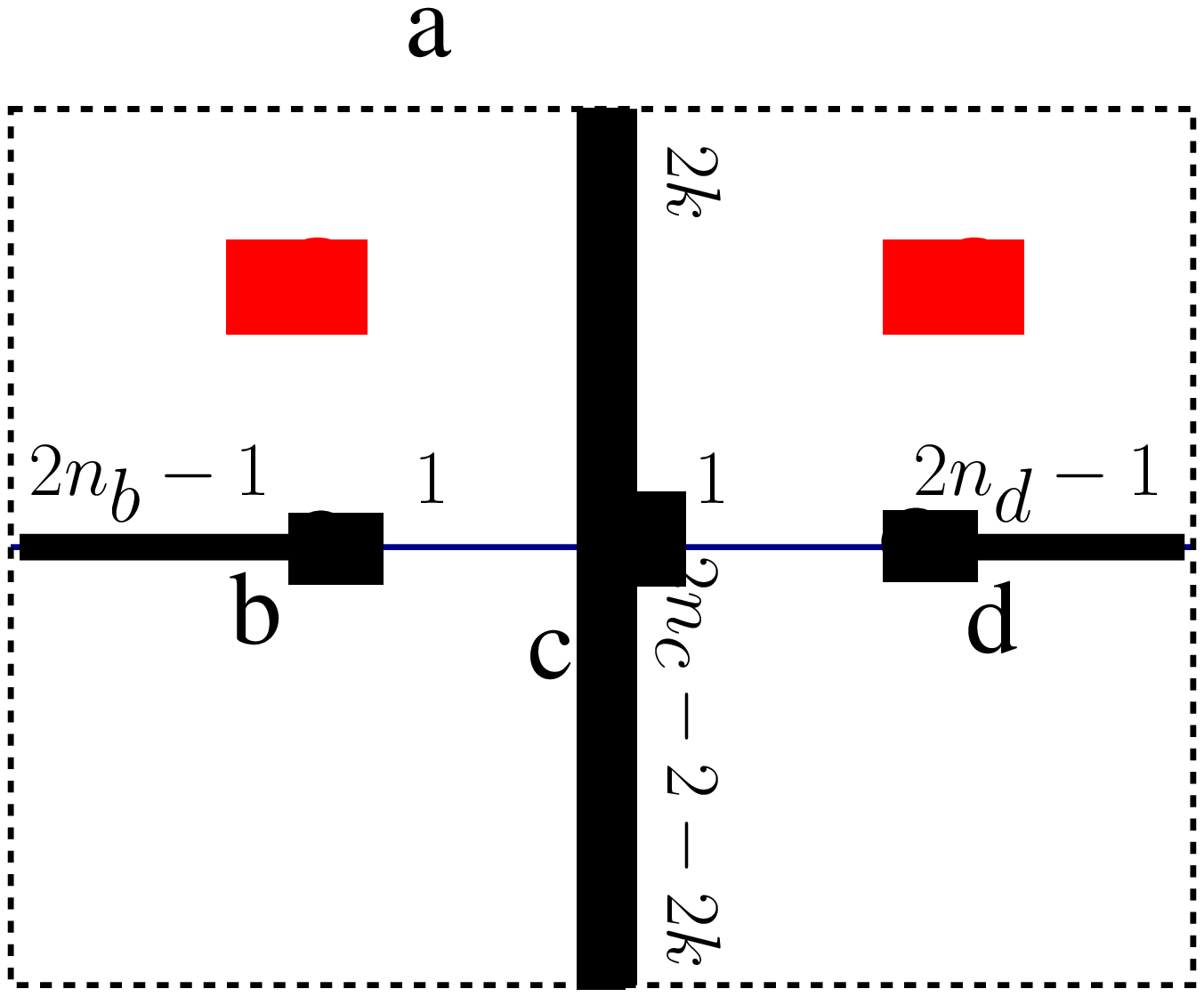,scale=0.3} & \epsfig{file=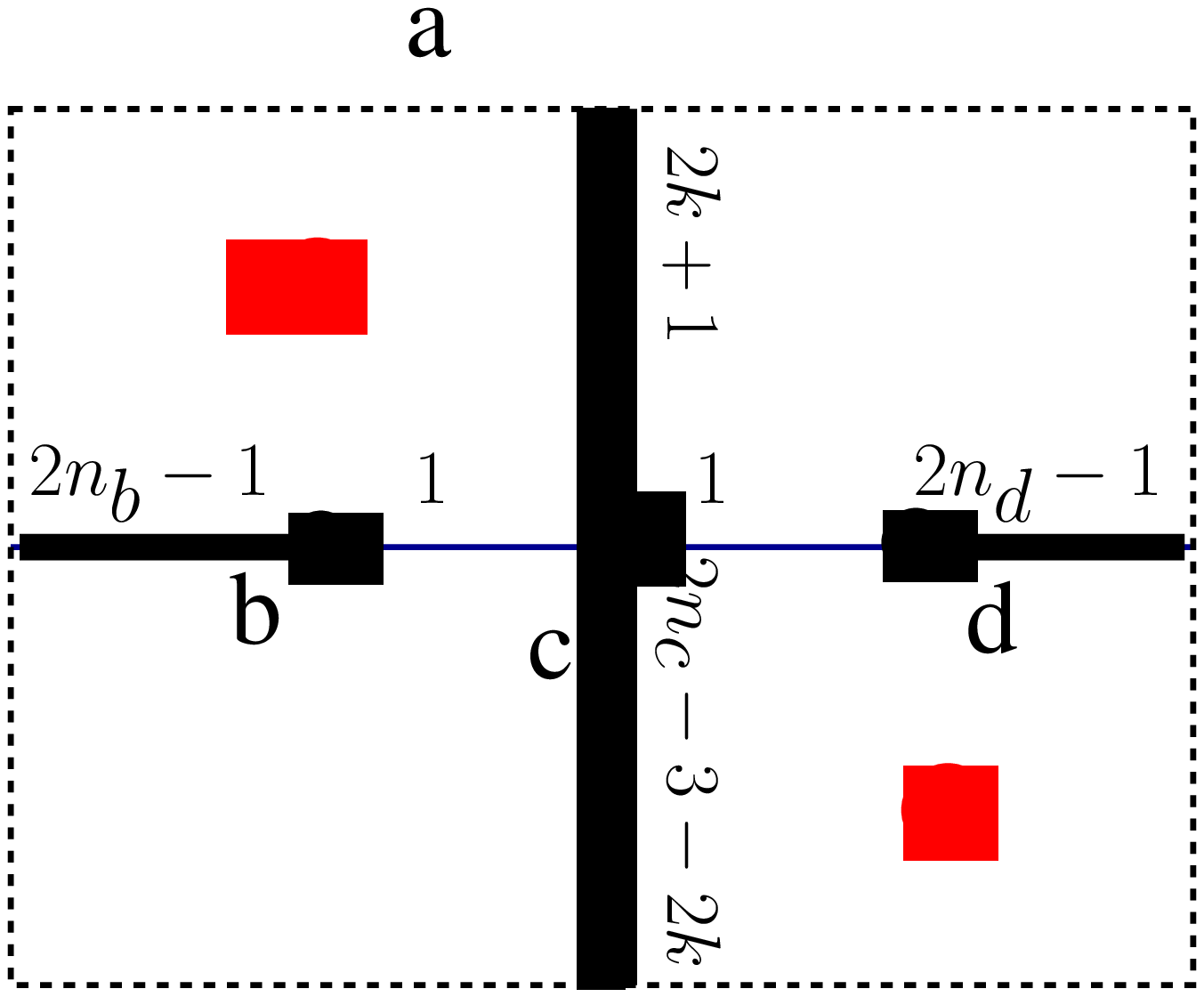,scale=0.3}&\epsfig{file=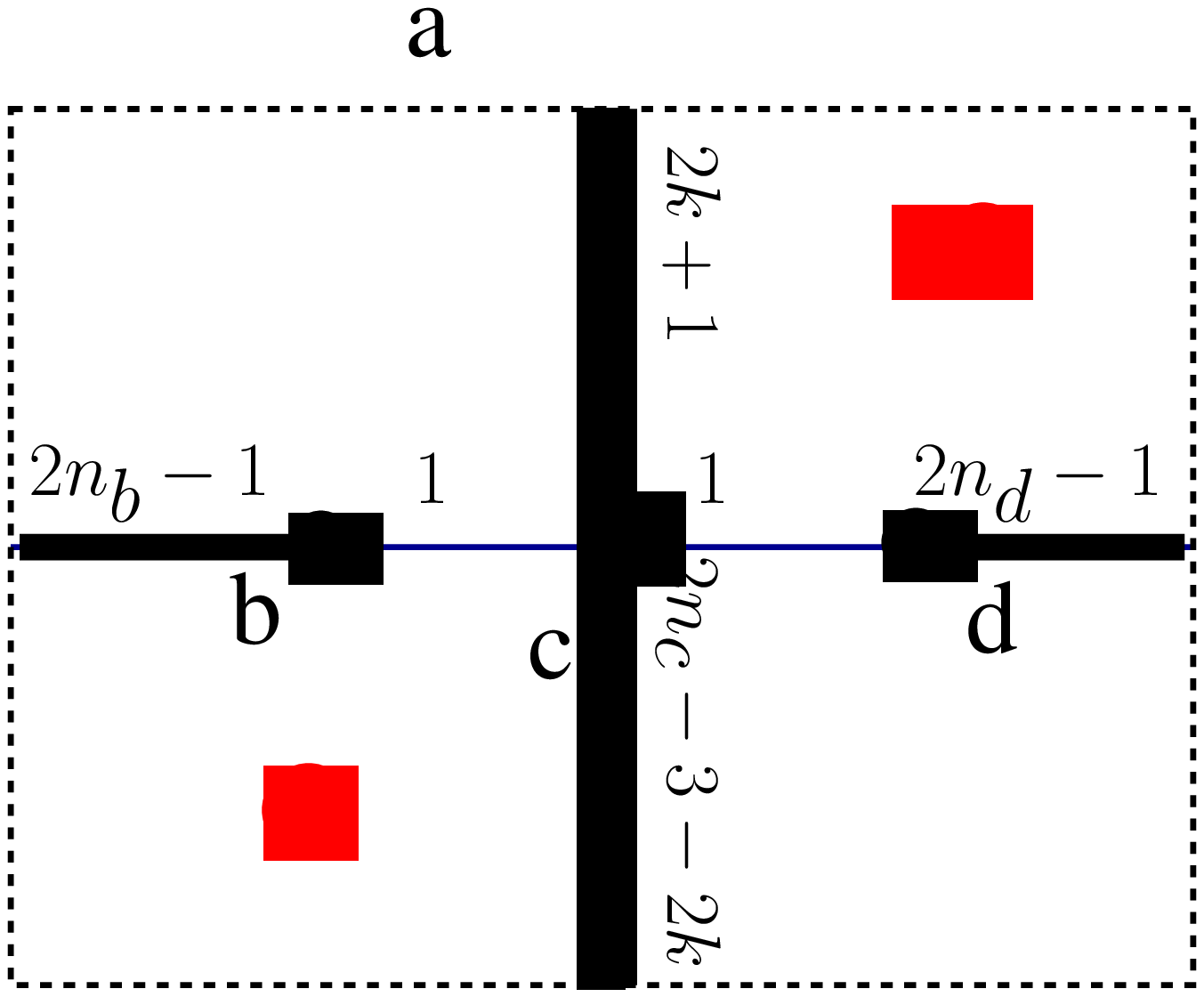,scale=0.3}\\
(I)\;:\;b\;c\;d\;a &(II)\;:\; \left[\,b,\,d\,\right]\;c\;a&(III)\;:\;c\;\left[\,b,\,d\,\right]\;a  
\\\end{array}$
\end{center}
 \begin{center}
\caption{The three classes of different diagrams that contribute to a generic polynomial four-point correlator. The number over each line is the number 
of diagram propagators joined. The four vertices are $a,b,c,d$, and $a$ is the vertex at infinity. Below each diagram we write the
ordering of the vertices inferred from it. The commutator denotes that the two vertices commute.
 } \label{genext4p}
\end{center}
\end{figure}    

 Diagrams for a polynomial correlator have very simple structure.
For definiteness we insert the cycles of lengths $n_1\leq n_2\leq n_3$  at finite points  on the base sphere
and the cycle $n_4=n_1+n_2+n_3-2$
 at infinity.
All  
propagators except two connect the cycles at finite position to the cycle at infinity. One can convince oneself 
that the two extra propagators must connect two different pairs 
of the cycles at finite positions --  otherwise the
orderings inferred from the diagram will not be consistent. This observation leaves only the three classes of diagrams
illustrated in Figure \ref{genext4p}.

Let us count the different diagrams. From diagrams of class $(I)$ with the $n_1$ cycle associated to position
 $b$, the $n_2$ cycle to $c$ and the $n_3$ cycle to $d$, we get 
the right ordering and the number of different diagrams is equal to the number of possible choices of $k$, which is $n_2$.
From diagrams of class $(II)$  with $n_1\to b,\,n_2\to d,\,n_3\to c$ we get the right ordering and the number of possibilities for $k$
is $n_3-1$. By choosing another assignment of the insertions, $n_1\to d,\,n_2\to b,\,n_3\to c$ we would get the same diagrams 
by graph symmetry and thus these should not be counted twice. Finally, from diagram of class $(III)$  
with $n_1\to c,\,n_2\to b,\,n_3\to d$ we get the right ordering and the number of possibilities for $k$
is $n_1-1$. Again, the second ordering in this case does not give rise to new diagrams. Finally, counting all the possibilities
we find  $n_2+(n_3-1)+(n_1-1)=n_4$ diagrams. As expected the number of diagrams equals the number of covering maps.

As an additional example of application of the diagrammatic techniques consider the following question:
 how many diagrams contribute in the OPE limit of say the $\sigma_{[n_1]}$  colliding with $\sigma_{[n_2]}$ cycle in the bosonic
 orbifold \eqref{bosaction}?
The OPE of twist operators can be singular only 
when the colliding cycles do not commute. In the 
polynomial case, for the $n_1$  and $n_2$  to not commute there has to be a single edge extended between them. All the diagrams in classes  $(I)$ and $(III)$ have this property but the diagrams of class $(II)$ do not. Thus
the number of diagrams contributing in this OPE limit is $n_1+n_2-1$.

\begin{figure}[htbp]
\begin{center}
$\begin{array}{c@{\hspace{0.25in}}c@{\hspace{0.25in}}c}
 \epsfig{file=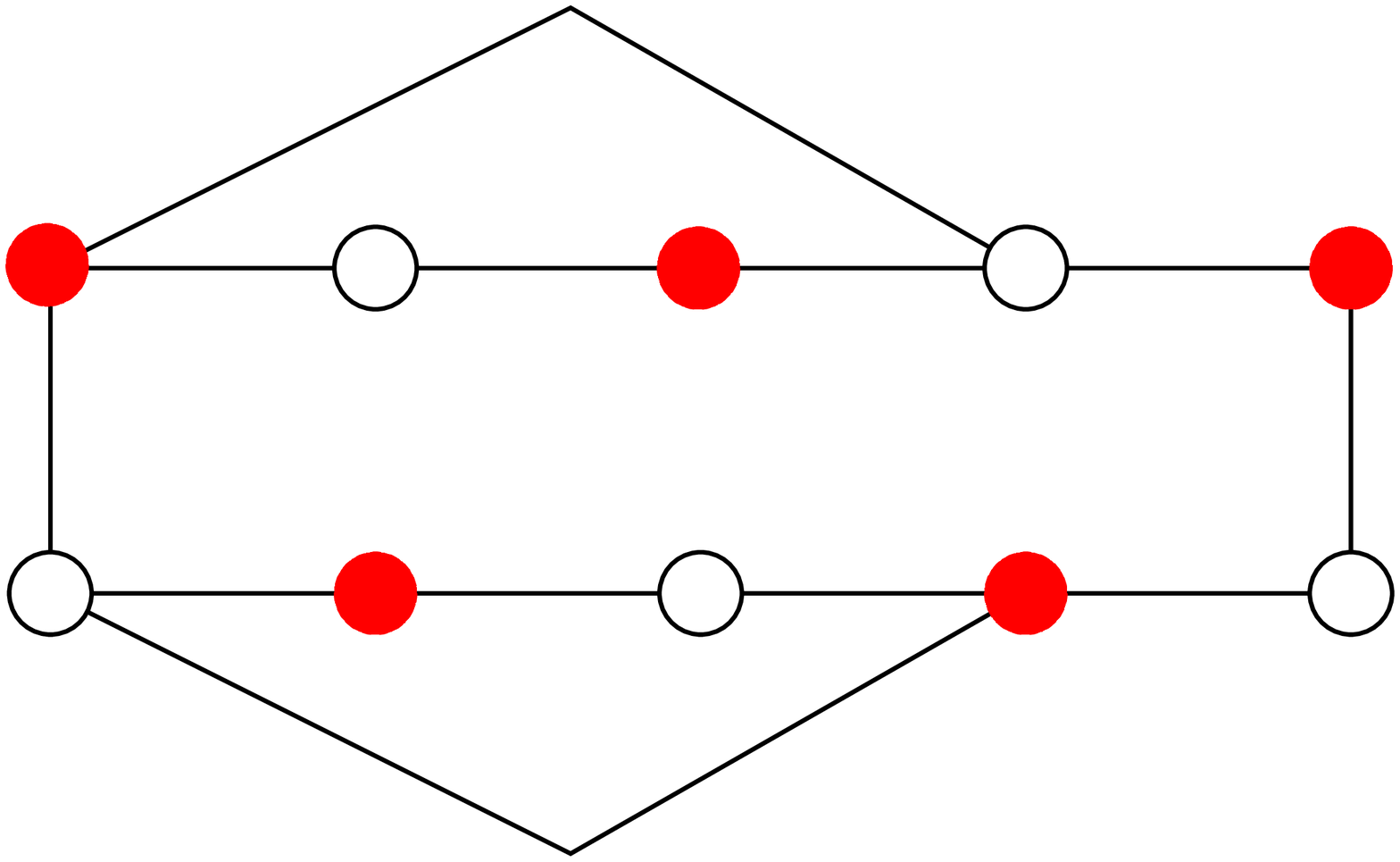,scale=0.1} & \epsfig{file=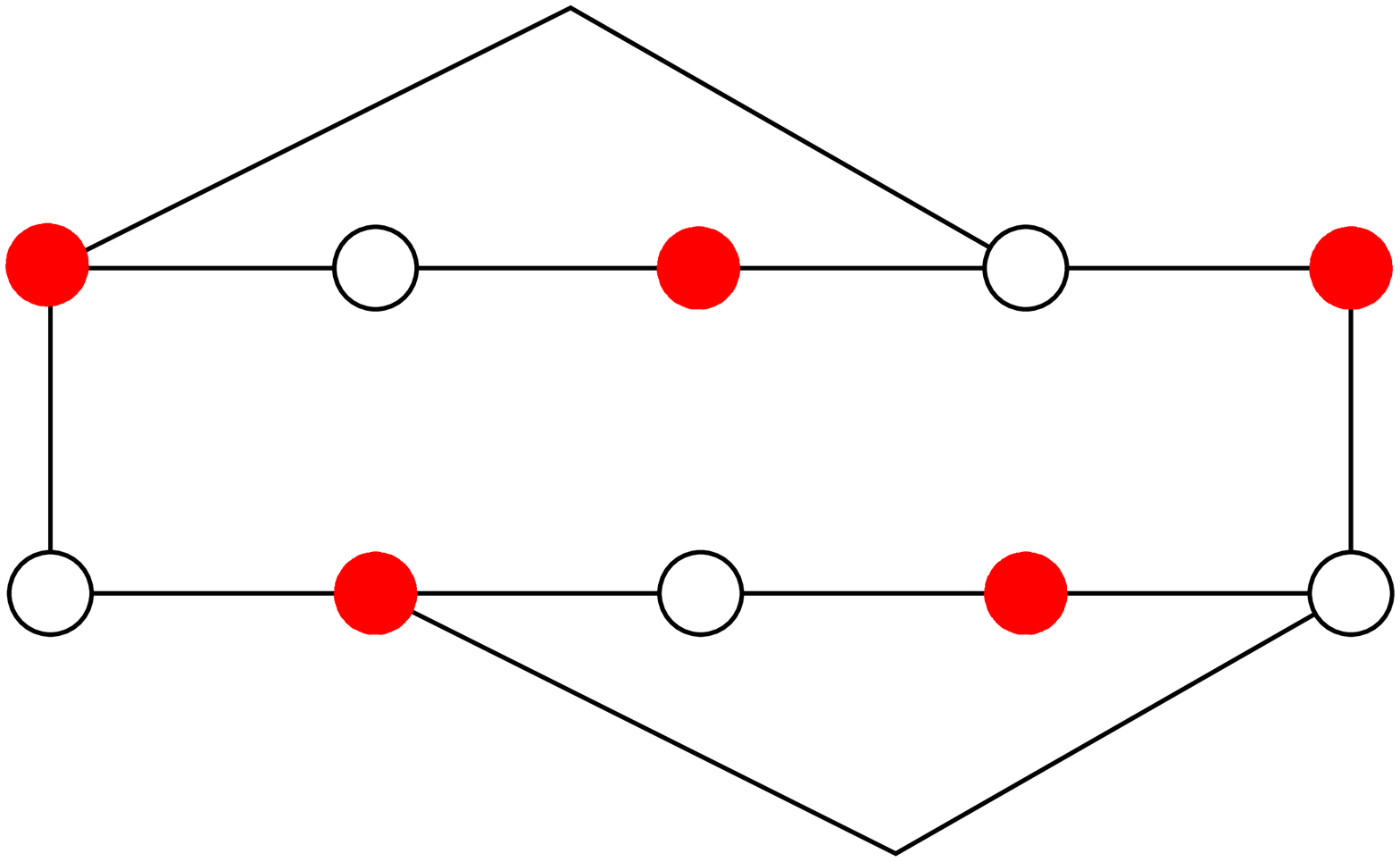,scale=0.1} & \epsfig{file=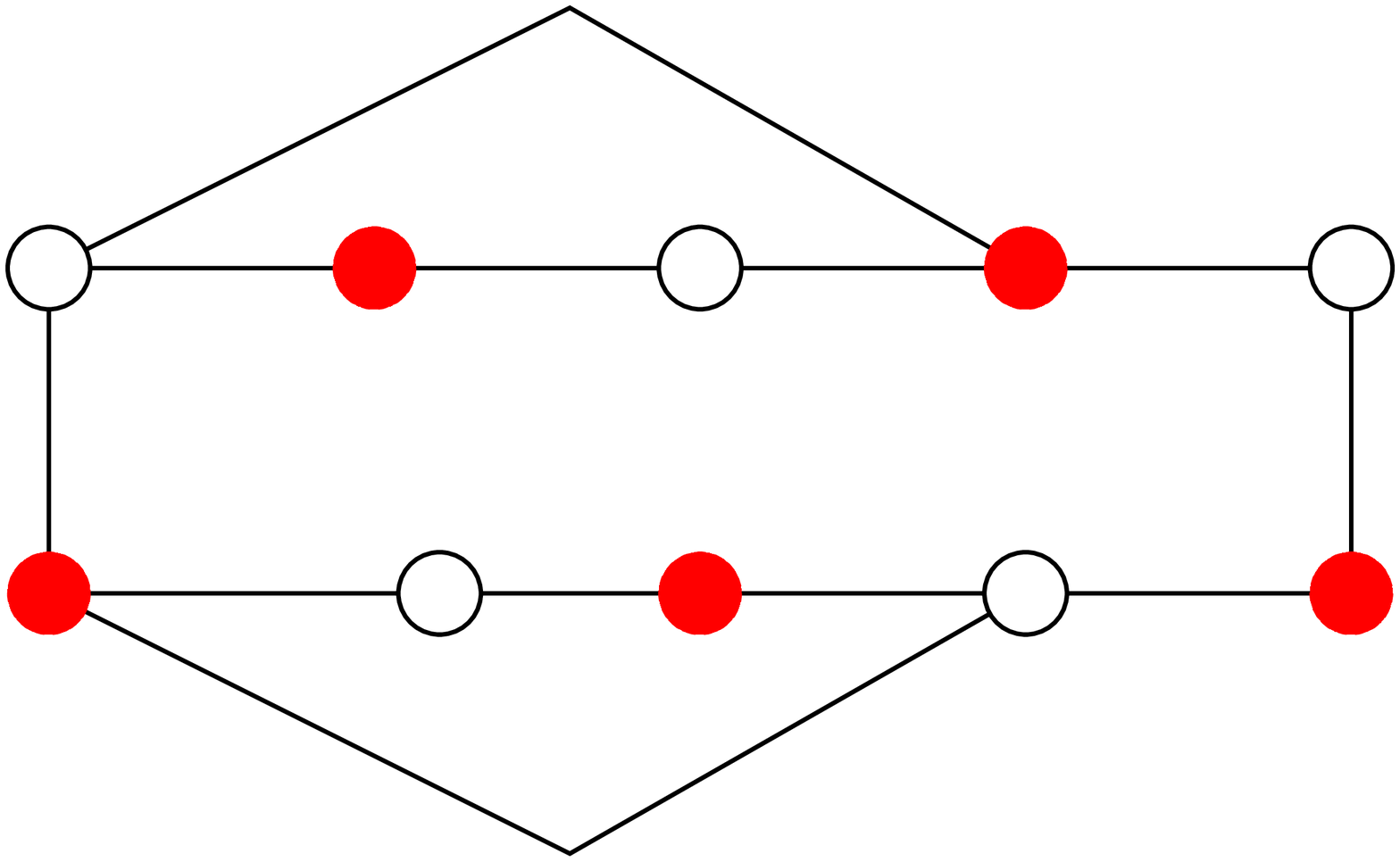,scale=0.1}\\
\a_1=(1\,2\,3\,4\,5)(2\,1)(5\,4)(5\,3\,2) &
\a_2=(1\,2\,3\,4\,5)(5\,4)(1\,5\,3)(2\,1) &
\a_3=(1\,2\,3\,4\,5)(4\,3\,1)(2\,1)(4\,5)\\
 \epsfig{file=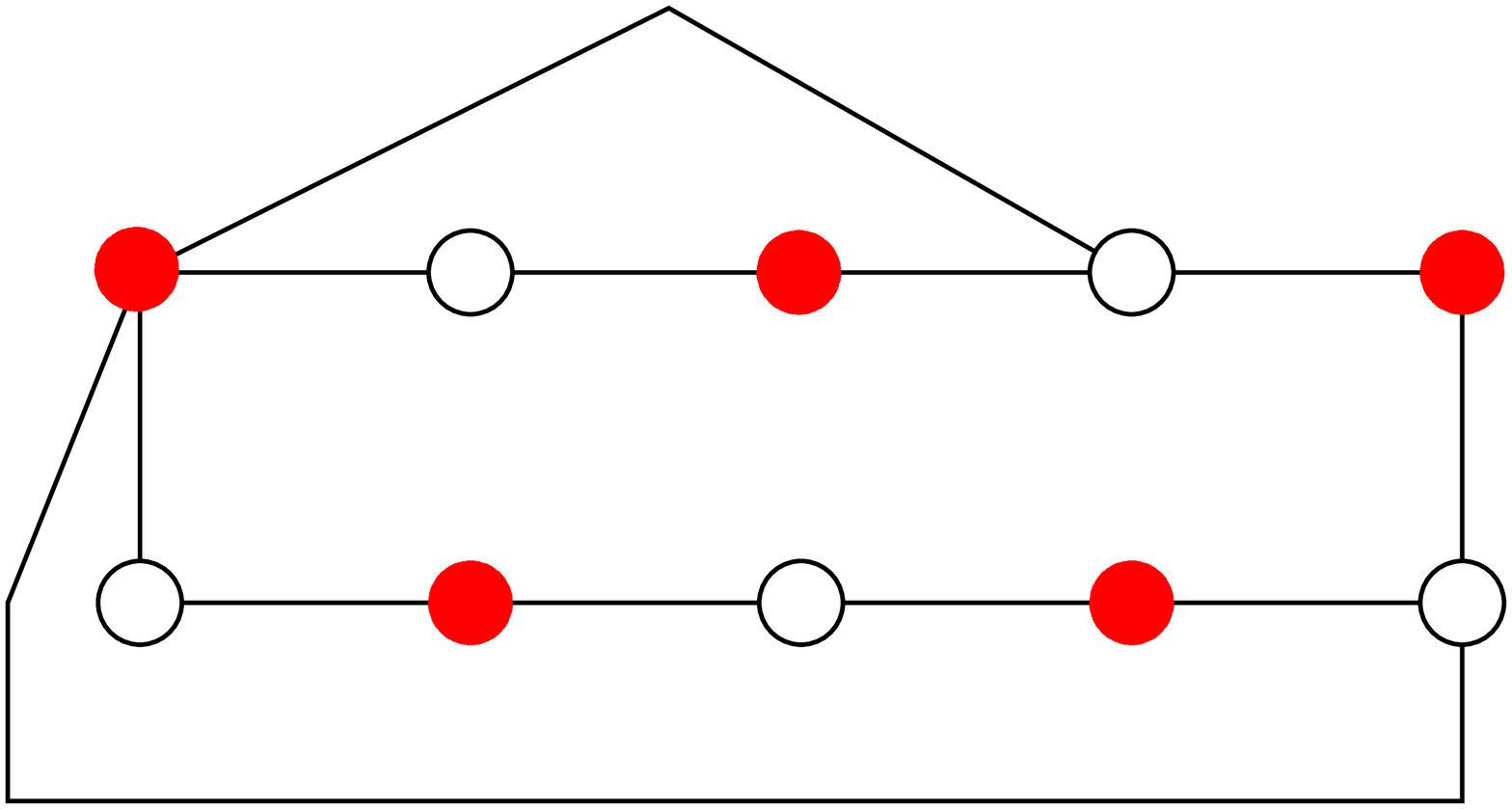,scale=0.1} & \epsfig{file=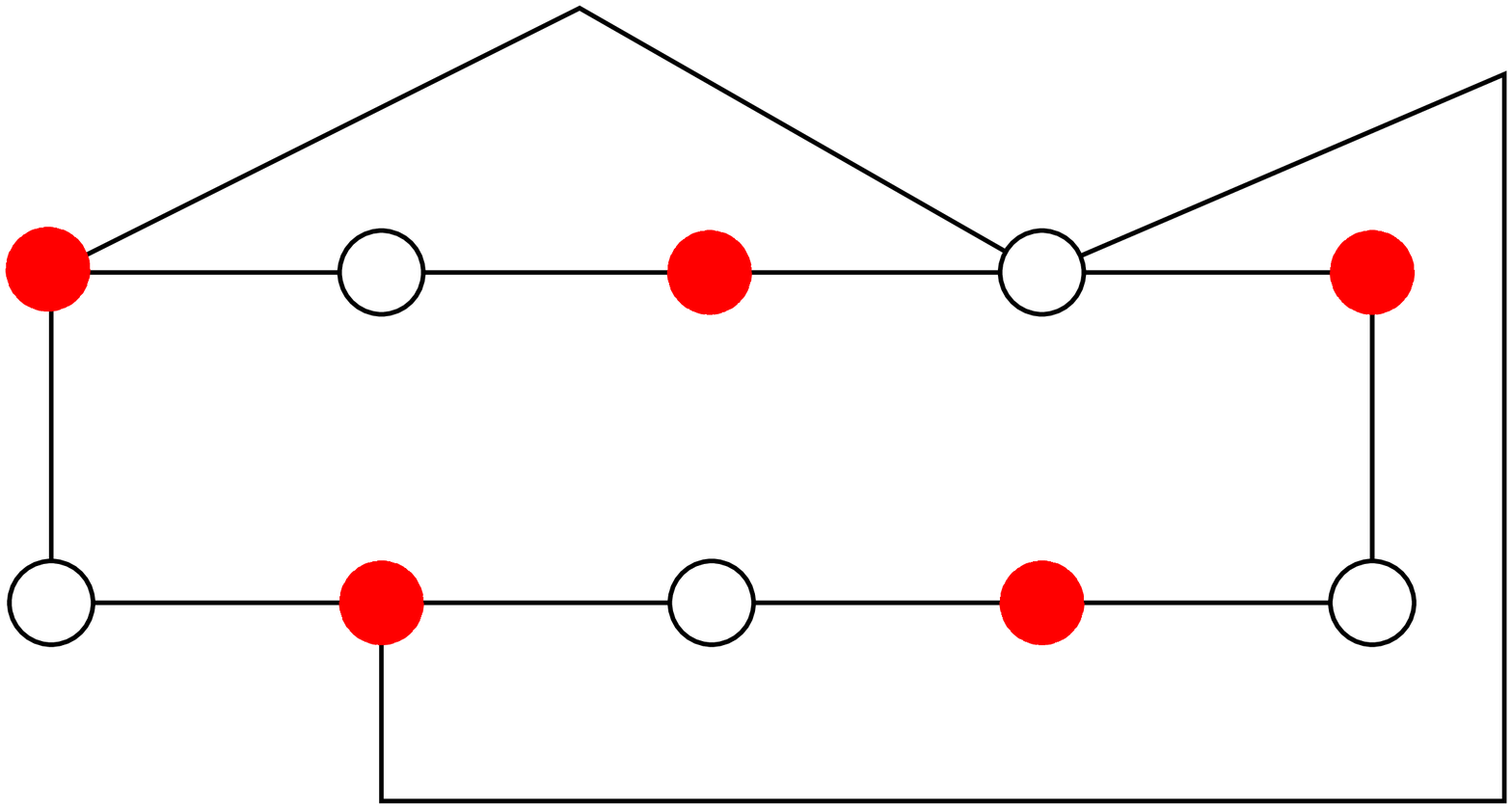,scale=0.1} & \epsfig{file=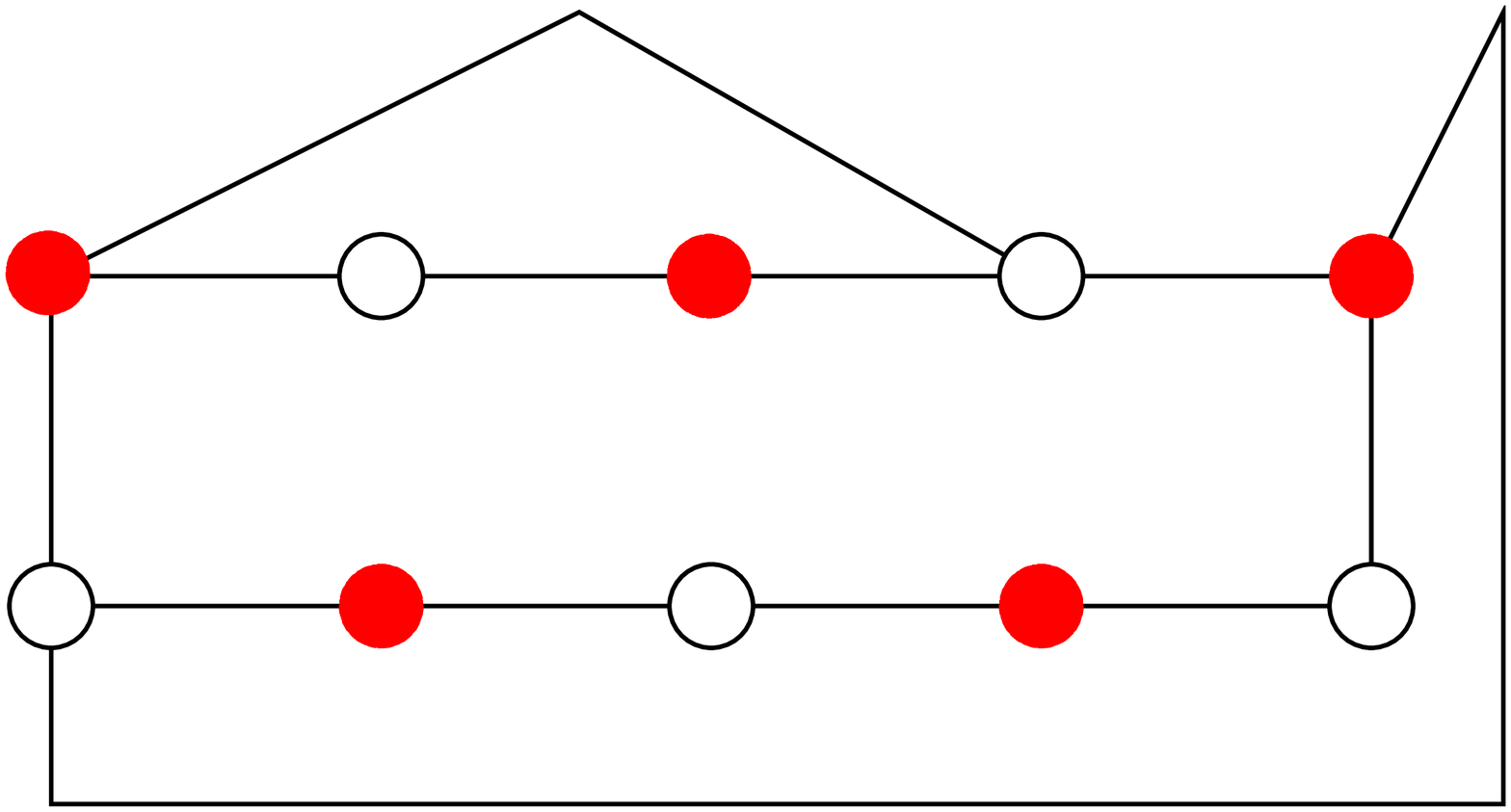,scale=0.1}\\
\a_4=(1\,2\,3\,4\,5)(5\,4)(5\,3)(5\,2\,1)&
\a_5=(1\,2\,3\,4\,5)(5\,4)(5\,1)(3\,2\,1)&
\a_6=(1\,2\,3\,4\,5)(5\,4)(3\,2\,1)(3\,5)\\
 \epsfig{file=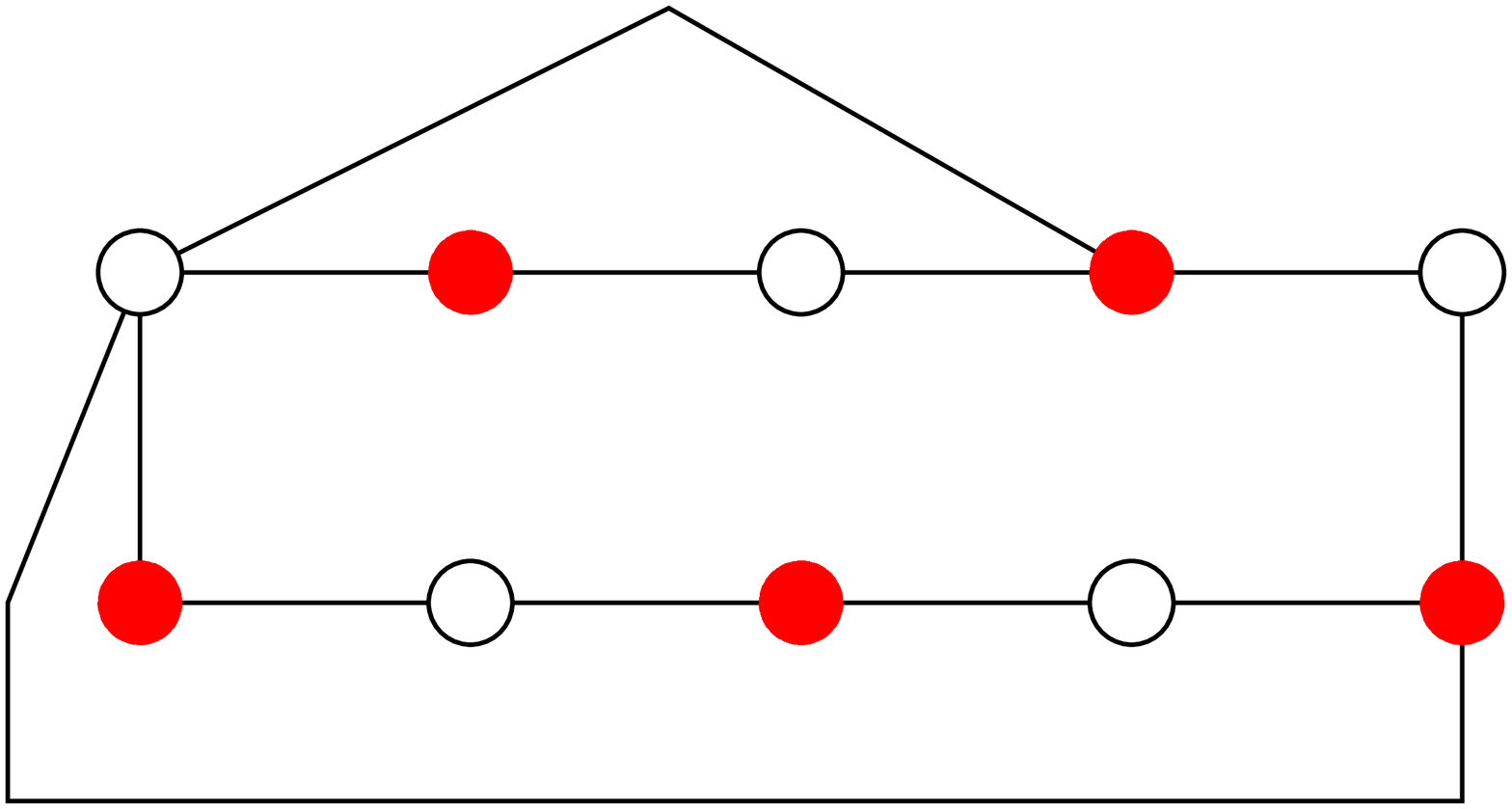,scale=0.1} & \epsfig{file=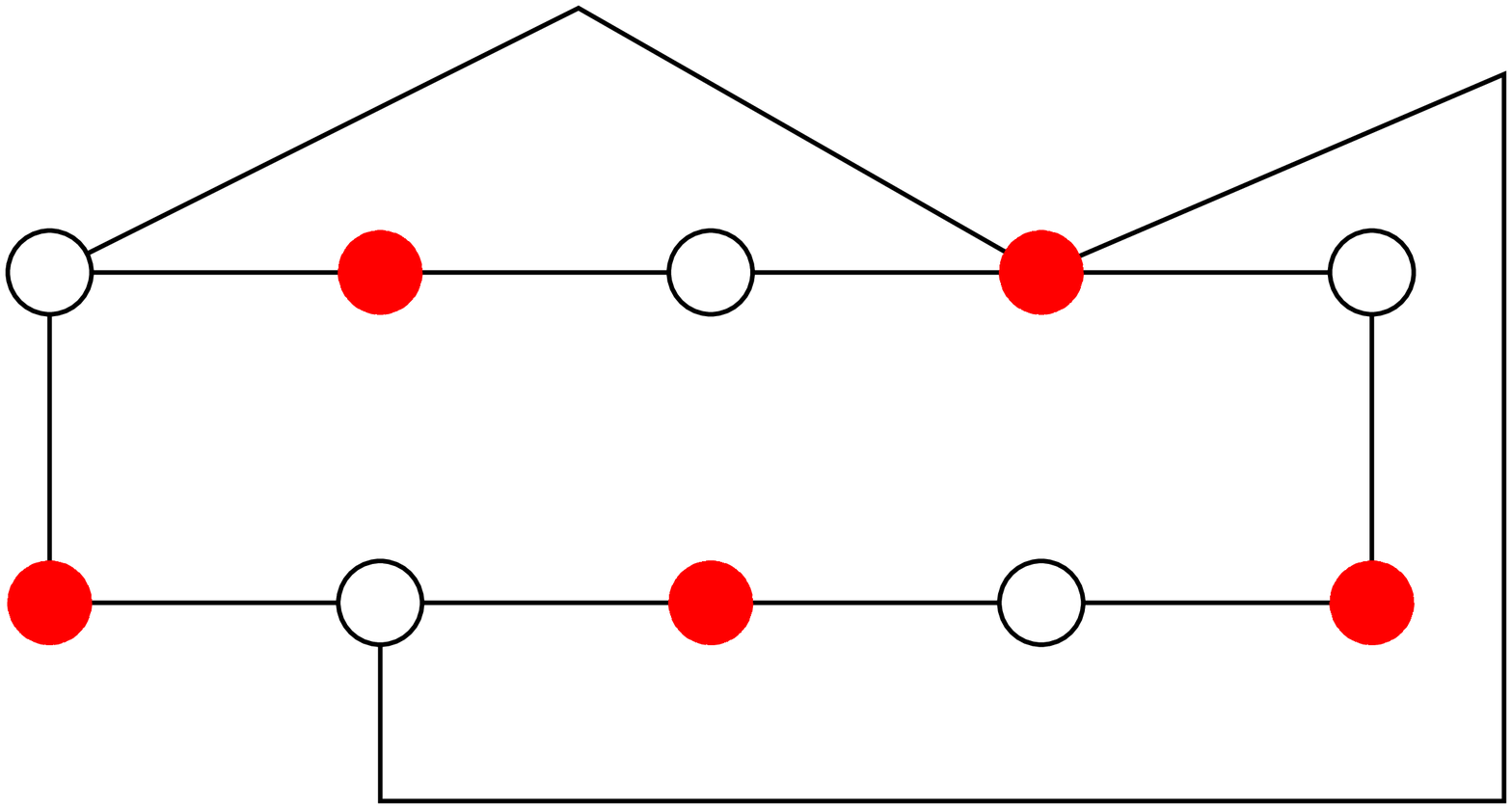,scale=0.1} & \epsfig{file=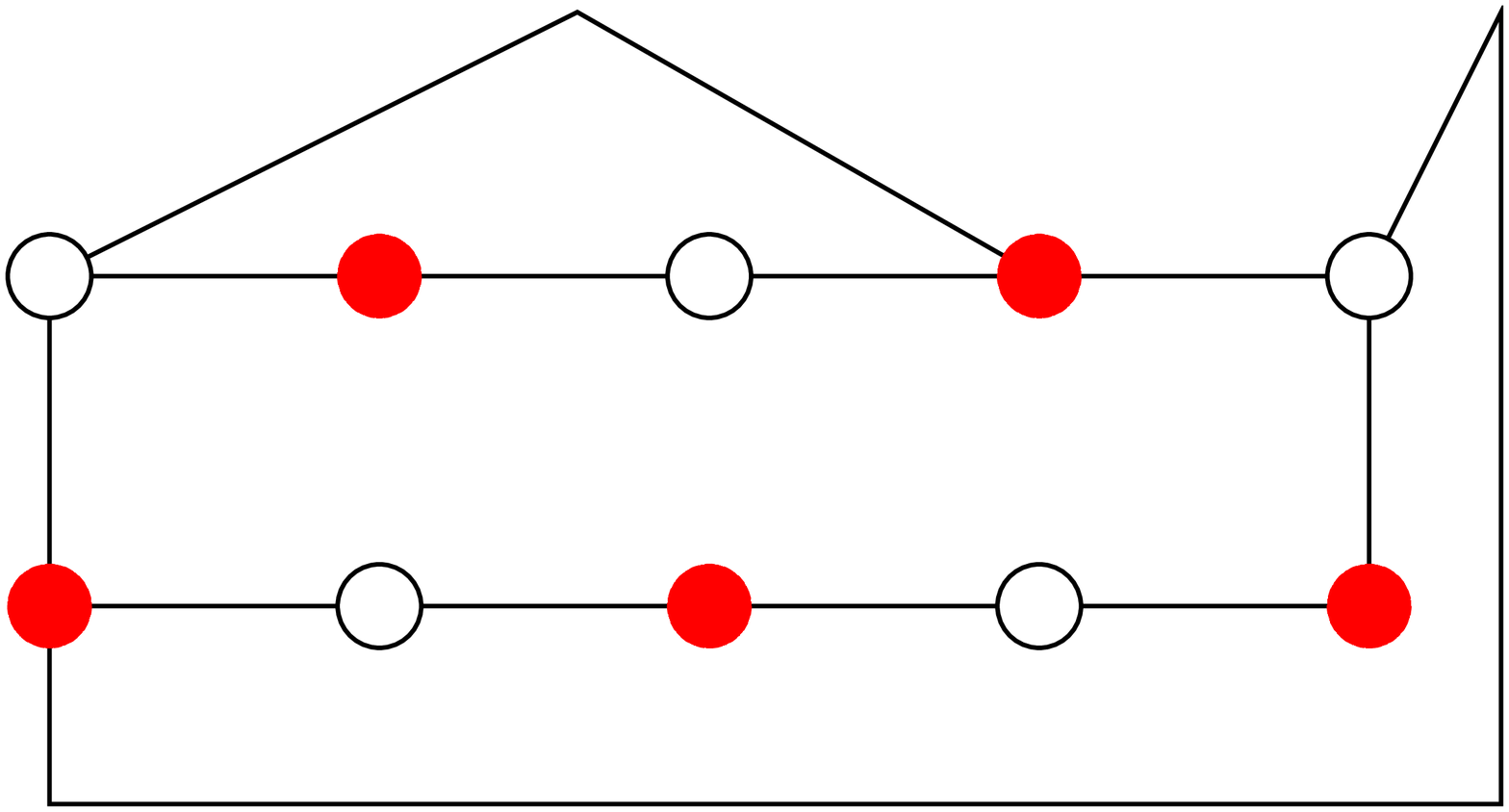,scale=0.1}\\
\a_7=(1\,2\,3\,4\,5)(3\,2\,1)(4\,3)(5\,4)&
\a_8=(1\,2\,3\,4\,5)(3\,2\,4)(4\,1)(5\,4)&
\a_9=(1\,2\,3\,4\,5)(4\,1)(3\,2\,1)(5\,4)\\
 \epsfig{file=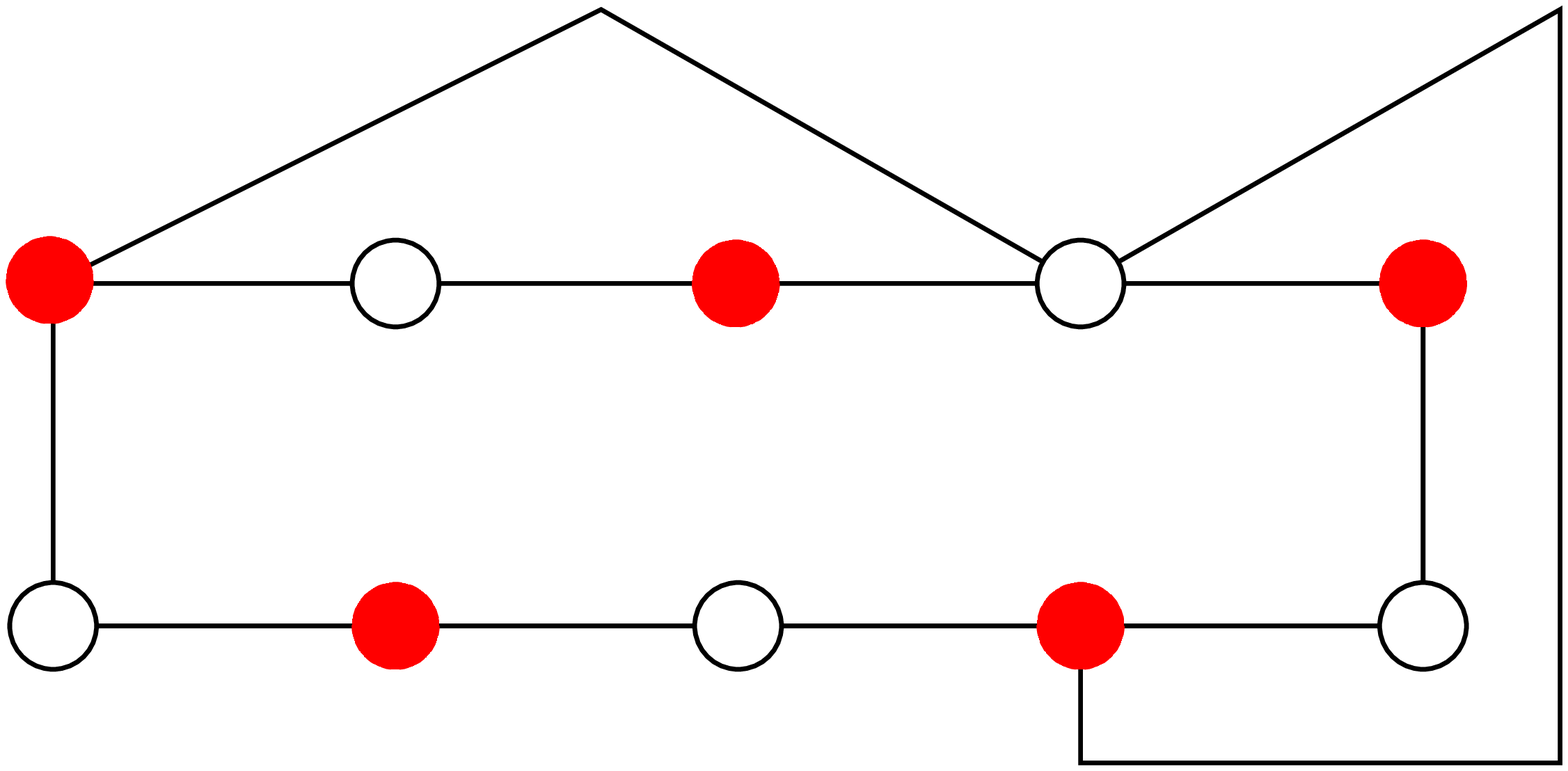,scale=0.1} & \epsfig{file=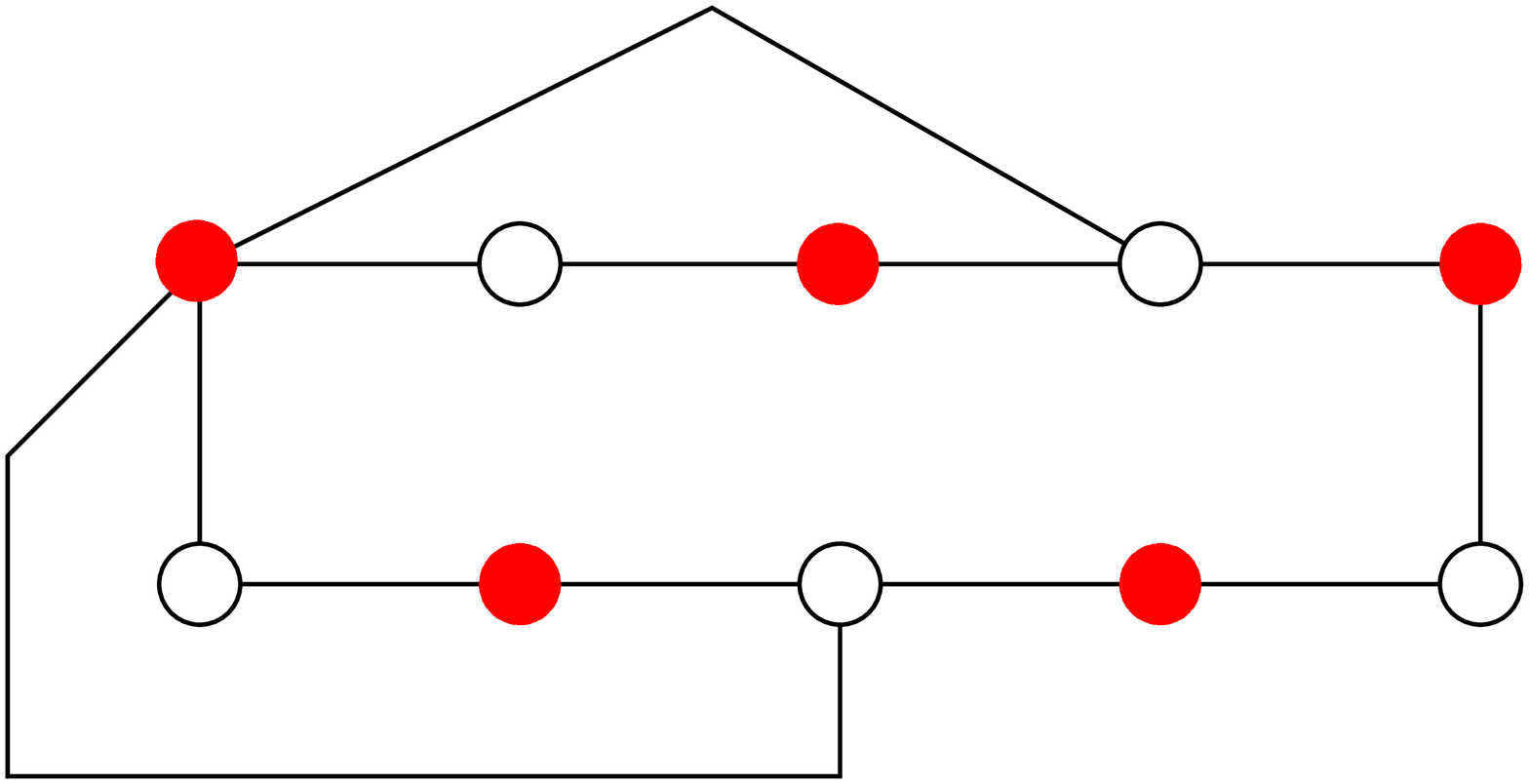,scale=0.1} &\\
\a_{10}=(1\,2\,3\,4\,5)(5\,4)(5\,2\,1)(2\,3)&
\a_{11}=(1\,2\,3\,4\,5)(5\,4)(3\,2\,5)(5\,1)&
\\ [0.2cm]
\end{array}$
\end{center}
 \begin{center}
\caption{``Unlabeled'' diagrams that could contribute to the polynomial correlator $\langle \sigma_{[2]} \sigma_{[2]} \sigma_{[3]} \sigma_{[5]} \rangle$.
Placing  $\sigma_{[3]}$ at $z=u$   and $\sigma_{[5]}$ at $z = \infty$
the restriction to radial ordering (with $|u|<1$) selects  diagrams $\a_{2,6,9,10,11}$.
Placing one of the $\sigma_{[2]}$s at $z=u$,  $\sigma_{[3]}$ at $z=1$, $\sigma_{[5]}$ at $z = \infty$
and restricting to radial   ordering (again with $|u|<1$) we get diagrams
 $\a_{4,5,9}$ and {\it two} contributions from $\a_1$ (the two $2$-cycles in this diagram commute and give two  distinct ``labeled'' diagrams, which cannot
 be related by a global $S_N$ transformation).
 } \label{extrem2D}
\end{center}
\end{figure}

As a more concrete example consider the polynomial correlator 
\begin{equation} \label{concrete}
\langle \sigma_{(2)} (0) \sigma_{(3)} (u) \sigma_{(2)} (1)  \sigma_{(5)} (\infty) \rangle \,.
\end{equation}
 The function $v(x)$
in this case is given by
\be
v_{2235}(x) = -\frac{ -5 + 2 x}{3 - 10 x + 10 x^2}\,x^4\,.
\ee
Upon solving the $v_{2235}(x)=u$ equation we get
five different solutions.
There are eleven different ``unlabeled''
diagrams (diagrams where the vertices have  not yet been assigned to a position on the base sphere)
that could contribute to the $2235$ case. They are shown in Figure \ref{extrem2D}.
In general, as was discussed in Section \ref{symmprodCFT}, the number of
diagrams is equal to number of maps only after we restrict to a given ordering of group elements.
Indeed, as one can see from Figure \ref{extrem2D}, there are only five diagrams satisfying
a given ordering. For the radial ordering of (\ref{concrete}) (with say $|u| < 1$), 
 these are diagrams
$\a_{2,6,9,10,11}$.

\subsection{Monodromies and channel-crossing}\label{modspace}

\begin{figure}[htbp]
\begin{center}
\epsfig{file=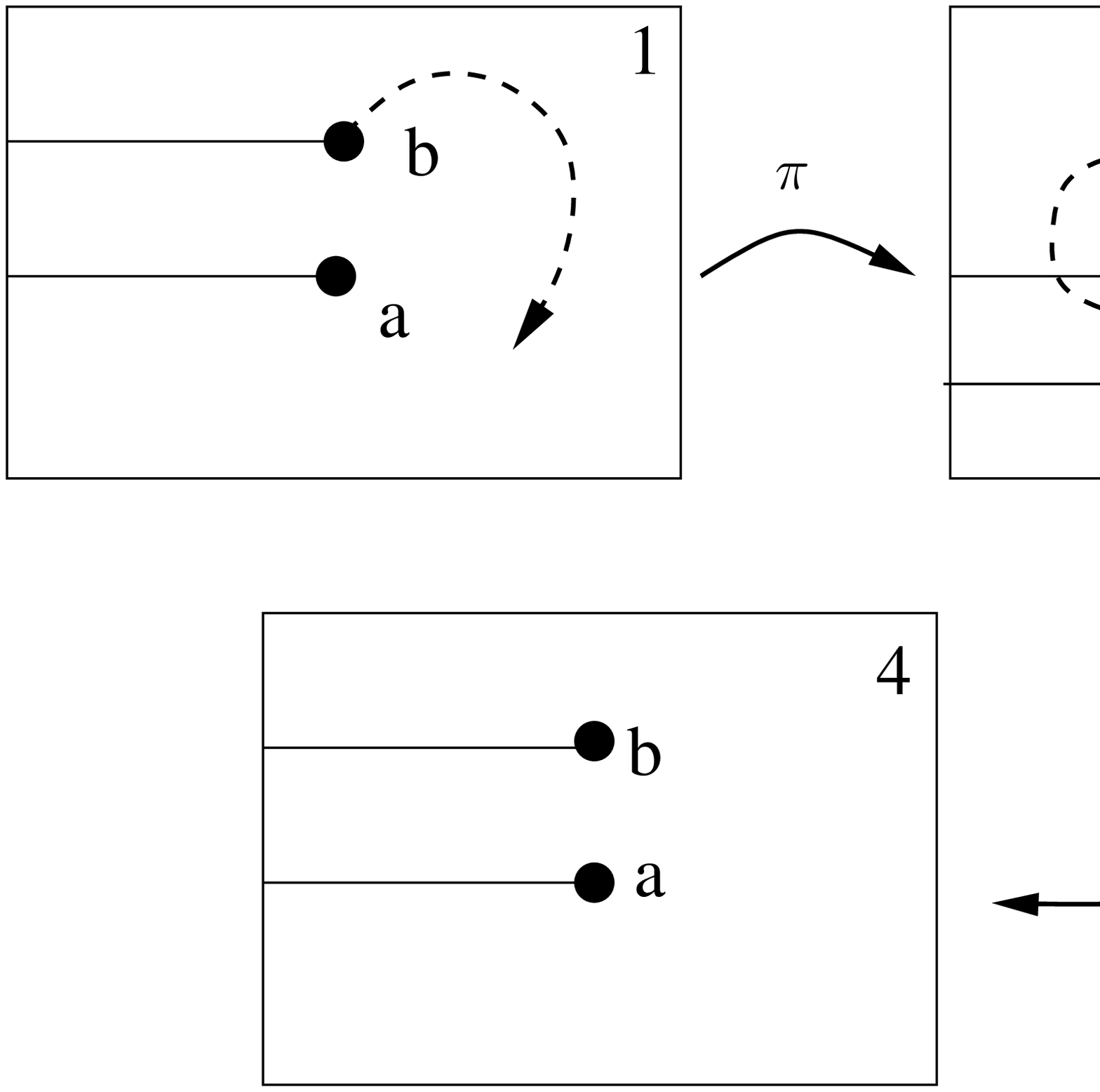,scale=0.45}
\caption{An illustration of a $2\pi$ rotation of one cycle around another. Here we 
rotate cycle $b=(1\, 2)$ around $a=(1\,4\, 3)$. After a rotation by $\pi$ the cycle $a$ crosses the branch cut of $b$
and becomes $a\to (2\,4\,3)$. After another $\pi$ rotation the cycle $b$ crosses a branch cut of $a$ and becomes $b\to (1\, 4)$.
} \label{pirot2}
\end{center}
\end{figure}

\begin{figure}[htbp]
 \begin{center} 
\epsfig{file=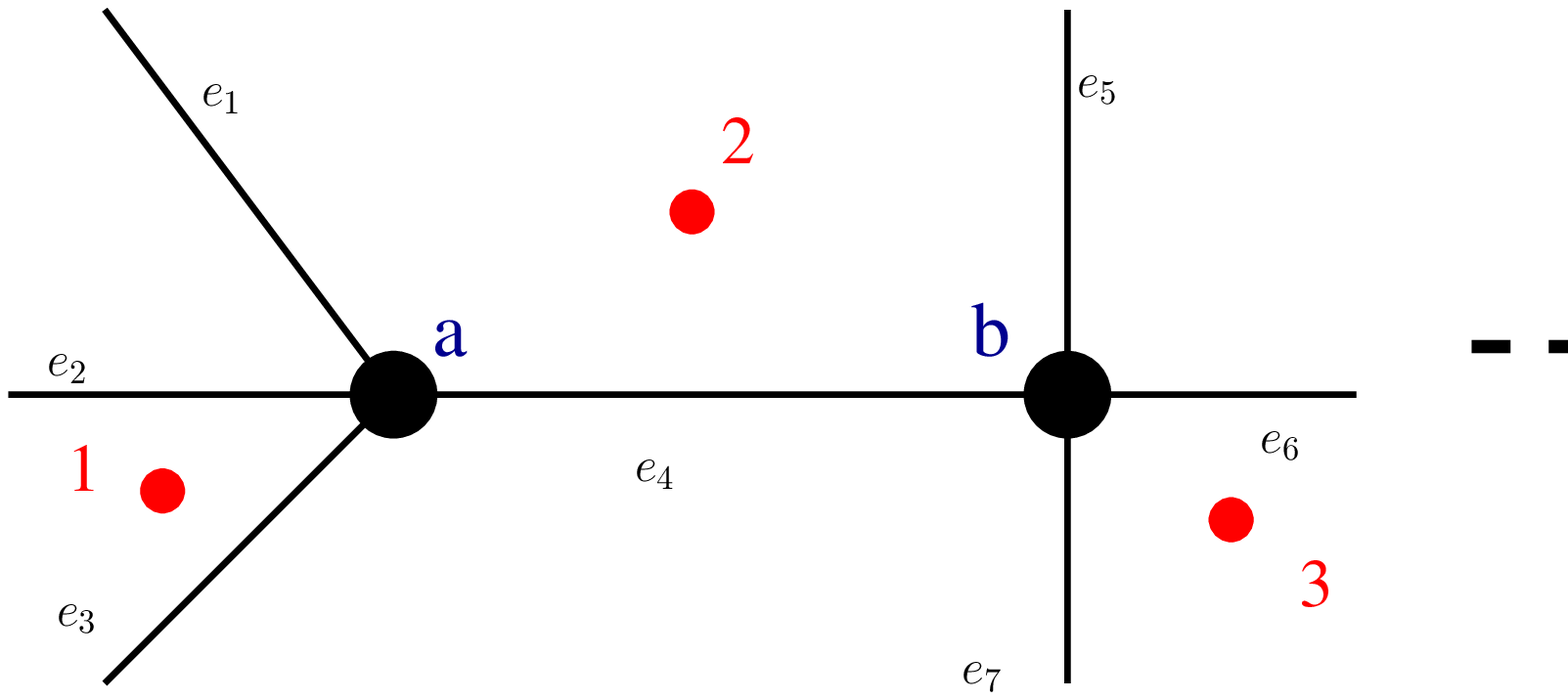,scale=0.45}
\caption{Channel-crossing exemplified. A propagator can  be shrunk as two vertices
are joined, 
and  expanded again by splitting the vertices in a different direction. One obstruction to the splitting  procedure
is that all the ``colors'' at a given vertex should be different (as a vertex corresponds to a cycle);
 the other obstruction
is that the cyclic orders of the vertices should be consistent after the channel-crossing.
 Thus not all splittings are allowed, unlike in a usual gauge theory.
The joining and splitting procedure simply corresponds to taking one vertex around the other on the base sphere.
In this figure the channel-crossing takes two $2$-cycles $(1\,2)_a$ and $(2\,3)_b$ and
transforms them into $(2\,3)_b$ and $(1\,3)_a$. This channel-crossing corresponds to a $\pi$ rotation of
$b$ around $a$. Note that a rotation by $\pi$ depends on the choice of cut picture and should be viewed as an intermediate
step in a $2\pi$ rotation (see Figure 13), which is unambiguous. } \label{channel} 
\end{center}
\end{figure}

We have given in Section 2.2 an algorithm to associate
diagrams to branched covering maps. We have repeatedly emphasized
the 1-1 correspondence between the diagrams 
and the branched coverings  contributing to a given correlator. To gain some more insight into
 this correspondence, we propose to look at the monodromies of the branched coverings
 as we make a full $ 2\pi$ rotation of a  ramification point around another ramification point. 
 To make the discussion concrete,
 let us  focus on polynomial four-point  correlators. As we have seen, in the polynomial case
 the different branched coverings with given ramification structure
 correspond to the different solutions of the equation
 \be\label{uuu}
v(x)=u \, .
\ee
As the insertion point $u$ encircles one of the other insertion points, the solutions of (\ref{uuu}) 
 are permuted into one another.  On the diagrammatic side, the same operation corresponds to a
 certain channel-crossing procedure, which we illustrate 
  in Figures \ref{pirot2} and \ref{channel}. The group of monodromies acts on the branched covering maps
  in the same way as a certain group of channel-crossings acts on the diagrams.
 For simple correlators we can use this isomorphism to determine the dictionary between diagrams and branched coverings, 
 confirming the rules of  of Section 2.2.

\begin{figure}[htbp]
\begin{center}
$\begin{array}{c@{\hspace{0.45in}}c}
 \epsfig{file=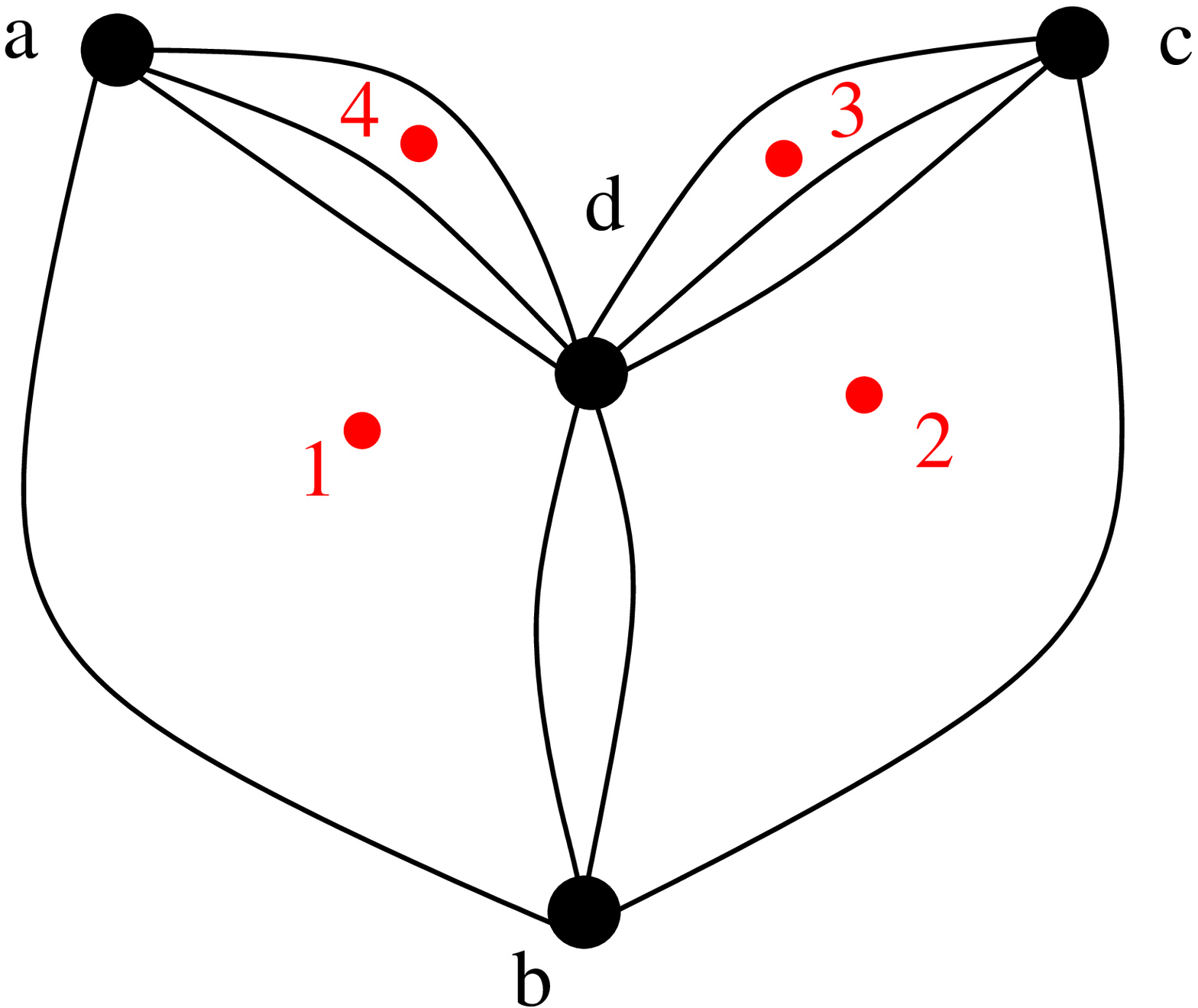,scale=0.2} & \epsfig{file=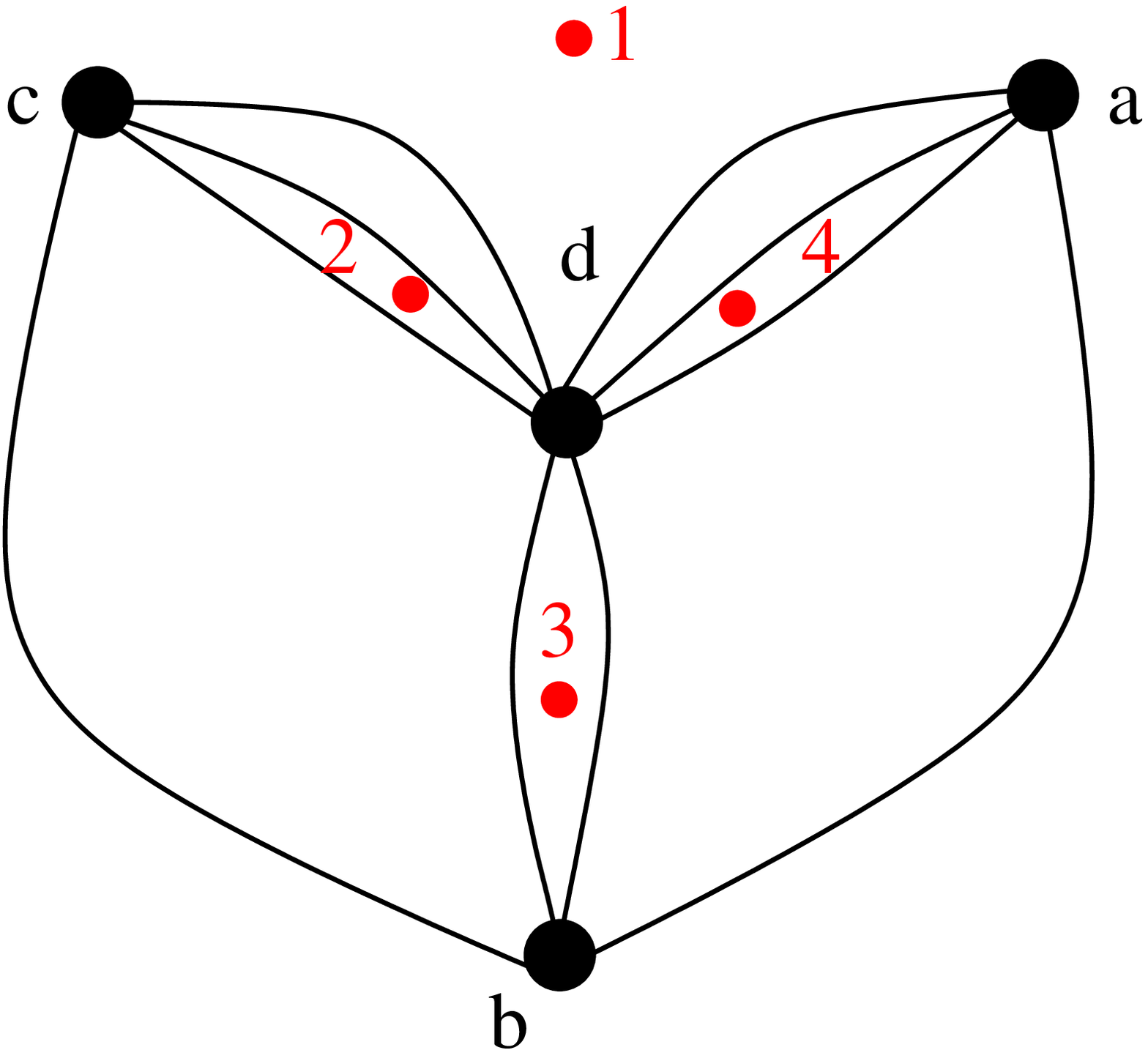,scale=0.2}\\
\a_1=(1\,4)_a(1\,2)_b(3\,2)_c (1\,2\,3\,4)_d&
\a_2=(1\,4)_a(1\,3)_b(2\,1)_c(1\,2\,3\,4)_d \\
  \epsfig{file=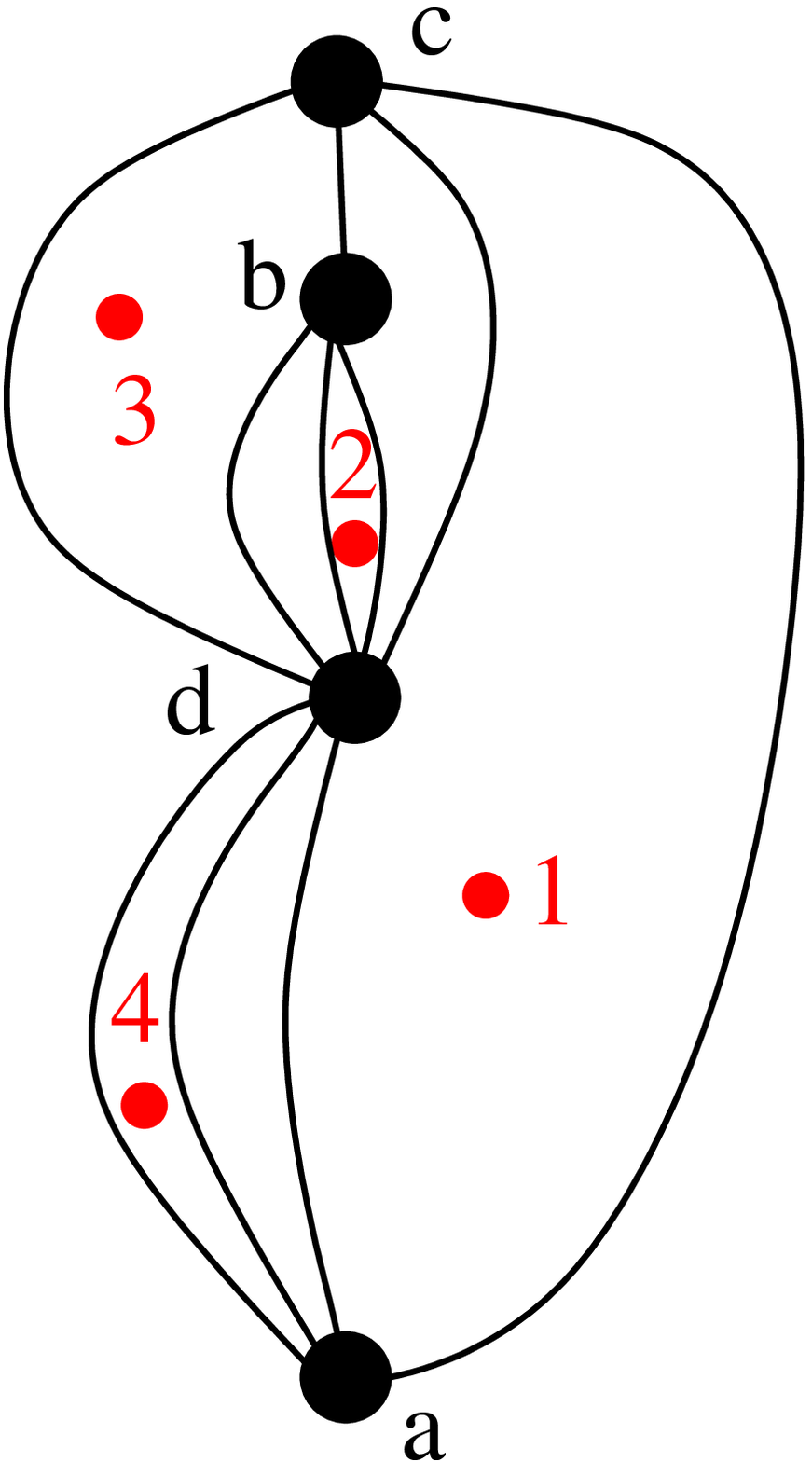,scale=0.2} &  \epsfig{file=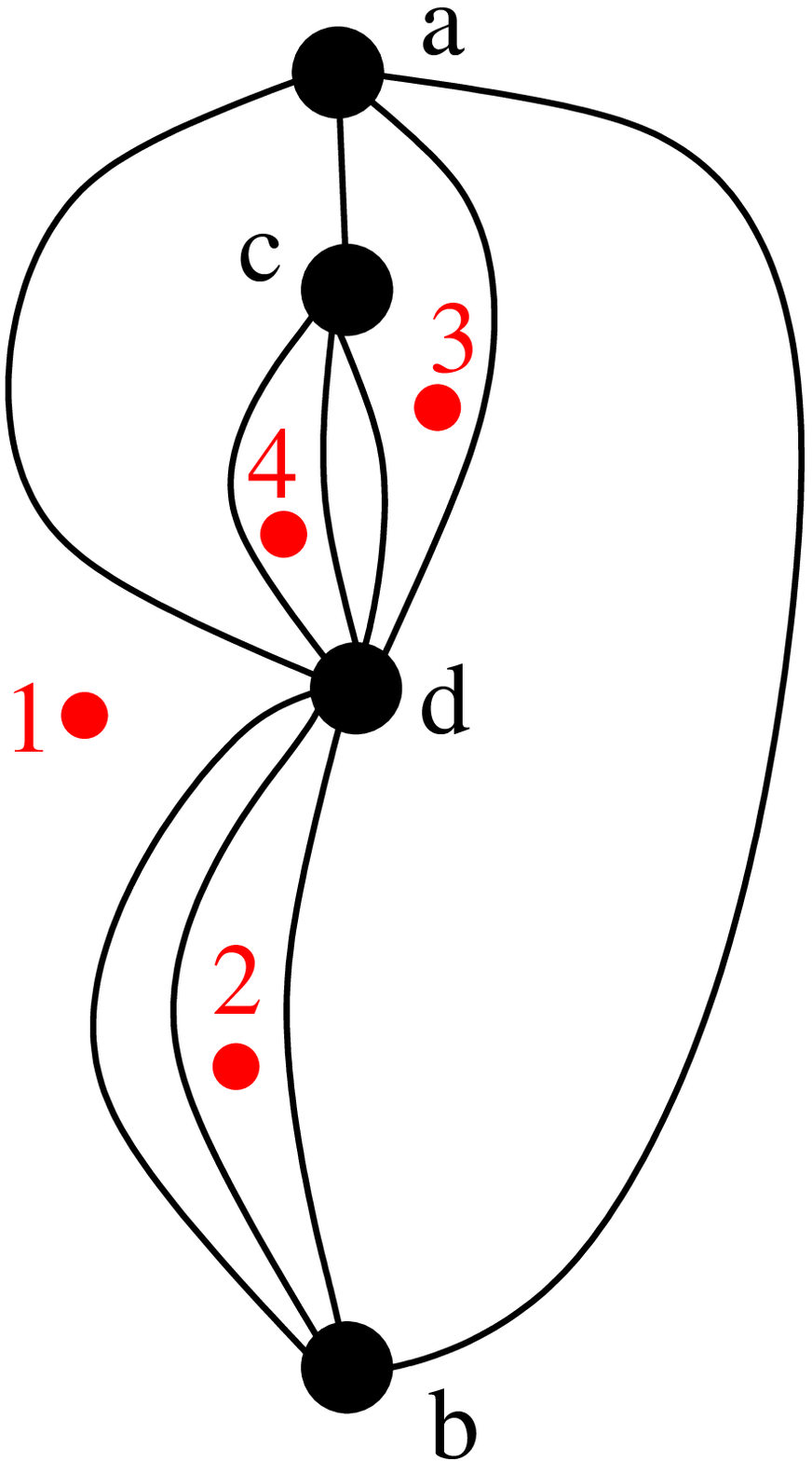,scale=0.2} \\
\a_3=(1\,4)_a(2\,3)_b(1\,3)_c(1\,2\,3\,4)_d&
\a_4=(3\,1)_a(2\,1)_b(3\,4)_c(1\,2\,3\,4)_d
\\ [0.2cm]
\end{array}$
\end{center}
 \begin{center}
\caption{
The figure shows the four contributing diagrams to 
(Note that in $\a_{3,4}$ two adjacent two-cycles
commute, but the two orderings are related by a simple color relabeling, so only one ordering should be counted as inequivalent.)
 These diagrams can be obtained from each other
by the channel-crossing procedure. 
} \label{extrem1}
\end{center}
\end{figure}

\begin{figure}[htbp]
\begin{center}
$\begin{array}{c@{\hspace{0.45in}}c}
 \epsfig{file=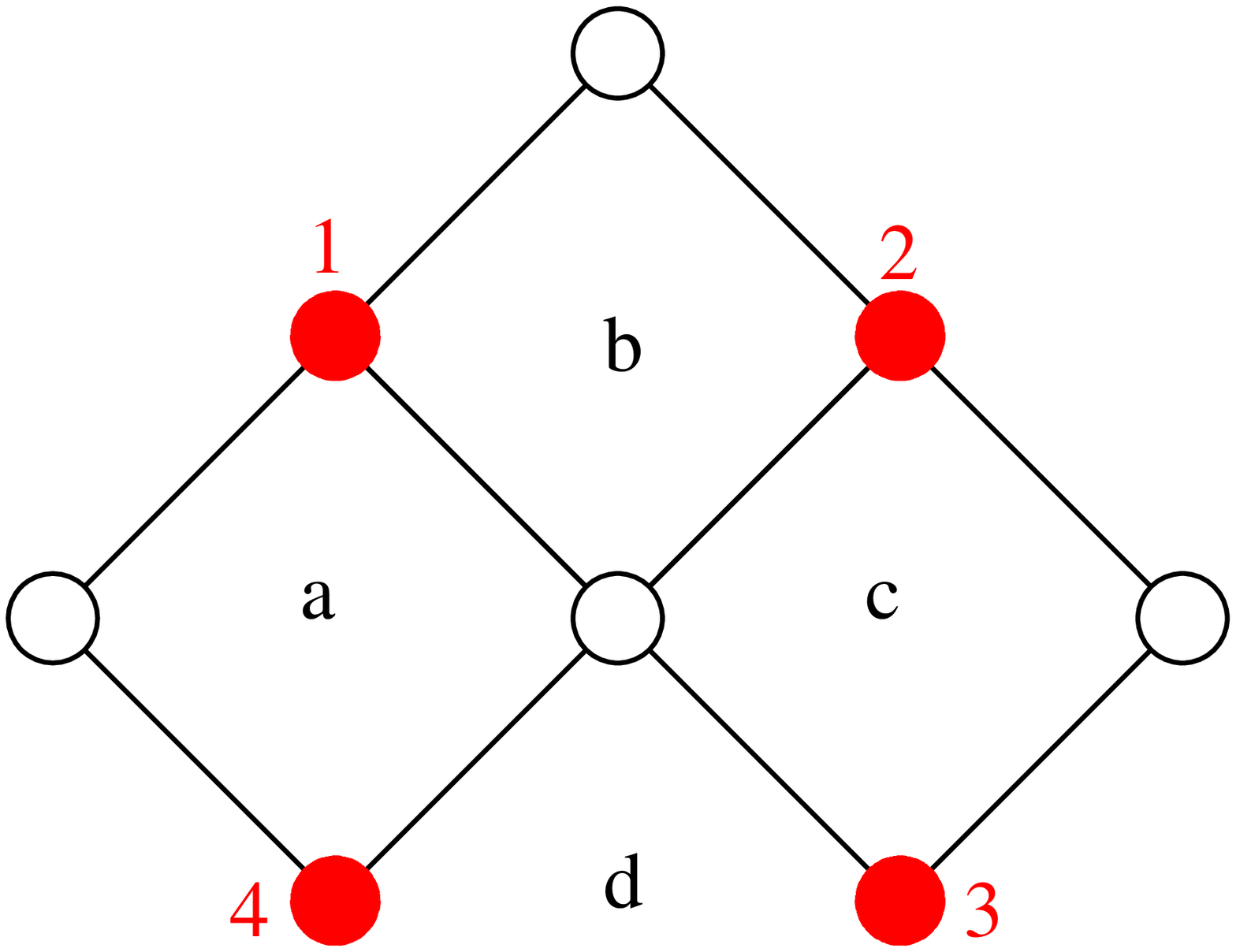,scale=0.21} & \epsfig{file=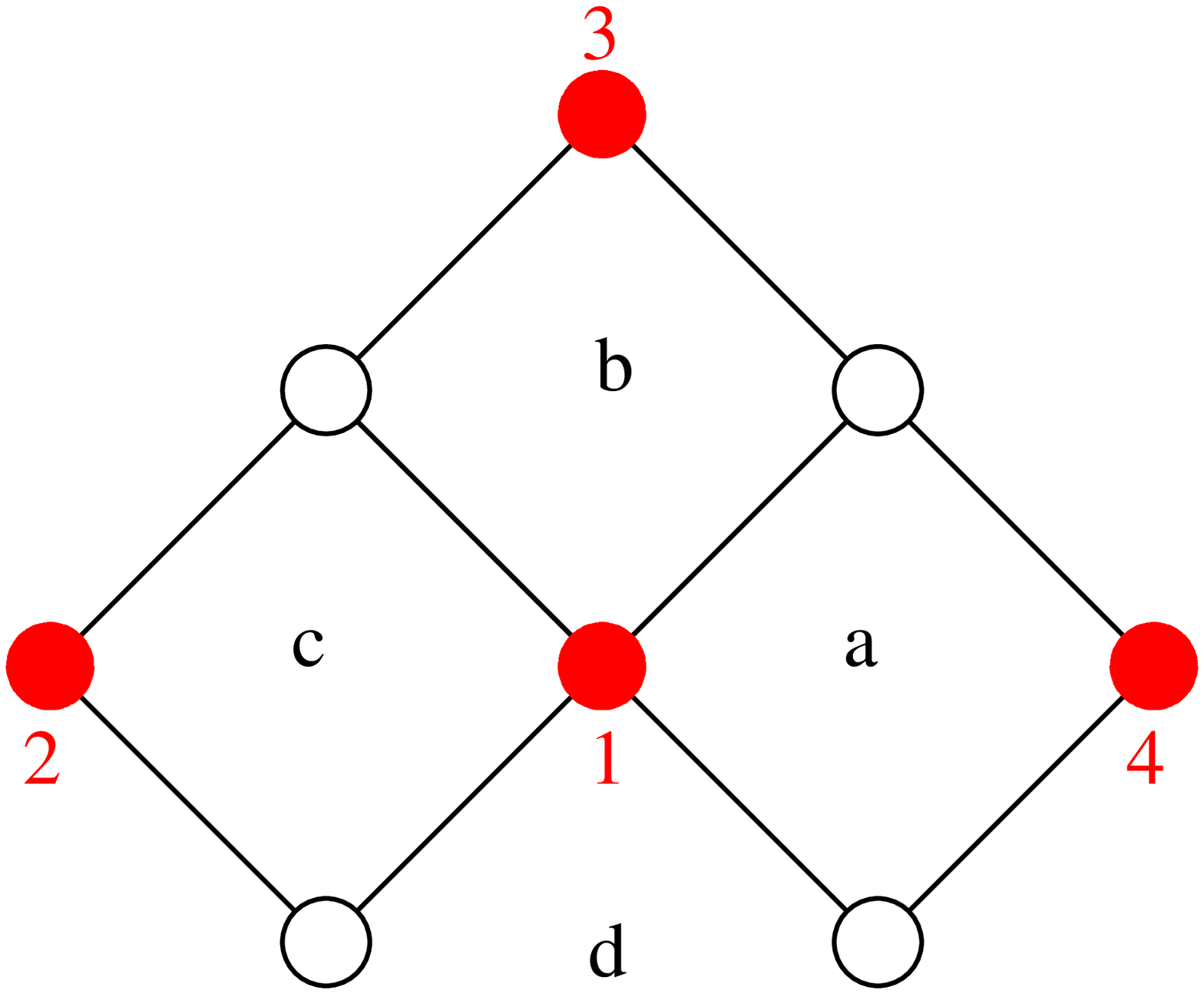,scale=0.21}\\
\a_1=(1\,4)_a(1\,2)_b(3\,2)_c(1\,2\,3\,4)_d &
\a_2=(1\,4)_a(1\,3)_b(2\,1)_c(1\,2\,3\,4)_d \\[0.1cm]
  \epsfig{file=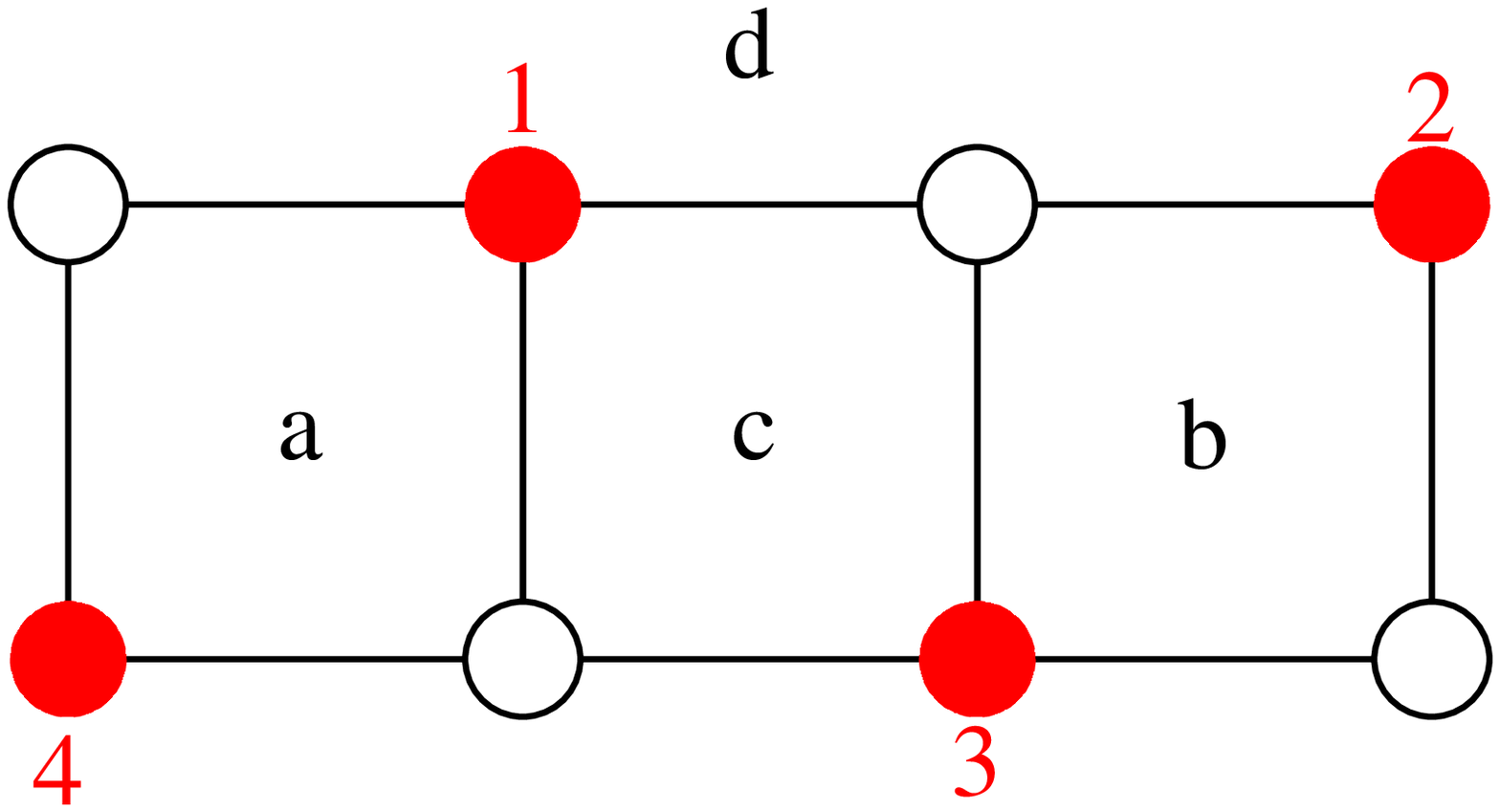,scale=0.21}& \epsfig{file=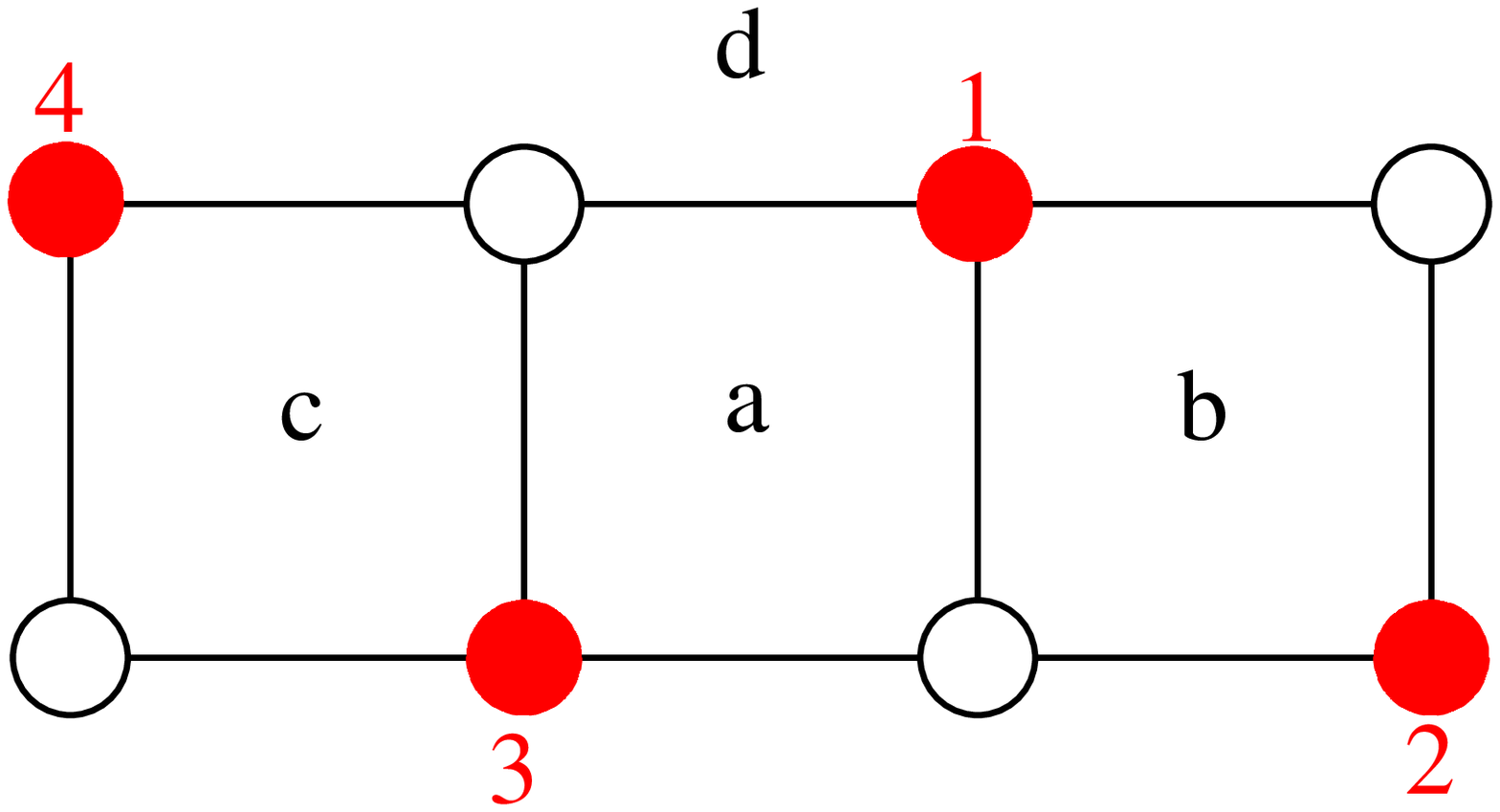,scale=0.21}  \\
\a_3=(1\,4)_a(2\,3)_b(1\,3)_c(1\,2\,3\,4)_d&
\a_4=(3\,1)_a(2\,1)_b(3\,4)_c(1\,2\,3\,4)_d
\\ [0.2cm]
\end{array}$
\end{center}
 \begin{center}
\caption{The graph-theoretic duals of the diagrams in Figure  15.
} \label{extremDual}
\end{center}
\end{figure}

Let us discuss  in complete detail the simple
polynomial correlator
\be
\langle\s_{[2]}(0) )\s_{[2]}(u)\s_{[2]}(1)\s_{[4]}(\infty)\rangle_{\gg=0} \, , \qquad |u| < 1 \,.
\ee 
We can easily draw all the different (four) diagrams contributing to this correlator following the Feynman rules of Section 2,
see  Figure \ref{extrem1}; the graph-theoretic dual diagrams are shown in Figure \ref{extremDual}. On the other hand,
we can work out explicitly the branched covering maps. We find
\be\label{2224ex}
 z(t; x) = t^2\frac{3t^2-4t(1+x)+6x}{2x-1} \, , \qquad v_{2224}(x)=\frac{(x-2)x^3}{1-2x} = u \,.
\ee 
For fixed $u$, there are four branched covering maps, corresponding to the four solutions
 to $u=v_{2224}(x)$,
\be\label{sols2224}
x_{\alpha\beta}&&=\\&&\half\left[1+\alpha\sqrt{1+2^{2/3}(u^2-u)^{1/3}}+\beta\sqrt{2-2^{2/3}(u^2-u)^{1/3}
+\alpha\frac{2-4u}{\sqrt{1+2^{2/3}(u^2-u)^{1/3}}}}\right],\nonumber
\ee where $\alpha,\beta=\pm$.

Changing perspective slightly, we can think of $x$
as parametrizing the ``moduli space''  of the maps $\{ z(t; x) \}$.\footnote{
For a discussion of moduli space of maps in the context of matrix string theory see~\cite{Bonelli:1998wx,Bonelli:1999qa}.}
 We write $x \in {\cal M}^{cover}$,
where in this case ${\cal M}^{cover}$ is the Riemann sphere.
 As we vary $u$ over the base sphere, the four roots $x_{\alpha \beta}(u)$ span
 ${\cal M}^{cover}$.
If we restrict to a given radial ordering
of the insertions, say specifying to $|u|<1$, the possible values of $x$ are restricted to a subspace of the moduli space,  $x\in v_{2224}^{-1}(|u|<1)\equiv {\mathcal M}^{cover}_{|u|<1}\subset
{\mathcal M}^{cover}$. 

We can now define a {\it cell decomposition} of ${\mathcal M}^{cover}_{|u|<1}$. In this example there are four cells,
spanned by the four roots $x_{\alpha \beta}(u)$ as we vary $u$. For a general polynomial correlator,
our construction associates to each
point of the moduli space  a unique diagram. For generic $x$,
 changing $x$ does not change the associated diagram as
the number of diagrams is finite: we can then define a top cell of  ${\mathcal M}^{cover}_{|u|<1}$ as a region associated 
to a particular diagram. 

\begin{figure}[htbp]
\begin{center}
\epsfig{file=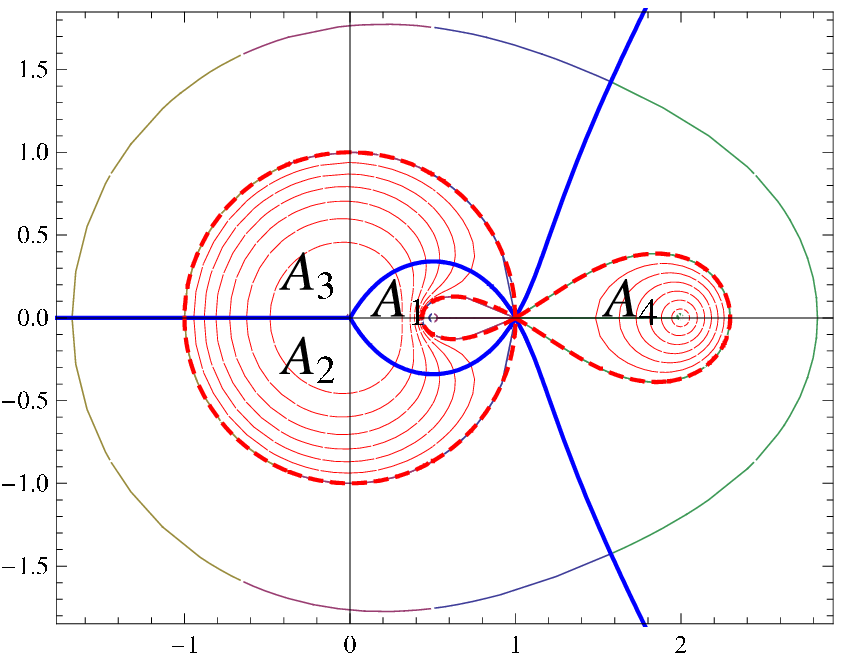,scale=1}
\caption{The structure of the moduli space ${\cal M}^{covering}_{2224}$
 (the $x$ sphere). The region in red is $v_{2224}^{-1} (\,|u|<1\,)$.
The blue lines delimit the different cells. The closed curves correspond to images of circles around $|u| = const$. 
} \label{moduli2224FF}
\end{center}
\end{figure}

 The cell decomposition of ${\mathcal M}^{cover}_{|v_{2224}|<1}$ is drawn in Figure \ref{moduli2224FF}.
Depicted  in this Figure is the $x$ sphere. The red region is given by ${\mathcal M}^{cover}_{|v_{2224}|<1}$.
 The point $u=0$ has two pre-images, $x=0^3$ and $x=2$. A $2\pi$ rotation around $x=2$ corresponds to $2\pi$ rotation around $z=0$. On the other hand,
a $2\pi$ rotation around $x=0$ corresponds to $6\pi$ rotation around $z=0$. The pre-images of $u=1$ are $x=1^3$ and $x=-1$, and the pre-images 
of $u=\infty$ are $x=\infty^3$ and $x=\half$.
 The blue lines delimit the different cells.

 The decomposition into cells can be understood  by looking at the monodromies of the solutions
(\ref{sols2224}) as $u$ goes around the point $z=0$, and at the corresponding 
channel-crossing operations on the diagrams. 
The goal is to associate  the four diagrams of Figures \ref{extrem1} and \ref{extremDual}, $\a_i$, with the four cells
of the moduli space (denoted by $A_1,A_2, A_3, A_4$ in Figure \ref{moduli2224FF}). 

First, consider the monodromies of the solutions. As  we encircle the point  $z=0$ with $u$
in region $A_4$, the solution goes back to itself; the other three cells are cyclically
permuted.  The ``monodromy'' structure of the regions of the moduli space is
\begin{equation} \label{mon}
A_1\to A_3\to A_2\to A_1,\qquad A_4\to A_4 \,.
\end{equation}
Next, we look at the channel-crossing operation on the diagrams.
Rotating $u$ around $z=0$ we get 
\be \label{diagrammon}
\a_1\to \a_3\to \a_2\to \a_1,\qquad \a_4\to\a_4.
\ee Explicitly starting with $\a_2$ we have
\be
\a_2&=&(1\,2\,3\,4)_d(1\,4)_a(1\,3)_b(1\,2)_c \to(1\,2\,3\,4)_d(1\,4)_a(2\,3)_c(1\,3)_b\to
(1\,2\,3\,4)_d(1\,4)_a(1\,2)_b(3\,2)_c=\a_1,\nonumber\\
\a_1&=&(1\,2\,3\,4)_d(1\,4)_a(1\,2)_b(3\,2)_c \to(1\,2\,3\,4)_d(1\,4)_a(1\,3)_c(1\,2)_b\to
(1\,2\,3\,4)_d(1\,4)_a(3\,2)_b(3\,1)_c=\a_3,\nonumber\\
\a_3&=&(1\,2\,3\,4)_d(1\,4)_a(2\,3)_b(1\,3)_c \to(1\,2\,3\,4)_d(1\,4)_a(1\,3)_c(1\,2)_b\to
(1\,2\,3\,4)_d(1\,4)_a(1\,3)_b(3\,2)_c=\a_2 \,.\nonumber\\ 
\ee
\begin{figure}[htbp]
\begin{center}
$\begin{array}{c@{\hspace{0.25in}}c@{\hspace{0.25in}}c}
 \epsfig{file=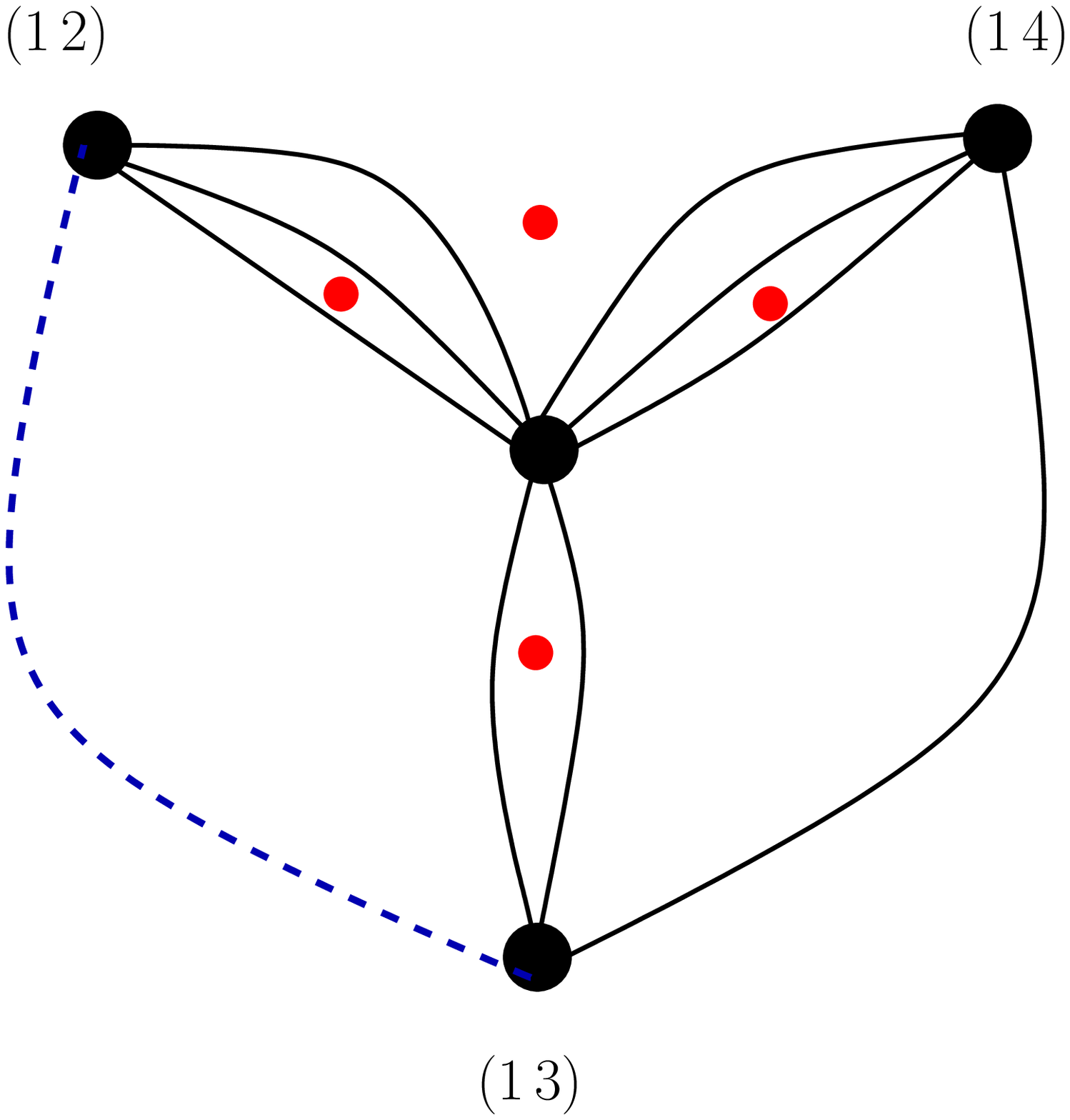,scale=0.25} & \epsfig{file=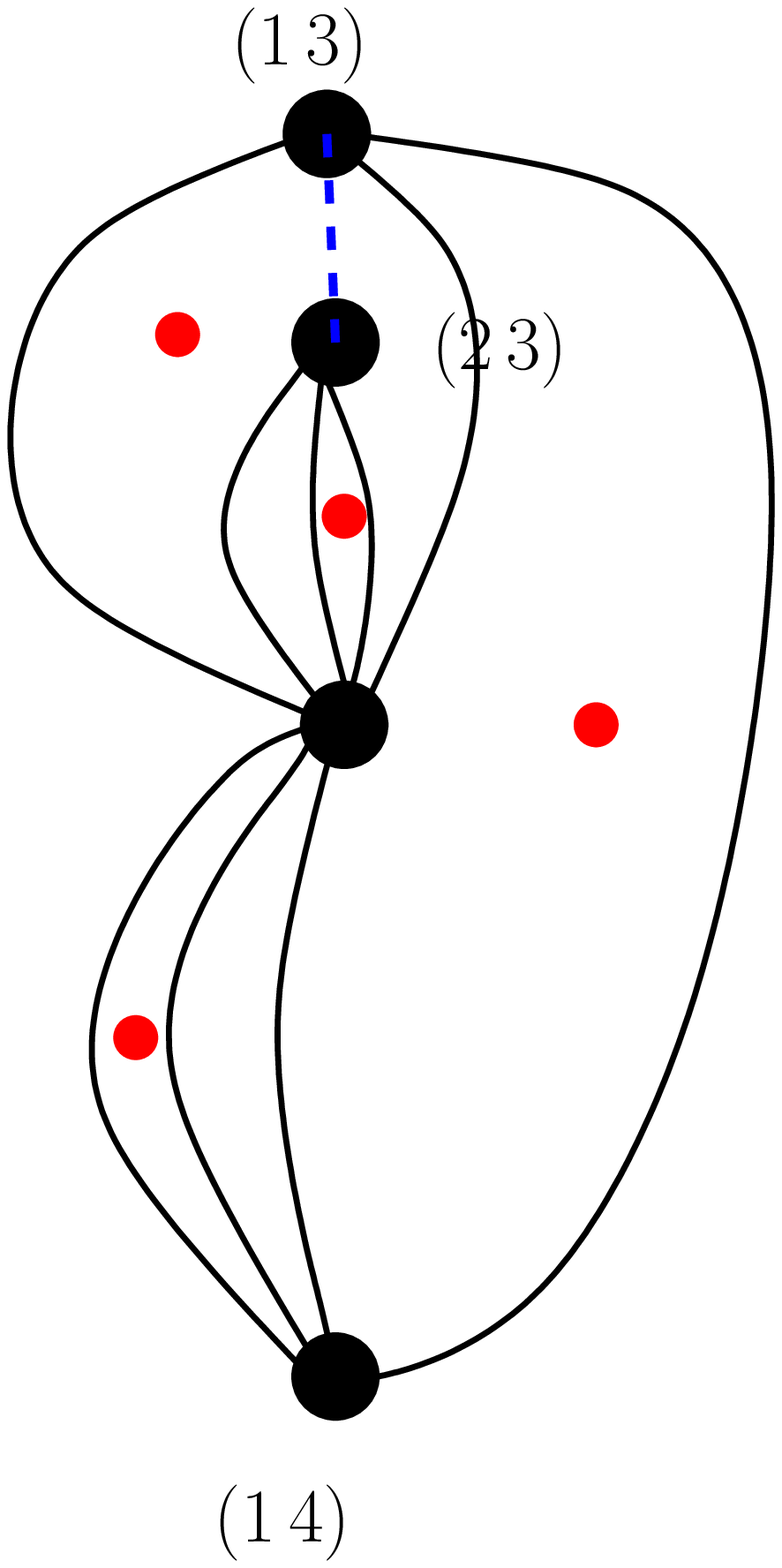,scale=0.25}&\epsfig{file=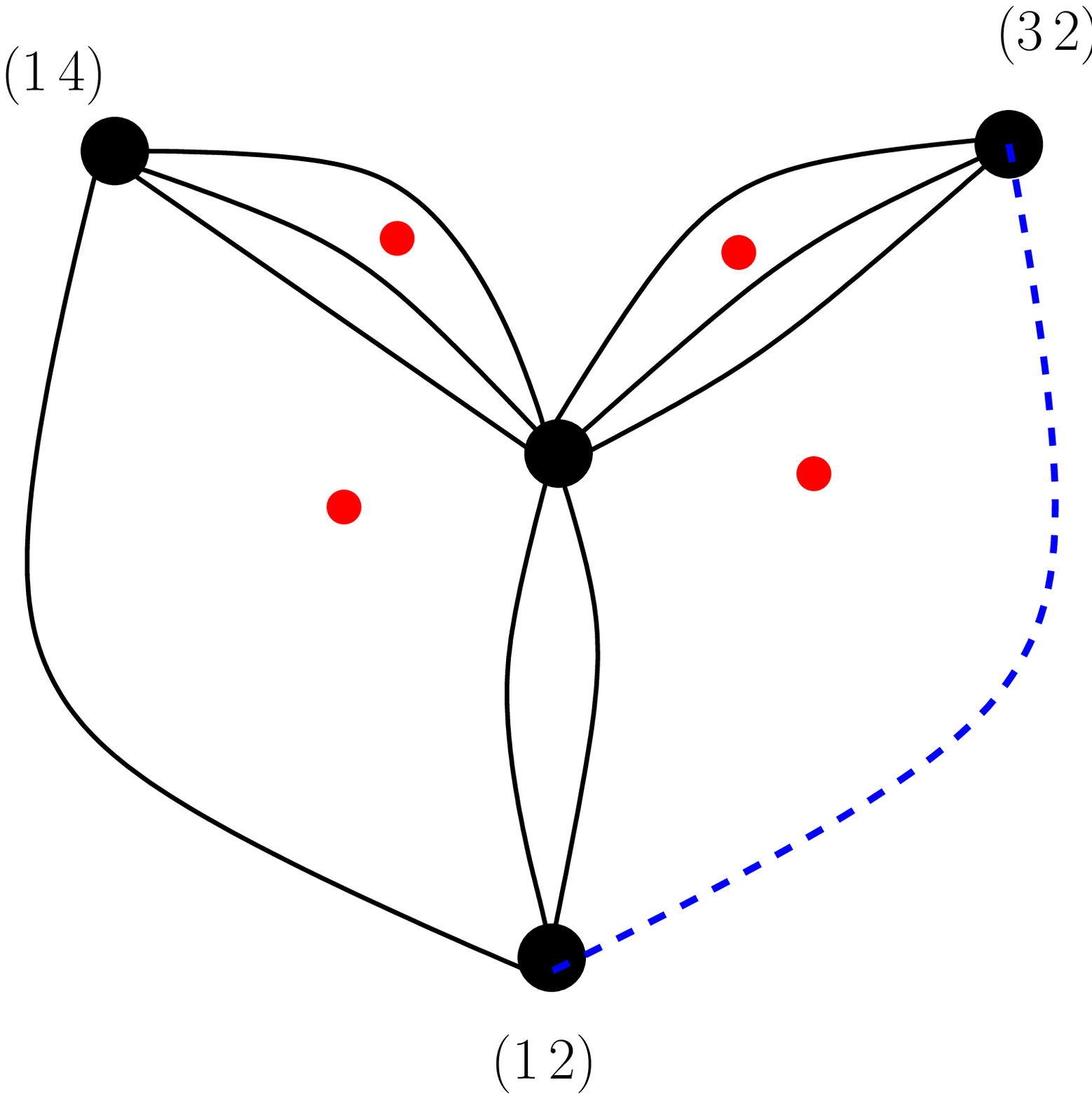,scale=0.25}\\
\end{array}$
\end{center}
\caption{channel-crossing of $\a_2$ to $\a_1$. The dashed blue line is the contracted propagator.} \label{chan2224} 
\end{figure}
 \begin{figure}[htbp]
\begin{center}
\epsfig{file=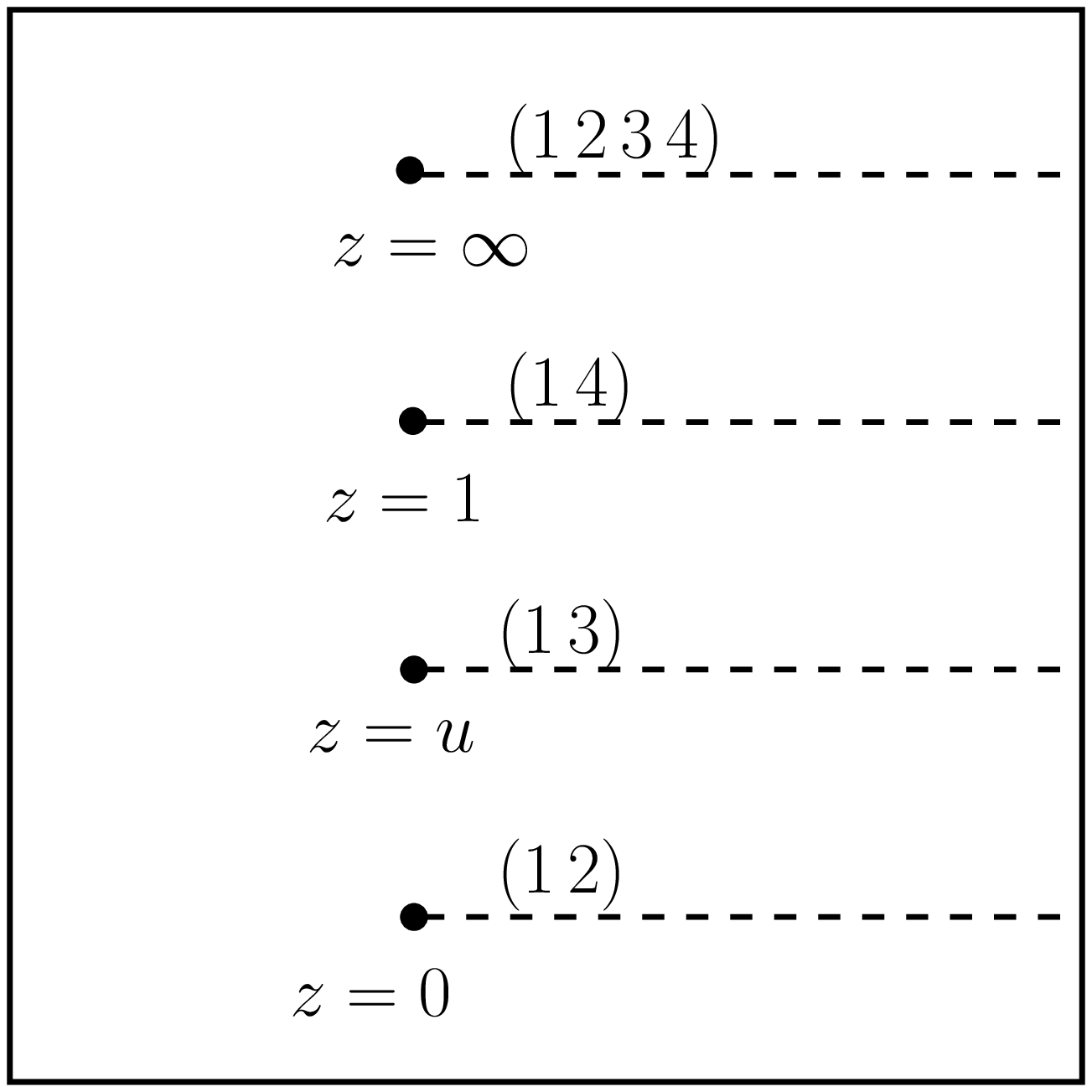,scale=0.45}
\caption{Convention for the cuts used in diagram $\a_2$.} \label{conven}
\end{center}
\end{figure} 	
Figures \ref{conven} and \ref{chan2224} illustrate the example of the channel-crossing between $\alpha_2$ and $\alpha_1$

We see that monodromy structure of the diagrams  (\ref{diagrammon}) is isomorphic to the monodromy structure of the maps (\ref{mon}).
 We must associate diagram $\alpha_4$ to region $A_4$, while diagrams $\alpha_1$, $\alpha_2$, $\alpha_3$ correspond
to regions $A_1$, $A_2$, $A_3$, up to an ambiguity which amounts to cyclic re-ordering of 1, 2 and 3 (the ambiguity could be
resolved following the conventions spelled out in Section 2.2).

A similar discussion applies to any polynomial  four-point correlator. In general the solutions of $v(x)=u$ cannot be found explicitly,
but the qualitative picture of ${\cal M}^{cover}$ can be understood by considering how the diagrams transform into one another through
channel-crossings. In the non-polynomial cases we have the additional parameter $q$ and to find the maps we have to fix both $q$
and $x$ by solving the two equations \eqref{3rel}. We can solve first for $q$ as a function of $x$ using \eqref{1rel}, and denote the solutions (which are a finite number)
as $q_i(x)$. Then, we insert  $q_i(x)$ in \eqref{2rel}  to obtain a finite number of equations
 of the form $v_k (x) = u$. We can think of
 ${\cal M}^{cover}$ for a general four point correlator as 
 consisting of several copies of the sphere, one for each equation $v_k (x) = u$,
 and we can repeat the discussion
above for each of the copies.

\subsection{Covering surfaces = dual worldsheets: a localization conjecture}

The genus expansion of correlators in a symmetric product orbifold is reminiscent
of the perturbative expansion of a closed string theory,  much like the genus (large $N$) expansion of a $U(N)$ gauge theory in the classic  analysis of `t Hooft \cite{'tHooft:1973jz}.
While historically the genus expansion of gauge theories was a motivation to search
for a dual string picture, for symmetric orbifolds 
 the duality with string theory came first \cite{Maldacena:1997re}.
Understanding systematically the genus expansion on the orbifold side of the duality was one of the motivations of this work.

An important difference between the cases of  $U(N)$ gauge theory and of symmetric product orbifold
 is that in the  latter the genus expansion does not quite correspond to the $1/N$ expansion.
 We have seen in Section 2.3 that while $1/N$ can be taken to leading order as the genus counting parameter,
 for given genus there is really an infinite sum over subleading powers of $1/N$.
 So it appears that the genus counting parameter on the dual string side
(the string coupling constant $g_s$)   should not be exactly identified with $1/\sqrt{N}$:
the relation $g_s \sim 1/\sqrt{N}$ is only valid to leading order for large $N$.
Instead the natural correspondence is between the genus expansion
 of the symmetric orbifold (as opposed to its large $N$ expansion) and the genus expansion
 of the dual string theory. 
 
 Thus we are led to directly identify the covering surfaces contributing
 to an orbifold correlator   with the {\it worldsheets} of the dual string theory.
 If this  is the correct dictionary, there should be a relation between
   the ``moduli space''   ${\cal M}^{cover}_{\gg, n}$  
of genus $\gg$ covering maps contributing to an $n$ point-correlator 
   and the familiar moduli space    ${\cal M}_{\gg, n}$ of genus $\gg$ Riemann surfaces
   with $n$ punctures, over which we are instructed to integrate to find the dual string amplitudes.
   This is particularly tempting for the genus zero contribution to  polynomial correlators, where both moduli spaces
   are the Riemann sphere. We may formally write
\be
G(u,\bar u)=\int_{{\cal M}^{cover}_{0,4}} d^2x\, {\mathcal F}(x;\,u, \bar u) \, .
\ee
For this expression to reproduce our algorithm,  the integration over $x$ should {\it localize} to the solutions of $u=v(x)$.
The conjecture is that if one were to evaluate the same amplitude on the dual string side, it
 would indeed localize to the solutions of $u=v(x)$. We may then literally identify the covering sphere $t$ as the worldsheet of the dual string,
and the different diagrams  with top cells of ${\mathcal M}_{0,4}$. 

Localization phenomena of this kind
are common in topological string theory, see {\it e.g.} \cite{Verlinde:1990ku, Distler:1989ax}. Recently a form of localization has been used in \cite{Belavin:2006ex} to compute
four-point correlators in minimal string theories (minimal models coupled to Liouville), which have been argued
to be closely related to the $AdS_3 \times S^3$ background with NSNS flux \cite{Rastelli:2005ph}. The symmetric orbifold 
${\rm Sym}^N {\cal M}_4$ lies at a very special point in the moduli space of string theory on  $AdS_3 \times S^3 \times {\cal M}_4$,
and it is indeed expected that this point would be ``topological'' in nature. In the related context of Gopakumar's approach to string duals of free field theories \cite{Gopakumar:2005fx}, 
a mechanism for the localization of worldsheet correlators
to points in moduli spaces was discussed in \cite{Razamat:2008zr}.

\section*{Acknowledgements}
We thank R.~Gopakumar for suggesting very useful references on ramified coverings.
SSR would like to thank the oragnizers of the Monsoon Workshop on String Theory and the HET group at
 the Weizmann Institute for hospitality during
different stages of this project. LR would like to thank the KITP, Santa Barbara and the Galileo Galilei Institute, Florence,
for hospitality during the completion of this work.
The work of AP was supported in part by DOE grant DE-FG02-91ER40688 and
NSF grant PHY-0643150. The work of LR and SSR is supported in part by DOE
grant  DEFG-0292-ER40697
and by  NSF grant PHY-0653351-001.  Any opinions, findings, and conclusions or recommendations
expressed in this material are those of the authors and do not necessarily reflect  the views of the National Science Foundation.

\appendix

\section{Deriving the polynomial map}\label{details}

In this Appendix we derive the map for the polynomial case discussed in Section \ref{extremcasesec}.

 From the local behavior of
the map near ramification points (\ref{b0}--\ref{branch-inf}) the derivative of the map is given by
\be
z'(y)&=&C\, (y+x)^{n_1-1}(y+x-1)^{n_2-1} y^{n_3-1}\\&=&C\, y^{n_3-1} \sum_{k=0}^{n_1-1}\sum_{l=0}^{n_2-1}
\left(
\begin{array}{c}
n_1-1  \\
k
\end{array}
\right)
\left(
\begin{array}{c}
n_2-1  \\
l
\end{array}
\right)
x^{n_1-1-k}(x-1)^{n_2-1-l}\, y^{k+l}\;,\nonumber
\ee where $y=t-x$. Integrating we get
\be
z(y)=
C\, \sum_{k=0}^{n_1-1}\sum_{l=0}^{n_2-1}\frac{1}{k+l+n_3}
\left(
\begin{array}{c}
n_1-1  \\
k
\end{array}
\right)
\left(
\begin{array}{c}
n_2-1  \\
l
\end{array}
\right)
x^{n_1-1-k}(x-1)^{n_2-1-l}\, y^{k+l+n_3}+v(x).\nonumber\\
\ee
We set $C$ by demanding $z(y=1-x)=1$,
\be
C^{-1}= \frac{1}{1-u(x)}\sum_{k=0}^{n_1-1}\sum_{l=0}^{n_2-1}\frac{ (-1)^{k+l+n_3}}{k+l+n_3}
\left(
\begin{array}{c}
n_1-1  \\
k
\end{array}
\right)
\left(
\begin{array}{c}
n_2-1  \\
l
\end{array}
\right)
x^{n_1-1-k}(x-1)^{n_2+n_3+k-1}.\nonumber\\
\ee Further, the relation between $x$ and $u$ is obtained by demanding  that $z(y=-x)=0$, 
\be
\centerline{$v(x)=$}\\=\frac{
\sum_{k=0}^{n_1-1}\sum_{l=0}^{n_2-1}\frac{(-1)^{k+l+n_3}}{k+l+n_3} 
\left(
\begin{array}{c}
n_1-1  \\
k
\end{array}
\right)
\left(
\begin{array}{c}
n_2-1  \\
l
\end{array}
\right)
x^{n_3+l}(x-1)^{-l}\,}
{\sum_{k=0}^{n_1-1}\sum_{l=0}^{n_2-1}\frac{(-1)^{k+l+n_3}}{k+l+n_3} 
\left(
\begin{array}{c}
n_1-1  \\
k
\end{array}
\right)
\left(
\begin{array}{c}
n_2-1  \\
l
\end{array}
\right)
\left[x^{n_3+l}(x-1)^{-l}-x^{-k}(x-1)^{n_3+k}\right]\,
}\nonumber
\ee

Specializing to $n_3=2$ we get a very simple expression for the derivative of $v(x)$,
\be\label{uofx}
\d_x v(x)=
(-1)^{n_2+1}n_1n_2\left(
\begin{array}{c}
n_1+n_2  \\
n_1
\end{array}
\right)\frac{ (x-1)^{n_2} x^{n_1}}{\left((n_1 + n_2)x-n_1\right)^2}.
\ee
Note that $v(x)$ is by itself a map to a sphere from a sphere with three ramification points, at $x = 0,\,1,\,\infty$,
with ramifications $n_1+1,\,n_1+n_2-1,\, n_2+1$ respectively. This map is called the {\textit{Belyi}} map in
the mathematical literature.

\section{Four-point functions from Lunin-Mathur}\label{luninmathur}
A general algorithm to obtain correlators of twist fields in a bosonic symmetric orbifold was discussed in
\cite{Lunin:2000yv} by Lunin and Mathur. These authors computed the correlators directly in the path integral formulation
of the theory by going to the covering surface and carefully taking into account the appropriate Liouville factor. 
In this Appendix we collect the results of \cite{Lunin:2000yv} for planar contributions to the four-point functions
 in a bosonic symmetric orbifold \eqref{bosaction} and discuss in detail a simple example. 

Given a four-point function with ramifications $n_1$ at $z=0$, $n_2$ at $z=1$, $n_3$ at $z=u$, $n_4$ at $z=\infty$,
we first compute the  genus zero branched  covering map, given as a ratio of two polynomials of order $d_1$ and $d_2$:
 $z(t)=\frac{P_{d_1}(t)}{Q_{d_2}(t)}$ (see Section \ref{heunsec}). We assume that $z=0$ has pre-image  $t=0$,  $z=1$ pre-image $t=1$,
$z=\infty$ pre-image $t=\infty$, and  $z=u$ pre-image $t=x$.
 Then we define
\be
a_0&=&\lim_{t\to 0} \frac{z(t)}{t^{n_1}},\qquad
a_1=\lim_{t\to 1} \frac{z(t)-1}{(t-1)^{n_2}},\qquad
a_u=\lim_{t\to x} \frac{z(t)-u}{(t-x)^{n_3}},\\
a_\infty&=&\lim_{t\to \infty} \frac{z(t)}{t^{n_4}} \, .\nonumber
\ee 
We denote by $t=q_i,\; i=1\dots d_2$ the zeros of the denominator,
which map to $z = \infty$.\footnote{Of course for polynomial maps
 there are no additional images of $z=\infty$.} At these points the  map behaves as
\be
z\sim \frac{C_i}{t-q_i}.
\ee With these notations in place, the
 four-point function (on the covering sphere) is given by 
\be\label{covcorrLM}
\ln G(x,\bar x)&=&-\frac{n_1-1}{12} \ln n_1a_0^{1/n_1} -\frac{n_2-1}{12} \ln n_2a_1^{1/n_2}
 +\frac{n_4-1}{12} \ln n_4a_\infty^{1/n_4}\\&&
-\frac{n_3-1}{12} \ln n_3a_u^{1/n_3}-\frac{1}{6}\ln \frac{{n_1}{n_2}{n_3}}{{n_4}}
-\frac{1}{6}\sum_{i=1}^{d_2}\ln C_i.\nonumber
\ee
To obtain the correlator we have to some over all the solutions $x_\a(u)$ to the equation $v(x)=u$, {\it i.e.} over
all the diagrams. We also have to appropriately normalize the operators as in Section \ref{largeNsec}. The final result is
\be 
G(u,\bar u)= \frac{\prod_{k=1}^4\sqrt{n_k(N-n_k)!}}{N!\,\left(N-\half(n_1+n_2+n_3+n_4)+1\right)!}\,
\,\sum_\a G(x_\a(u),\bar x_\a(u)) \, .
\ee

 Let us discuss in detail the example  $\langle \sigma_{[n]} (0) \sigma_{[2]} (u) \sigma_{[2]} (1) \sigma_{[n+2]}(\infty) \rangle$, which is a polynomial correlator. 
 The map as obtained in Appendix \ref{details} is given by
\be
v(x)=x^{1 + n} \frac{2 + n - n x}{(n+2) x-n },\quad
z(t; x)=t^n \frac
{n(n+1)\,t^2-n(n+2)(1+x)\,t+(n+2)(n+1)x}{(n+2) x-n}.\nonumber\\
\ee
 Computing the coefficients $a_i$ and plugging them into the general formula \eqref{covcorrLM} we get
\be \label{app_4p}
\ln G(x,\bar x)&=&-\frac{1}{12}\biggl[  \ln|1 - x| + \frac{-2 + n + n^2}{2n} \ln |x|  -
  \frac{-2 + n (2 + n)}{n (2 + n)}  \ln\left|n  - (2+n) x\right|+\nonumber\\
&&- \frac{1+n+n^2}{n} \ln (n+2) +\frac{ 1 + n^2}{n}\ln  n
+\frac{n^2+2n-2}{n(n+2)}\ln (n+1)+5\ln2\biggr].\nonumber\\
\ee In the OPE limit $u\to 0$ we have the following $n+2$ solutions to the equation $v(x)=u$,
\be \label{uxapp}
x\sim \left(-\frac{n}{n+2}\,u\right)^{\frac{1}{n+1}}, \qquad x\sim \frac{2+n}{n}+O(u).
\ee 
Note that \eqref{app_4p} is singular only for the first $n+1$ solutions and thus only these contribute to the
singular terms in this OPE limit. The contribution of each of these $n+1$ solutions to the four-point function is 
\be
\ln G(x_\a(u),\bar x_\a(u))&\sim&-\frac{1}{24}\biggl[ \frac{-2 + n + n^2}{n(n+1)} \ln |u|  +(1 +2 n +\frac{ 2}{1 + n}) \ln n-\\
  &&- (3 + 2 n + \frac{2}{1 + n})\ln (n+2) 
+(2 -\frac{2}{n} +\frac{ 2}{2 + n}) \ln  (1 + n)  \biggr] -\frac{5}{12}\ln 2.\nonumber
\ee
The expression for the un-normalized three-point functions in the  $(n+1)2(n+2)$ and $(n+1)2n$ cases as obtained in \cite{Lunin:2000yv} are 
\be
\ln|C_{n+1,2,n+2}|^2&=&
-\frac{8 + 7 n + 2 n^2}{24 (2 + n)}\ln(n+1)
+\frac{5+5n+2n^2}{24(n+1)}\ln(n+2)-\frac{5}{24}\ln 2.\nonumber\\
\ln|C_{n,2,n+1}|^2&=&
-\frac{3+n (3 +2 n)}{24 (1+n)}\ln n+\frac{2 +  n + 2 n^2}{24 n}\ln(n+1)-\frac{5}{24}\ln 2.
\ee
Combining the above results we see that
\be
\ln G(x_\a(u),\bar x_\a(u))&\sim&-2\biggl[\Delta_n+\Delta_2-\Delta_{n+1}\biggr] \ln |u|+\ln|C_{n+1,2,n+2}|^2+\ln|C_{n,2,n+1}|^2,
\nonumber\\
\ee where \be \Delta_n=\frac{1}{24}\left(n-\frac{1}{n}\right)\ee is the conformal dimension of an $n$-cycle.
To complete the calculation we have to take into account the normalization of the gauge invariant twist fields (see Section \ref{largeNsec}). For
the four-point function the normalization is
\be
{\mathcal N}_{n,2,2,n+2}=\sqrt{\frac{4n(n+2)(N-n)(N-n-1))}{N^2(N-1)^2}}
\ee For the three-point functions we get
\be
{\mathcal N}_{n+1,2,n+2}=\sqrt{\frac{ 2 (n+1) (n+2) (N-n-1) }{ N (N-1) } },\qquad
{\mathcal N}_{n,2,n+1}=\sqrt{\frac{ 2n (n+1) (N-n) }{ N (N-1) } }.\nonumber\\
\ee Thus we learn
\be
{\mathcal N}_{n+1,2,n+2}{\mathcal N}_{n,2,n+1}=(n+1)\,{\mathcal N}_{n,2,2,n+2}.
\ee Combining the above results and summing over the  roots \eqref{uxapp} we conclude that in the $u\to 0$ OPE limit we get the expected answer
\be
G(u,\bar u)=|u|^{-2\left[\Delta_n+\Delta_2-\Delta_{n+1}\right]} |\hat C_{n+1,2,n+2}|^2\, |\hat C_{n,2,n+1}|^2 \, ,
\ee where the hatted $C$s represent properly normalized three-point functions.

For the OPE limit to be consistent with the three-point functions it is important that we count
every map (or equivalence class, or diagram) exactly once.
In general we should expect agreement only at leading $1/N$ order but here we get an exact equality because
the correlator is polynomial and there are only planar contributions.

Let us just briefly mention the other OPE limits of \eqref{app_4p}. The single image of $u\to 0$
with $x\sim\frac{2+n}{n}$ corresponds to the $n$-cycle and the 2-cycle joining into 
a double-cycle (two-particle state). In this case the OPE limit is not singular. 
There are three images of $u\to 1$ corresponding to $x\to 1$ and
this corresponds to the two $2$-cycles joining to a 3-cycle. The single image of $u\to\infty$ satisfying
$x\to\frac{n}{n+2}$ corresponds to the $n+2$-cycle and $2$-cycle joining into a double-cycle consisting of a $2$-cycle and
an $n$-cycle. The $n+1$ images of the limit of $u\to\infty$ satisfying $x\to\infty$ correspond to the $n+2$-cycle
and the $2$-cycle joining to form an $n+1$ cycle.

\bibliography{h3bib}

\bibliographystyle{JHEP}

\end{document}